\documentclass{aa}  

\usepackage{graphicx}
\usepackage{hyperref}
\hypersetup{linkcolor=blue,citecolor=blue,colorlinks=true,pdfborderstyle={}}
\usepackage{natbib}
\usepackage{xcolor}
\usepackage{xspace}
\usepackage{xcolor}
\usepackage{amsmath}
\usepackage{amssymb}
\usepackage{multirow}
\usepackage{graphicx}
\usepackage{booktabs}
\usepackage{adjustbox}
\usepackage{subcaption}
\usepackage{enumerate}
\usepackage{enumitem}
\usepackage{float}
\usepackage{soul}

\bibpunct{(}{)}{;}{a}{}{,} 
\usepackage{txfonts}
\begin{document}

   \title{Multi-epoch leptohadronic modeling of neutrino source candidate blazar PKS 0735+178}


   \author{ A. Omeliukh
          \inst{1}\fnmsep\thanks{\email{omeliukh@astro.rub.de}}
          \and S.~Garrappa \inst{2,1}
          \and V.~Fallah Ramazani \inst{3,4,1}
          \and A.~Franckowiak \inst{1}
          \and W.~Winter \inst{5}
          \and E.~Lindfors \inst{3,6} 
          \and K.~Nilsson\inst{3} 
          \and J.~Jormanainen \inst{3,6,4} 
          \and F.~Wierda \inst{6} 
          \and A.~V.~Filippenko \inst{7}
           \and W.~Zheng \inst{7}
          \and M.~Tornikoski \inst{4} 
          \and A.~Lähteenmäki \inst{4,8} 
          \and S.~Kankkunen \inst{4,8} 
          \and J.~Tammi \inst{4} 
}

   \institute{
   Ruhr University Bochum, Faculty of Physics and Astronomy, Astronomical Institute (AIRUB),  Universitätsstraße 150, 44801 Bochum, Germany
         \and
    Weizmann Institute of Science, Department of Particle Physics and Astrophysics, Herzl Street 234, 76100 Rehovot, Israel
        \and
    Finnish Centre of Astronomy with ESO (FINCA), University of Turku, FI-20014 Turku, Finland
        \and 
    Aalto University Metsähovi Radio Observatory, Metsähovintie 114, 02540 Kylm\"al\"a, Finland
        \and
    Deutsches Elektronen-Synchrotron DESY, Platanenallee 6, D-15738 Zeuthen, Germany
        \and
    Department of Physics and Astronomy, University of Turku, FI-20014, Turku, Finland
    \and
    Department of Astronomy, University of California, Berkeley, CA 94720-3411, USA
    \and
    Aalto University Department of Electronics and Nanoengineering, P.O. BOX 15500, FI-00076 AALTO, Finland
  }

   \date{Received XX; accepted XX}

 
  \abstract
   {The origin of the astrophysical neutrino flux discovered by IceCube remains largely unknown. Several individual neutrino source candidates were observed. Among them is the gamma-ray flaring blazar TXS 0506+056. A similar coincidence of a high-energy neutrino and a gamma-ray flare was found in blazar PKS 0735+178.}  
   {By modeling the spectral energy distributions of PKS 0735+178, we expect to investigate the physical conditions for neutrino production during different stages of the source activity.}
   {We analyze the multi-wavelength data during the selected periods of time. Using numerical simulations of radiation processes in the source, we study the parameter space of one-zone leptonic and leptohadronic models and find the best-fit solutions that explain the observed photon fluxes. }
   { We show the impact of model parameter degeneracy on the prediction of the neutrino spectra. We show that the available mutli-wavelength data are not sufficient to predict the neutrino spectrum unambiguously. Still, under the condition of maximal neutrino flux, we propose a scenario in which 0.1 neutrino events are produced during the 50 days flare.
   }
   {}


   \maketitle
%

\section{Introduction}

The detection of a diffuse high-energy astrophysical neutrino flux \citep{doi:10.1126/science.1242856} started a new era of neutrino astronomy. However, the nature of the extragalactic neutrino flux remains unclear.

Blazars are a subclass of active galactic nuclei (AGNs) with a relativistic jet pointing close to the observer's line of sight. The relativistic boosting of the jet radiation and the overall high power of the jet emission make them prominent neutrino source candidates \citep[see, e.g., review by][]{2021Univ....7..492G}. The multi-wavelength spectral energy distribution (SED) of blazars typically shows a two-bump structure. While the low-energy bump likely originates from synchrotron radiation of relativistic electrons in the blazar jets, the origin of the high-energy bump is still debated, as contributions from both leptonic and hadronic processes are possible. 

Blazars can be probed as high-energy neutrino emitters in searches for temporal and spatial associations of blazar flares with high-energy neutrino events. The most promising neutrino blazar candidate is the blazar TXS~0506+056, which was in a gamma-ray flaring state during the arrival of a 300\,TeV neutrino detected by IceCube, resulting in a significance at the $3\, \sigma$ level \citep{2018Sci...361.1378I}. An additional flare of TeV neutrinos at the $3.5\sigma$ level was found from the direction of TXS~0506+056 during a gamma-ray quiet state \citep{2018Sci...361..147I}.

In addition to TXS~0506+056, IceCube has also detected high-energy events in spatial coincidence with other individual blazars of different classes, among which are PKS~1424-418 \citep{2016NatPh..12..807K, 2017ApJ...843..109G}, GB6~J1040+0617 \citep{2019ApJ...880..103G}, PKS~1502+106 \citep{Franckowiak_2020,Rodrigues_2021}, and
PKS~0735+178 \citep{10.1093/mnras/stac3607}. No statistical correlation between gamma-ray blazars and neutrinos was found in the IceCube analysis \citep{2017ApJ...835...45A, 2023ApJ...954...75A}. The correlations of neutrino events with blazar positions were claimed by \cite{2022ApJ...933L..43B, 2023arXiv230511263B} but disfavored by \cite{2023ApJ...955L..32B}. Additionally, a statistical correlation between neutrinos and radio blazars was claimed by \cite{2021ApJ...908..157P, 2023MNRAS.523.1799P} but was not confirmed by \cite{2023ApJ...954...75A}.


In early December 2021, multiple neutrino events detected by IceCube \citep{2021GCN.31191....1I}, Baikal-GVD \citep{2021ATel15112....1D},
the Baksan Underground Scintillation Telescope \citep{2021ATel15143....1P}, and KM3NeT \citep{2022ATel15290....1F} were in temporal and spatial coincidence with the largest ever observed flare of the blazar PKS~0735+178 in the gamma-ray band. The source was also flaring in the optical, infrared, ultraviolet, and X-ray bands. Among neutrino source candidates, PKS~0735+178 is the only source for which multiple neutrino events from different detectors were observed.

Blazar PKS~0735+17 was among the first sources to be designated as ``classical BL Lac'' \citep{Carswell1974}, owing to the fact that its spectrum does not display strong, broad emission lines. Therefore, the determination of the distance of the host galaxy is a challenge. \citet{Falomo2021}, assuming that the host galaxy belongs to a small group, estimated the redshift of PKS~0735+178 as $z\approx0.65$. \citet{Nilsson2012} performed deep optical imaging and derived $z=0.45\pm 0.06$ in agreement with the absorption redshift of 0.424 \citep{Carswell1974} within the uncertainties. We adopt $z=0.45$ as the redshift of PKS~0735+178 throughout this work. 

The possible neutrino emission from PKS~0735+178 was already discussed by \citet{10.1093/mnras/stac3607}, \citet{2023ApJ...954...70A}, and \citet{10.1093/mnras/stad3804}. The source emission in these works was modeled either during the flare in December 2021 that lasted over one month or based on all available data nonsimultaneous for this source. Multi-epoch leptonic modeling was performed by \cite{2024MNRAS.529.3503B} but does not characterize the source behavior in a multi-messenger context. In this work, we describe the multi-wavelength and multi-messenger emission from PKS~0735+178 during both quiescent and active states, which is motivated by the fact that the blazar activity is highly variable. This allows us to consider a scenario where physical conditions for neutrino production vary, not requiring sustaining the same extreme conditions (e.g., high proton injection power) over months of the enhanced source activity. 

The previous works that utilize highly advanced numerical modeling methods \citep[e.g.,][in addition to works already mentioned above]{2013ApJ...768...54B, 10.1093/mnras/stu2691, 2020ApJ...891..115P, 2019MNRAS.483L..12C, 2019MNRAS.489.4347O} usually propose only a single set of radiation model parameters to explain the blazar emission. Radiation models typically have a high number of free parameters with the possible degeneracy of some of them which is also often acknowledged in the literature but not studied in detail. In the current study, we aim to address this issue and explore the parameter space of one-zone leptonic and leptohadronic models in a more complete way than current state-of-the-art models. We apply a novel approach to search for the best-fit solutions.

The paper is structured as follows. Section \ref{sec:data} presents the available multi-wavelength and neutrino data used for this study. In Section \ref{sec:method} we describe the radiative models that were used to explain the multi-messenger emission as well as our method for searching the best-fit values. Section \ref{sec:results} provides the results of the modeling, which are further discussed and summarized in Section \ref{sec:discuss}. 

For the calculations in the paper, we adopt a flat $\Lambda$CDM cosmological model with parameters H$_0 = 70$ km s$^{-1}$ Mpc$^{-1}$ and $\Omega_{\textrm{m}} = 0.3$.

\section{Data collection and data analysis} \label{sec:data}

The detection of multiple neutrinos from the direction of PKS~0735+178 triggered a campaign of multi-wavelength observations of this source. We analyze the available data and build multi-wavelength light curves as well as SEDs for this source. 

\subsection{Neutrino emission}

On 8 December 2021 at 20:02:51.1 UTC, the IceCube real-time alert system \citep{Aartsen_2017} detected a 172 TeV track-like event, IceCube-211208A, with a 30\% probability to be an astrophysical neutrino \citep{2021GCN.31191....1I}. Four hours after the IceCube event, Baikal-GVD reported the detection of a 43~TeV cascade event with a probability of $\sim$ 50\% of being an astrophysical neutrino \citep{2021ATel15112....1D}. Additionally, four days prior to the IceCube event, the Baksan Underground Scintillation Telescope detected a muon neutrino with energy larger than 1 GeV \citep{2021ATel15143....1P} consistent with the location of PKS~0735+178 and the directions of the neutrinos detected by IceCube and Baikal-GVD.

\begin{figure}[htbp!]
\centering
 \includegraphics[width=6cm]{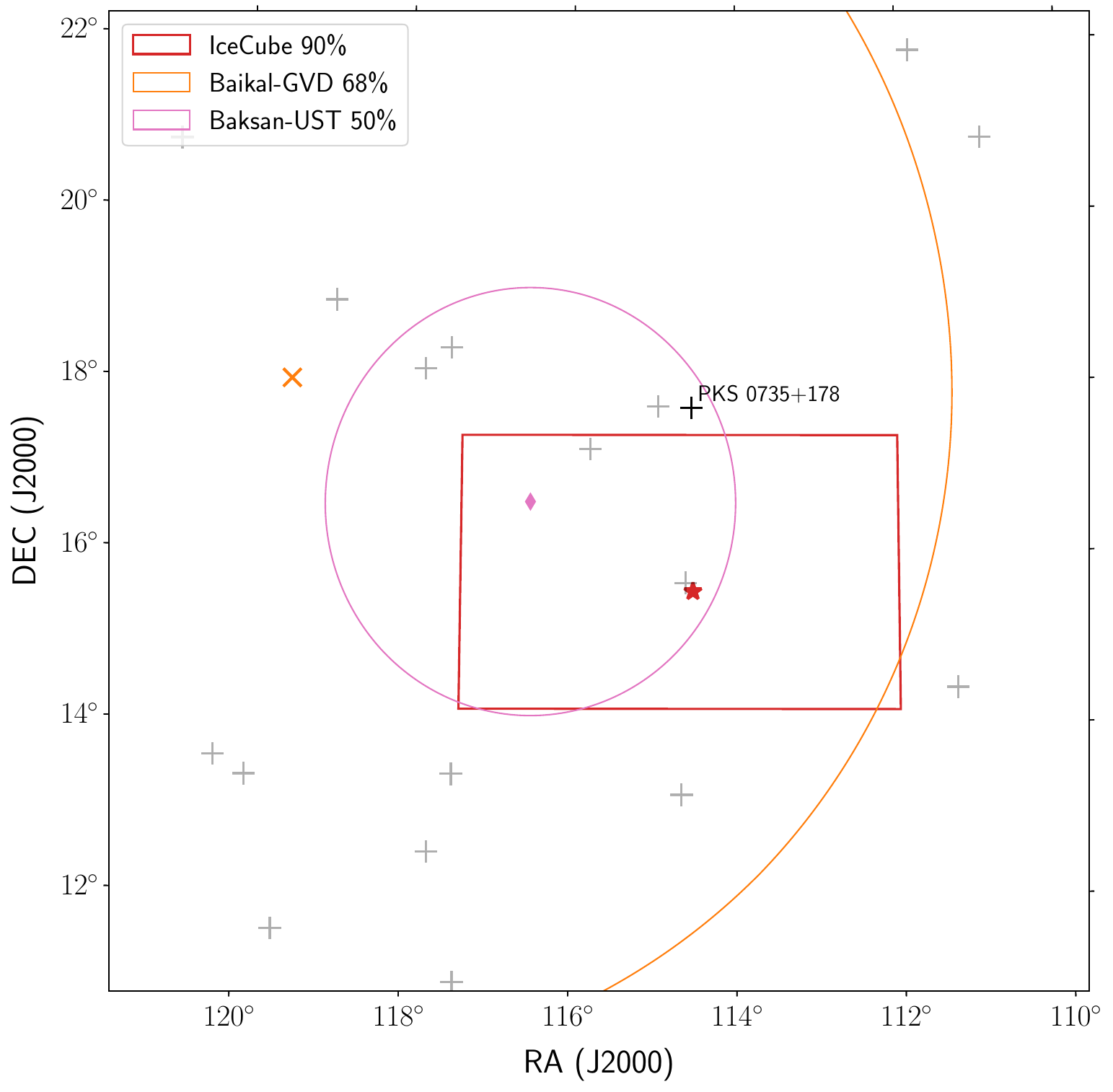}
  \caption{Localization of neutrino arrival directions and the position of PKS 0735+178. The gray markers indicate the position of \textit{Fermi}-LAT sources in the 4FGL.}
     \label{fig1:neutrino_contours}
\end{figure}

Figure \ref{fig1:neutrino_contours} shows the spatial uncertainty contours of the reconstructed neutrino detections in all three neutrino telescopes. Blazar PKS~0735+178 lies just outside of the IceCube-211208A 90\% uncertainty region. The neutrino still can be associated with the blazar owing to the possible existence of IceCube systematic uncertainties and the fact that 10\% of the real counterparts are expected to be outside of it as argued by \citet{10.1093/mnras/stac3607}.

The KM3NeT neutrino detector found no neutrinos from the direction of PKS~0735+178 within a time window of one day before and after the IceCube event, but an additional search over the entire month of December 2021 revealed an 18 TeV neutrino event in ARCA, a component designed for the detection of high-energy neutrinos, with a weak association ($p$-value 0.14) and no neutrinos in ORCA, a component targeted to low-energy neutrinos \citep{2022ATel15290....1F}. 

\subsection{Broad-band light curves}

The multi-wavelength light curves of PKS~0735+178 from 2006 to 2022 are illustrated in Fig. \ref{fig2:lightcurves}. In December 2021, the source showed the highest activity in gamma rays over the time of its monitoring with \textit{Fermi}-LAT. Simultaneously with the gamma-ray flare, the source was flaring in optical and X-ray bands. 


\subsection{Spectral energy distributions}

Based on the source activity in the gamma-ray band (see Section \ref{subsec:gamma}), we define the following states that will be used for further studies: quiescent state, beginning of the flare (neutrino arrival), flare peak, and end of the flare. The dates of the selected time periods are shown in Table \ref{table1:time_periods}. The corresponding SEDs were composed based on the observations made within those periods.
   \begin{table}
      \caption[]{Selected time periods used for the SED generation.}
         \label{table1:time_periods}
     \[
         \begin{array}{p{0.25\linewidth} p{0.35\linewidth} p{0.3\linewidth}}
            \hline
            \noalign{\smallskip}
            Blazar state     &  Date & MJD\\
            \noalign{\smallskip}
            \hline
            \noalign{\smallskip}
            Quiescent & Jan. 23 -- Feb. 2, 2010   & 55219 -- 55233  \\
            Neutrino arrival       & Dec. 8 -- 11, 2021 & 59556 -- 59559\\
           flare peak & Dec. 17 -- 19, 2021  &  59565 -- 59567        \\
            Post flare     &  Dec. 25, 2021 -- Jan. 6, 2022&  59573 -- 59600         \\
            \noalign{\smallskip}
            \hline
         \end{array}
    \]
   \end{table}

\subsubsection{Gamma rays} \label{subsec:gamma}
We use the gamma-ray observations of PKS 0735+178 from the Large Area Telescope (LAT) on board of the \textit{Fermi} satellite collected between August 2008 and May 2022 presented by \citet{Garrappa2024}. The bottom panel of Figure \ref{fig2:lightcurves} shows the adaptively binned light curve of the gamma-ray flux integrated between 100 MeV and 1 TeV. The continuous all-sky coverage of \textit{Fermi}-LAT shows a detailed picture of the temporal activity of the source in the 14 years, with the flaring activity temporally coincident with the arrival of IC211208A being by far the brightest activity observed in gamma rays. During this period, from the Bayesian Blocks \citep{ScargleBB} representation of the light curve in \citet{Garrappa2024}, the shortest significant ($\geq$ 2$\sigma$) variability timescale can be derived as the minimum of the quantity (e.g., \citealt{2019ApJ...877...39M})

\begin{equation}
    t_{{\rm var},ij} = \frac{F_i + F_j}{2}\left \| \frac{t_i - t_j}{F_i - F_j} \right \| ,
\end{equation}
\noindent
where $F_{i}$ and $F_{j}$ are the fluxes in the bins at the times before ($t_{i}$) and after ($t_{j}$) each block edge that denotes a significant variation. We find the shortest variability timescale at $\sim 14.4$ days. This timescale will be considered to constrain the radius of the emission region in Sec. \ref{sec:method}.

In addition, we use the gamma-ray SEDs from \citealt{Garrappa2024} calculated in different time windows selected from the \textit{Fermi}-LAT light curve. Three periods have been chosen with respect to the bright gamma-ray flaring activity of the source: the first time window is simultaneous with the IceCube neutrino arrival (MJD 59556--59559), the second one includes the peak of the gamma-ray flare (MJD 59565--59567), and the third one is during part of the decaying phase of the flare (MJD 59573--59600). As comparison to the period of flaring activity, we consider also the quiescent state SED from MJD 55219-55233 (see also Table \ref{table1:time_periods}).

VERITAS and H.E.S.S. observations shortly after the IceCube neutrino detection put upper limits on the very-high-energy gamma-ray emission from PKS 0735+178 \citep{2023ApJ...954...70A}.

\subsubsection{X-rays}

\begin{figure*}[h]
\centering
 \includegraphics[width=\textwidth]{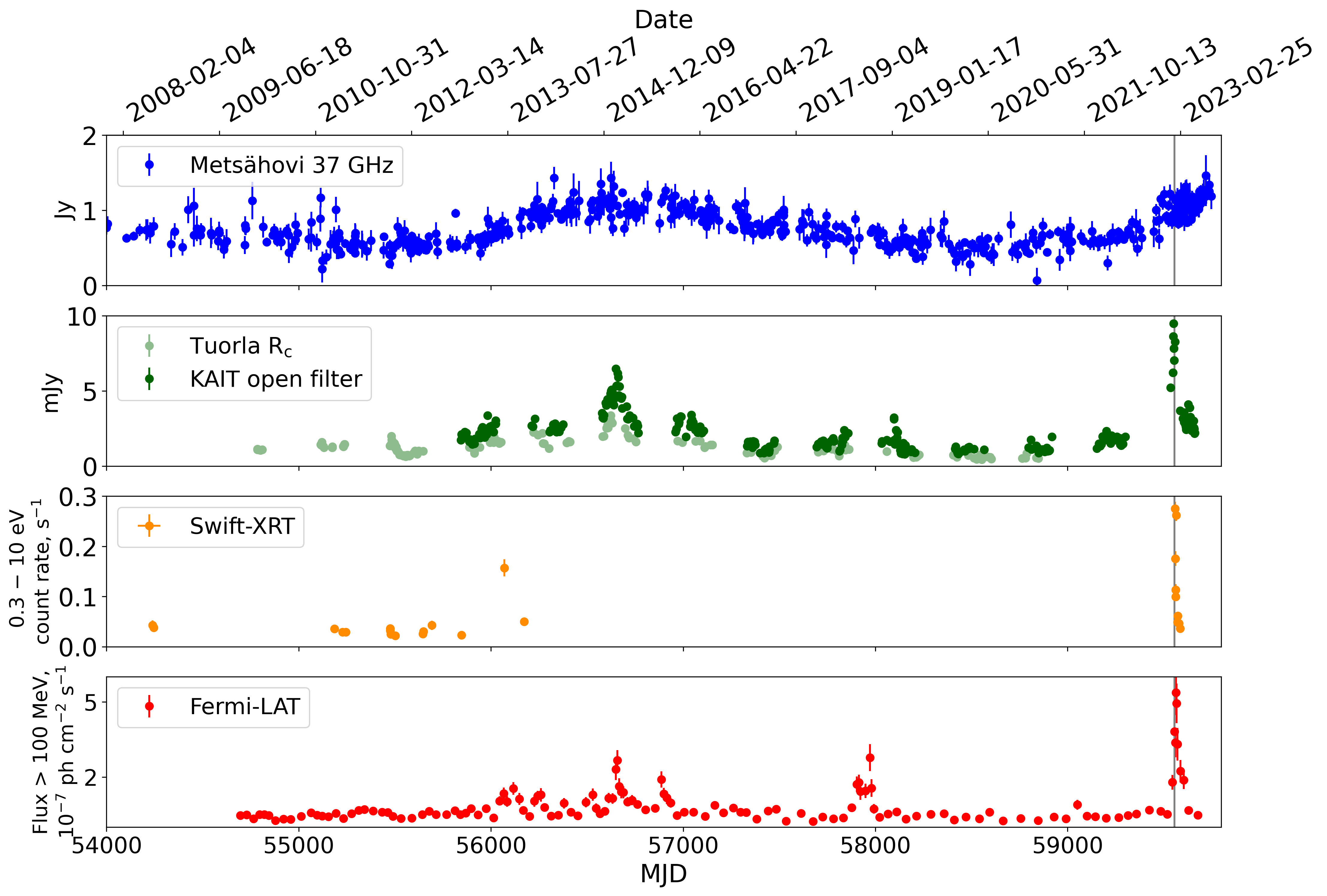}
  \caption{The multi-wavelength light curves of PKS 0735+178 from 2006 to 2022. The gray vertical line corresponds to the time of the IceCube neutrino detection. The light curves show that the source was in an exceptionally high state in all bands during the neutrino arrival.}
     \label{fig2:lightcurves}
\end{figure*}

The X-ray Telescope \citep[XRT;][]{2004SPIE.5165..201B} onboard the \textit{Neil Gehrels Swift Observatory (Swift)} observed the source 24 times between 20 December 2009 and 6 January 2022 in photon-counting (PC) mode. The multi-epoch event lists were downloaded from the publicly available SWIFTXRLOG (\textit{Swift}-XRT Instrument Log)\footnote{\url{https://heasarc.gsfc.nasa.gov/W3Browse/swift/swiftxrlog.html}}. Following the standard \textit{Swift}-XRT analysis procedure described by \citet{2009MNRAS.397.1177E}, the data were processed using the configuration described by \citet{2017A&A...608A..68F} for blazars and assuming a
Galactic column density of hydrogen of $4.42\times10^{20}$\,cm$^{-2}$ reported by \citet{2013MNRAS.431..394W}. The spectra of each observation were fitted using the maximum-likelihood-based statistic for Poisson data \citep[Cash statistics;][]{1979ApJ...228..939C}. The X-ray integral fluxes between 0.3 and 10\,keV of these observations are shown in Fig.~\ref{fig2:lightcurves}.

No spectral variability was observed within the \textit{Swift}-XRT data between 31 January 2010 and 17 February 2010 (observation ID 00090099001 and 00090099002). These two observations were combined to produce an average X-ray spectrum of the source during the quiescent state.  Similarly, no spectral variability was observed between 25 December 2021 and 6 January 2022 (observation ID 00036372023, 00036372024, and 00036372025). These three observations were combined to produce an average X-ray spectrum of the source after the flaring state. In order to produce the X-ray spectra for the periods of the neutrino arrival and the gamma-ray peak (as defined above), the \textit{Swift}-XRT data on 10 December 2021 (observation ID 00036372014) and 17 December 2021 (observation ID 00036372021) were used.

The X-ray spectrum above 2 keV from NuSTAR observation on December 11, 2021 was taken from \citet{2023ApJ...954...70A}.

\subsubsection{Ultraviolet and Optical}

Simultaneously with \textit{Swift}-XRT observations, PKS 0735+178 was observed with the UVOT instrument \citep{2005SSRv..120...95R} onboard \textit{Swift}. The source was observed in the $W1$ band on 31 January 2010, and in the $U$, $B$, $W1$, $M2$, and $W2$ bands on 10, 17, 25, 30 December 2021 and 6 January 2022. In addition, it was observed with the $V$ filter on 6 January 2022. The data were reduced using the analysis pipeline from the Space Science Data Center (SSDC\footnote{\url{https://www.ssdc.asi.it/}}).

PKS 0735+178 has been monitored in the optical $R$ band as part of the Tuorla blazar monitoring program\footnote{\url{http://users.utu.fi/kani/1m}} since 2005. The monitoring observations were performed using a 35\,cm Celestron telescope coupled to the KVA (Kunglinga Vetenskapsakademi) telescope located on the Canary Island of La Palma, Spain. Data analysis was performed using standard procedures with the semi-automatic pipeline developed in Tuorla \citep{nilsson_kva}.

The optical light curve is complemented by data obtained with the 76~cm Katzman Automatic Imaging Telescope (KAIT) as part of the Lick Observatory Supernova Search \citep[LOSS;][]{2001ASPC..246..121F} in the $Clear$ (open) band \citep[close to the $R$ band; see][]{Li_2003}. All images were reduced using a custom pipeline\footnote{\url{https://github.com/benstahl92/LOSSPhotPypeline}} detailed by \cite{2019MNRAS.490.3882S}. Several nearby stars were chosen from the Pan-STARRS1\footnote{\url{http://archive.stsci.edu/panstarrs/search.php}} catalog for calibration; their magnitudes were transformed into Landolt magnitudes 
using the empirical prescription presented by Equation 6 of \cite{2012ApJ...750...99T}. The light curve is shown in Figure \ref{fig2:lightcurves}.

For the quiescent state SED, we used the average $R$-band flux on MJD 55231, 55232, and 55233 where no significant variability was detected in the data obtained by the Tuorla blazer monitoring. During the neutrino arrival time, the optical measurement by KAIT on MJD 59559 was used for the SED modeling. For the SED modeling of the gamma-ray flare peak, the optical and near-infrared (NIR) data are obtained from \citet{2021ATel15136....1L}. Finally, for the post-flare period, the NIR data from \citet{2021ATel15148....1C} and optical data from KAIT on MJD 59585 and 59600 were used.

The observed UV, optical, and NIR fluxes are corrected for Galactic extinction using the values from \citet{2011ApJSchlafly}. The contribution of the host-galaxy flux in UV, optical, and NIR bands is negligible when compared with the statistical uncertainty of the flux measurements \citep{2012A&A...547A...1N}.


\subsubsection{Radio}

The Mets{\"a}hovi radio telescope (located at Kirkkonummi, Finland), operating at 37\,GHz, has been observing the source for three decades. We selected radio data obtained after mid-2006 for our study. The instrument and data-reduction procedures are described by \citet{1998A&AS..132..305T}. The long-term radio light curve is presented in Figure \ref{fig2:lightcurves}. Additionally, we use MOJAVE \citep{2018ApJS..234...12L} measurements of the 15 GHz radio flux from two days after the arrival of IceCube-211208A.

\section{Method} \label{sec:method}

\subsection{Simulation framework of one-zone models}

Within the one-zone framework, we assume that the observed photon and neutrino fluxes originate from the emission of highly energetic particles inside a compact zone (blob) in the jet. Radio fluxes cannot be explained within the one-zone model framework since the compact region  is optically thick to low-frequency radio emission owing to synchrotron self-absorption. In addition, the typical size of the emission region obtained from the gamma-ray variability timescales is two to four orders of magnitude smaller than the size of the radio core which is the dominant emitter of the radio flux. For our modeling, radio fluxes will be treated as upper limits.

We utilize the open-source code AM$^3$ \citep{klinger2023am3} for simulating the radiation processes and interactions undergone by relativistic  electrons and protons. AM$^3$ numerically solves the system of coupled differential equations that describes the evolution of the particle spectra, in a fully time-dependent and self-consistent manner. In this paper we produce spectra using a steady state approximation. We set characteristic escape time equal to light-crossing time for all particles. We evolve the kinetic equations over several escape timescales to ensure that the steady state is reached. The absorption of high-energy gamma rays due to extragalactic background light (EBL) is accounted for in all models, based on \cite{2011MNRAS.410.2556D}.

\subsection{Search for the best solutions}

Self-consistent modeling of the radiation processes in one-zone models utilizes a high number of free parameters (usually between seven and fourteen). Some of the parameters can be tentatively constrained from the observables, such as an upper limit on the size of the emission region (from the gamma-ray variability timescale) or the blob Lorentz factor (from radio observations). The radiation power of the jet is fundamentally limited by the Eddington luminosity. Still, in most cases, those constraints leave large regions of possible parameter values and the problem of finding the best-fit values arises. 

The radiation processes are highly sensitive to the values of the model parameters. A small change in a parameter value can result in an SED whose fluxes differ by orders of magnitude or has a drastically different shape due to a different radiation process starting to contribute significantly. Thus, the goodness-of-fit is not a smooth function of the model parameters, which introduces challenges to numerical differentiation. Additionally, common algorithms based on gradient descent methods encounter performance problems in high-dimensional parameter spaces \citep{doi:10.1080/10556780512331318290}.

We propose here a simple and effective two-step algorithm for searching for the best-fit solutions. As a first step, we generate a large number of different SEDs with values of the model parameters that are equally spaced between their boundary values. When choosing an equal number of points for each parameter, we get $N^m$ different models, where $m$ is the number of model parameters (dependent on the model) and $N$ is the number of points per dimension (dependent on the available computational resources). This gives a rough localization of the best solutions. As a second step, we locally minimize the goodness-of-fit value within a narrow region in the parameter space. Since we know that all neighbor points of the current best-fit solution have a lower goodness-of-fit value, we adopt their parameter values as new boundaries for searching of the new best fit. We apply the \texttt{migrad} and \texttt{symplex} algorithms from \texttt{Minuit} \citep{1975CoPhC..10..343J, iminuit} for this local minimization because the parameter space is now better constrained, ensuring the convergence of the algorithms.

A search of the parameters with algorithms like Markov Chain Monte Carlo (MCMC) or genetic algorithm typically requires simulating $10^6$--$10^7$ SEDs for finding one best-fit solution. The proposed grid scan requires only a single generation of a set of SEDs (in the source frame) which can be reused for different sources, thus using the computational resources only to recalculate the goodness-of-fit. The local minimization is computationally effective as well, requiring only $\sim 200$--300 SED generations. The disadvantage of the method is an exponential increase in the required computational resources when the number of model parameters increases.

\subsection{Leptonic models} \label{lep_model}

We start with the simplest models, which assume that all radiation from the NIR to gamma rays originates solely from leptonic processes.  

We assume that electrons are pre-accelerated to a simple power-law spectrum\footnote{Parameters with or without prime refer to the values in the jet or observer's frame, respectively.} $dN/d{\gamma'}_\mathrm{e} = N_0 {\gamma'}_\mathrm{e}^{-\alpha_\mathrm{e}}$ with spectral index $\alpha_\mathrm{e}$, spanning a range of Lorentz factors from  ${\gamma'}_\mathrm{e}^\mathrm{min}$ to ${\gamma'}_\mathrm{e}^\mathrm{max}$. The energy spectrum of the electrons is normalized to the total electron luminosity parameter $L'_{\mathrm{e}}$ as $L'_{\mathrm{e}} = 4/3 \pi R^3 m_{\mathrm{e}} c^2 \int_{{\gamma'}_\mathrm{e}^\mathrm{min}}^{{\gamma'_\mathrm{e}}^\mathrm{max}} \gamma'_\mathrm{e} dN/d{\gamma'}_\mathrm{e} d\gamma$. These particles are then injected into a single spherical blob of size $R_{\textrm{b}}'$ (in the comoving frame of the jet) moving along the jet with Lorentz factor $\Gamma$, where they encounter a homogeneous and isotropic magnetic field of strength $B'$. As electrons are continuously injected into the radiation zone, they lose energy via synchrotron cooling before leaving the blob, thus a break in the spectral index of electrons occurs self-consistently. The gamma-ray emission is produced in the inverse Compton scattering with the synchrotron photons being a target photon field. We assume the jet is observed at an angle $\theta_{\textrm{obs}} = 1/\Gamma_{\textrm{b}}$ relative to its axis, resulting in a Doppler factor of $\delta_{\textrm{D}}=\Gamma_{\textrm{b}}$.

To explore the seven-dimensional parameter space of the leptonic models, we do a simple grid scan probing 10 points per dimension resulting in 10 million simulated models. The parameters of the leptonic models and the boundaries of the parameter space are given in Table \ref{tab:lep_parspace}.

We find the best-fit parameters in two steps. First, for each of the four SEDs (each corresponding to a different period of blazar activity)  we evaluate the reduced $\chi^2$ using the 10  million simulated models. It is calculated as

  \begin{equation} \label{eq:chi2}
 \chi^2 =\frac{1}{N-N_\mathrm{par}+1} \sum\limits_i\frac{(F_i^\mathrm{data}-F_i^\mathrm{model})^2}{\sigma_i^2} ,
\end{equation}  
\noindent
where $N$ is the number of data points, $N_{\mathrm{par}}$ the number of free parameters in the model, $F^\mathrm{data}$ are the observed fluxes, $F^\mathrm{model}$ are the predicted fluxes by the model, $\sigma$ is the observed flux measurement error, and $i$ is the summation index which corresponds to the observed frequency values. The models that predict higher flux values than \textit{Fermi}-LAT sensitivity-based upper limits are rejected because the model fluxes are expected to significantly decrease in the corresponding energy range. 
 
For the blazar states during the 2021 flare, an additional constraint on the radius of the emission region applies:

\begin{equation}
    R_{\textrm{b}}' \; \leq \; \frac{\delta_{\textrm{D}} \; \mathrm{c} \; t_{\mathrm{var}}}{1+z} .
    \label{var_eq}
\end{equation}
\noindent
We use the value of $t_{\rm var}$ = 14.4 days for the 2021 flare from Sect. \ref{subsec:gamma}, and the values of $R_{\textrm{b}}'$ and $\delta_{\textrm{D}}$ are free model parameters. 

We select the best-fitting model for a particular SED based on the value of its reduced $\chi^2$. Afterward, we locally minimize the reduced $\chi^2$ with the parameter boundaries corresponding to the neighbor parameter values on the grid to  obtain a final solution. If the values of $R_{\textrm{b}}'$ and $\delta_{\textrm{D}}$ do not satisfy Eq. \ref{var_eq}, the model is rejected.

\subsection{Leptohadronic models} \label{method_lephad}

In an alternative scenario, we assume that both electrons and protons are pre-accelerated in the source to power-law spectra  $dN/d{\gamma'}_\mathrm{e,p} = N_{0_{\mathrm{e,p}}}{\gamma'}_\mathrm{e,p}^{-\alpha_\mathrm{e,p}}$ with spectral indices $\alpha_\mathrm{e,p}$, spanning a range of Lorentz factors from  ${\gamma'}_\mathrm{e,p}^\mathrm{min}$ to ${\gamma'}_\mathrm{e,p}^\mathrm{max}$. The energy spectra of the electrons and protons are normalized to the corresponding total electron and proton luminosities, $L'_{\mathrm{e}}$ and $L'_{\mathrm{p}}$ defined as as $L'_{\mathrm{e,p}} = 4/3 \pi R^3 m_{\mathrm{e,p}} c^2 \int_{{\gamma'}_{\mathrm{e,p}}^\mathrm{min}}^{{\gamma'}_{\mathrm{e,p}}^\mathrm{max}} \gamma'_\mathrm{e,p} dN/d{\gamma'}_\mathrm{e,p} d\gamma$. Similarly to the leptonic case, electrons and protons undergo interactions and radiate inside of the spherical blob of size $R'$ with a homogeneous and isotropic magnetic field of strength $B'$ moving along the jet with Lorentz factor $\Gamma$.

The application of the same two-step algorithm (grid scan followed by a local minimization) described in Section \ref{lep_model} would now require significantly more computational resources, since the number of model parameters increased from seven to ten.  

To circumvent this issue, we follow the approach adopted by \cite{rodrigues2023leptohadronic, 2024arXiv240606667R} in which IR, optical, UV, and gamma-ray fluxes are first fit with a purely leptonic model followed by the selection of hadronic parameters to fit the X\nobreakdash-rays. Most of the contribution to optical and GeV gamma-ray emission typically originates from leptons, even in the models where protons are added to the emission zone, as shown by leptohadronic models for several neutrino candidate blazars \citep[e.g.,][]{2019MNRAS.483L..12C, Rodrigues_2021, 10.1093/mnras/stac3607}. We fit first IR, optical, UV, and gamma rays via the grid scan method (using the same boundaries as in Table \ref{tab:lep_parspace}) and locally minimize them using \texttt{Minuit}. X-ray fluxes at this step are treated as upper limits. Afterward, we add protons to the AM$^3$ simulation. All leptonic parameters are fixed to the previously found values, while the four hadronic parameters are varied according to the grid probing $10^4$ combinations. The boundaries of parameters in the hadronic parameter space used for the grid scan are shown in Table \ref{tab:had_parspace}. Adding $L_p=0$ to the parameter space ensures a possible convergence to the purely leptonic solutions. As a final step, all ten parameters of the leptohadronic models are locally minimized with \texttt{Minuit} allowing for slight variations (with the boundaries being parameter value $\pm20\%$) in each of them.

\begin{figure*}[htpb!]
\centering
\includegraphics[width=0.8\textwidth]{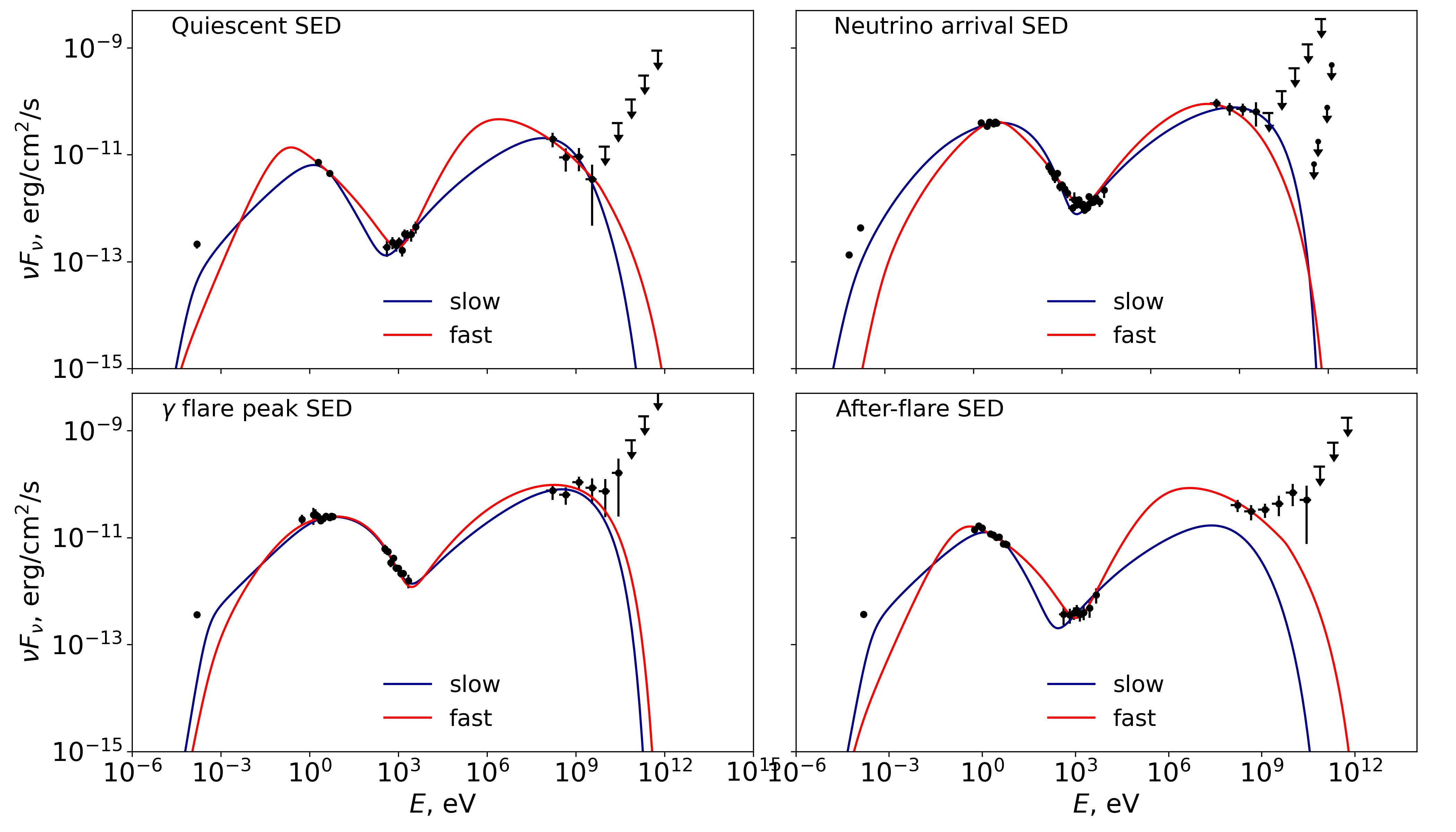}
  \caption{Purely leptonic best fits for the SEDs of PKS 0735+178 during the selected periods. Slow and fast solutions are named according to their Lorentz factors and correspond to two models with different parameters. The parameters are given in Table \ref{table:lep_pars}}.
     \label{fig:lep_models}
\end{figure*}

\section{Results} \label{sec:results}

We apply the above described method of the analysis of the parameter space of a one-zone model to the SEDs of PKS 0735+178 during the selected periods of observations.

\subsection{Purely leptonic fits}

When analyzing the reduced $\chi^2$ values resulting from the grid scan, we conclude that many different combinations of parameters yield similar goodness-of-fit values for a particular dataset. We select two physically different solutions for each SED and explore the parameter space around them to illustrate model parameter degeneracy. The number of solutions that provide a similarly good fit to the observed data can be higher, especially if the boundaries of the parameters are expanded beyond what was selected for this study.

The results of the leptonic modeling are shown in Fig.\ref{fig:lep_models}. The corresponding parameter values are presented in Table  \ref{table:lep_pars}. We show two best-fit results that correspond to distant regions in the parameter space also implying physically different solutions. For each epoch, we refer to the two different solutions as ``slow'' and ``fast'' based on their Lorentz factors. The slow solutions have Lorentz factor values between two and six, while for the fast solution those values span between twelve and eighteen. The different values of the magnetic field, electron luminosities, and Lorentz factors lead to the different shapes of the SED peaks. Still, both slow and fast solutions explain the observed data well, with the values of reduced $\chi^2$ being 1.4 and 1.1, respectively. In the period of neutrino arrival, the fast solution is characterized by a substantially higher magnetic field (4 G compared to 0.7 G for the slow model), a softer electron energy spectrum, a smaller emission-region size, and a lower electron luminosity. Despite the fact that model parameters lie in the different regions of the parameter space, both models explain well the observed photon fluxes (the major contribution to the large value of $\chi^2$/n.d.f. comes from the five UV data points with $\sim 1\%$ error bars). Similar differences between slow and fast solutions are observed for the gamma flare peak SED. For the post-flare SED, on the contrary, our method failed to explain the gamma-ray fluxes. Both slow and fast solutions fit optical, UV, and X-rays. While the slow solution produces a significantly lower gamma-ray flux than observed, the fast solution matches the flux level but fails to reproduce the correct shape.


\begin{table*} 
\centering
\caption{Best-fit leptonic model parameters.}
\begin{adjustbox}{width=0.9\textwidth}
    \begin{tabular}{l cccc cccc}
\toprule
 & \multicolumn{4}{c}{Slow solutions} & \multicolumn{4}{c}{Fast solutions} \\
\cmidrule(lr){2-5} \cmidrule(lr){6-9}
Parameters     & quiescent   & neutrino arrival  & $\gamma$ flare peak & post flare & quiescent   & neutrino arrival  & $\gamma$ flare peak & post flare  \\
\midrule
$\log_{10}$(R$'_{\mathrm{b}}$[cm])       & 16.94 & 16.91 & 16.67   & 16.96 & 15.86 & 15.22  & 15.57  & 15.45  \\

$B'$ [G] & 0.79 & 0.65 & 2.66   & 0.60 & 0.40 & 4.44  & 2.27  & 0.72  \\

$\Gamma_{\mathrm{b}}$   & 2.85 & 6.21 & 3.08   & 5.98 & 16.08 & 16.02  & 12.05  & 18.49  \\
$\log_{10}\gamma'^{\textrm{min}}_{\mathrm{e}}$   & 3.90 & 3.13 & 3.60   & 3.30 & 3.06 & 3.37  & 3.00 &  3.00 \\
$\log_{10}\gamma'^{\textrm{max}}_{\mathrm{e}}$   & 4.78 & 4.37 & 4.38   & 4.68 & 5.08 & 4.50  & 4.11  & 4.62  \\

$\alpha_{\textrm{e}}$   & 3.45 & 0.81 & 1.23   & 2.50 & 3.15 & 2.72  & 1.16  & 2.79  \\

$\log_{10}L'_{\textrm{e}}$[erg/s]& 44.79 & 44.15 &  45.29 &  43.67&  42.62& 42.47 &43.07& 42.51 \\

$\chi^2$/n.d.f.& 1.4 & 9.6 &  2.0&  3.8 &  1.1& 10.8 &2.2& 1.4 \\
\bottomrule
\end{tabular}
\end{adjustbox}
\tablefoot{Parameters: R$'_{\mathrm{b}}$ is the radius of the blob, $B'$ is the magnetic field strength in the emission region, $\Gamma_{\mathrm{b}}$ is blob Lorentz factor; $\gamma'^{\textrm{min}}$ and $\gamma'^{\textrm{max}}$ are the minimum and maximum Lorentz factor of the electrons respectively, $\alpha_{\textrm{e}}$ is the spectral index, $L'_{\textrm{e}}$ is electron luminosity, and $\chi^2$/n.d.f. is a value of the reduced $\chi^2$ function}
    \label{table:lep_pars}
\end{table*}

\begin{figure*}[htpb!]
\centering
 \includegraphics[width=0.75\textwidth]{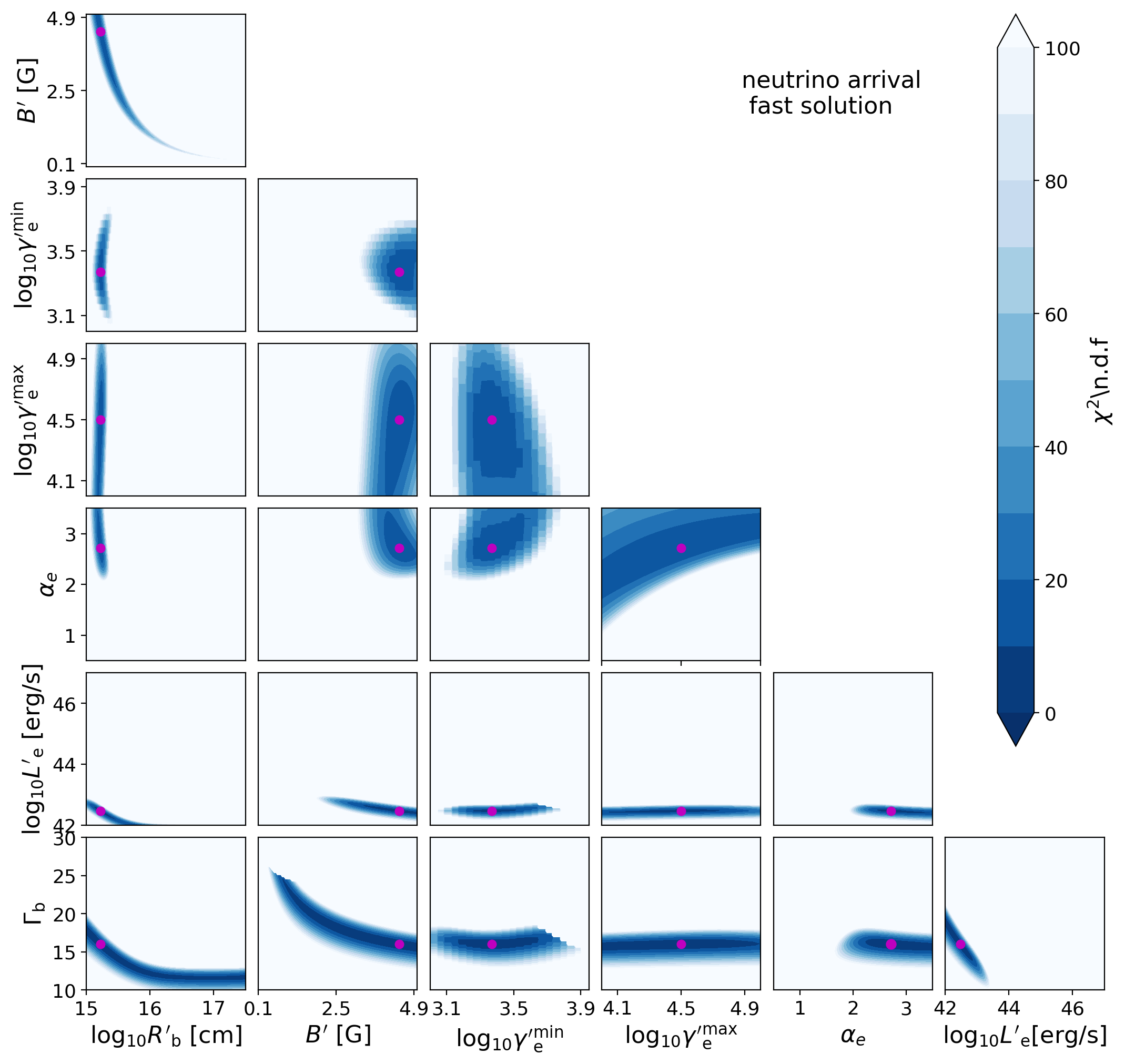}
  \caption{Values of the reduced $\chi^2$ around the best-fit fast solution for the neutrino arrival SED. Plots for other models and SEDs are shown in Appendix \ref{appB}.}
\label{fig:sed2_fast_cornerplots}
\end{figure*}

Small deviations in values of certain model parameters can still keep the model valid. To explore how the change of each pair of parameters influences the goodness of fit (the rest of the parameters are fixed to their best-fit values), we build contour maps of the projections of the parameter space. Fig. \ref{fig:sed2_fast_cornerplots} shows the contour maps constraining the parameters of the neutrino arrival SED fast solution. Similar plots for the remaining leptonic models can be found in Appendix \ref{appB}. Since our original grid resolution (ten points per dimension) is insufficient to draw the contours, we perform additional scans in which all parameters but two are fixed to their best-fit values, probing 100 points per dimension. The color bar corresponds to the values of reduced $\chi^2$ with values over 100 being saturated. 

The best-fit fast solution for the period of neutrino arrival suggest that the sychrotron emission is produced by a population of electrons with a soft spectrum (with a power-law index of 2.7). Therefore, the spectrum is dominated by low-energy electrons and the high-energy end of the spectrum is not well constrained, which can be seen in the third row of plots in Fig. \ref{fig:sed2_fast_cornerplots}. Since the peak of the synchronton emission cannot be estimated from the SED (which was also noticed by \citealt{2023ApJ...954...70A}), the minimum electron energy also has a wide range of possible values. The Lorentz factor, magnetic field, and electron luminosity, on the contrary, produce good fits only in narrow regions of the parameter space.

\subsection{Leptohadronic fits}

Since the leptonic grid scan revealed the existence of multiple solutions with close $\chi^2$/n.d.f value, we modeled the leptohadronic SEDs according to the grid-scan-based procedure described in Section \ref{method_lephad} also for the slow and fast solutions.

Before the local minimization, we selected the models with the lowest reduced $\chi^2$ value for each SED and each solution type (slow or fast). In the leptonic case, the slow and fast solutions had similar values of reduced $\chi^2$. When adding protons to the radiation zone, the values of reduced $\chi^2$ for the slow leptohadronic solutions became more than twice worse than that for the fast leptohadronic solutions. We also notice that among the best-fit slow leptohadronic solutions, the hadronic contribution was either zero or heavily suppressed owing to favoring the lowest proton energies or the lowest proton luminosities. When searching for the leptohadronic solutions, we started with the assumption that hadrons mostly contribute to the observed X-ray fluxes as described in Sec. \ref{sec:method}. Generally, higher values of Lorentz factors produce a deeper gap in the X-rays, which can be filled with the emission from hadronic processes. Therefore, the slow solutions naturally suggest a leptonic origin of the X-rays, with any additional contribution from hadronic processes leading to overshooting the observed X-ray fluxes. The slow leptohadronic models converge to purely leptonic models with the lowest proton energies or luminosities producing a better fit than any other combination of the hadronic parameters. Since we are interested in the leptohadronic models, all the following steps were carried out only with the fast leptohadronic solutions for all selected periods of the blazar activity. 

As for the fast leptohadronic solutions, the analysis revealed the degeneracy of the hadronic contributions in all fast leptohadronic models. We find that along with the best-fit solution, there exist many other solutions with the same leptonic parameters and different hadronic contributions; all these solutions have very similar goodness-of-fit values (less than 2\% difference in the reduced $\chi^2$ values). 

\begin{figure}[htpb!]
\centering
 \includegraphics[width=0.5\textwidth]{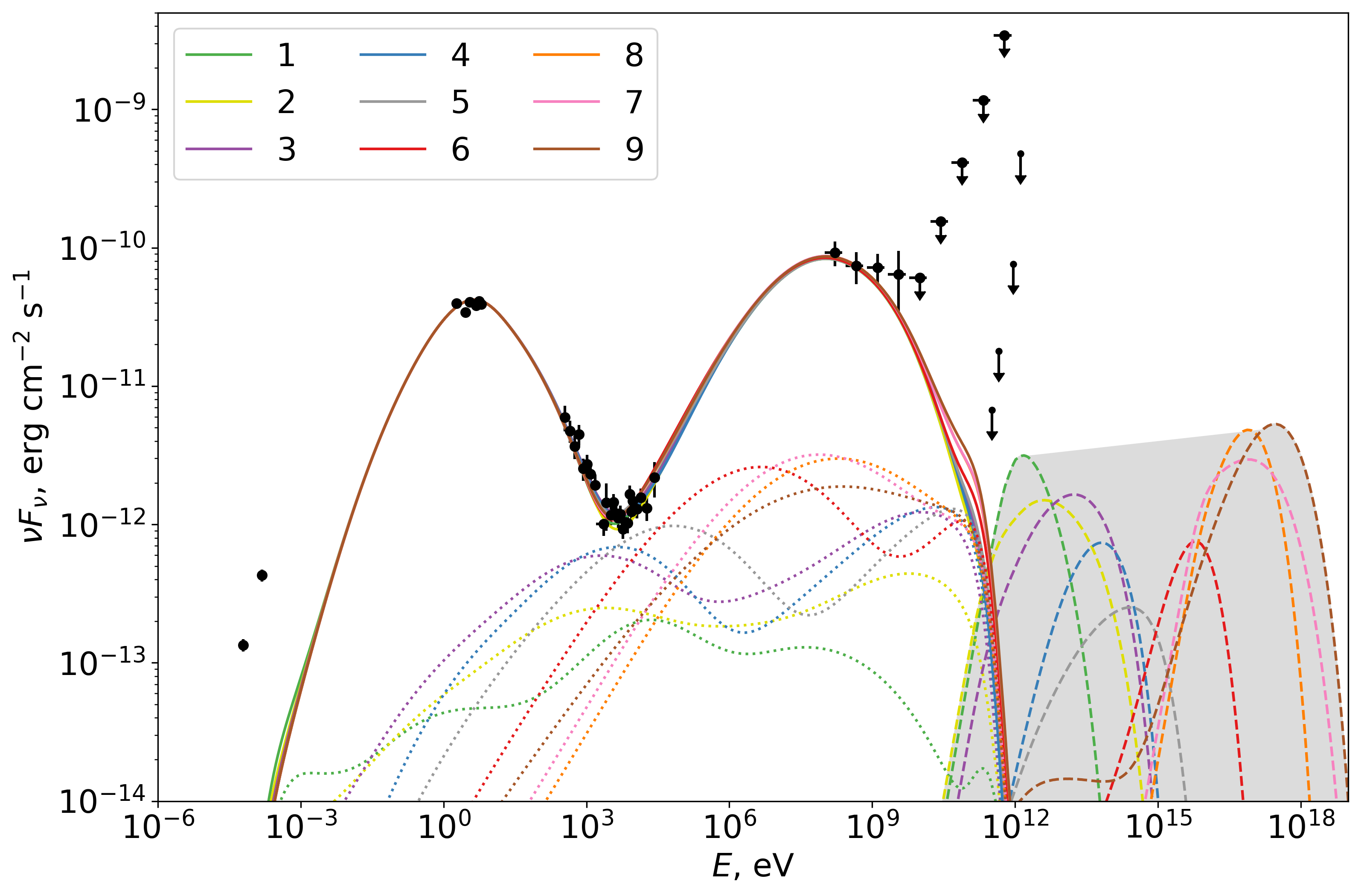}
  \caption{Multiple leptohadronic models that explain the observed photon fluxes during the neutrino arrival. The solid curves correspond to the electromagnetic radiation. The dotted curves represent the contribution of the hadronic process in total photon fluxes, and the dashed curves represent the predicted all-flavor neutrino spectra. The values of the hadronic parameters corresponding to each model number from the legend are shown in Table \ref{tab:nu_degeneracy}.}
     \label{fig:nu_spectra}
\end{figure}

An example of the degeneracy of the hadronic contributions is shown in Figure \ref{fig:nu_spectra}.  Although all shown leptohadronic models produce similar levels of photon fluxes that agree with the data, the hadronic contributions in each model are subdominant and differ both in spectral shapes and in energies. This results in a possible range of neutrino spectra located in the shadowed gray area in Fig. \ref{fig:nu_spectra}. The parameters of the leptohadronic models shown in Fig. \ref{fig:nu_spectra} are presented in Table \ref{tab:nu_degeneracy}. The degeneracy comes from the fact that the total energy of the protons can be deposited into the emission zone in different ways. For example, a high number of low-energy protons can produce a similar amount of radiation as a small number of high-energy protons. The neutrino spectra for models 1--9 in Fig. \ref{fig:nu_spectra} span from lower energies to higher energies as the maximum proton energy for each model increases. Models 2, 3, and 4 all have the same minimal and maximal proton energies and differ only by the value of spectral index, which results into the difference in the required proton luminosity being two orders of magnitude between models 2 and 4. Models 1--3 with lower proton energies have high ratio $L'_{\mathrm{p}}/L_{\mathrm{Edd}}$, while models 7 and 8 have $L'_{\mathrm{p}}/L_{\mathrm{Edd}} \approx 10^{-4}$. The hadronic interactions of the proton population in model 5 produce a photon spectrum that peaks around the X-ray gap in the SED, which limits the possible proton luminosity and leads to lower neutrino fluxes than other models. 

\begin{table}[H]
    \centering
    \caption{Parameters of the leptohadronic models shown in Fig. \ref{fig:nu_spectra}.}
    \begin{tabular}{cccccc}
    \toprule
    Model number & $\gamma'^{\textrm{min}}_{\mathrm{p}}$ & $\gamma'^{\textrm{max}}_{\mathrm{p}}$ & $\alpha'_{\mathrm{p}}$ &$\log_{10}L'_{\mathrm{p}}$ & $\chi^2$/n.d.f\\
    \midrule
   
   1& 3.0 & 4.0 & 3.0 & 48.1 & 13.33 \\
   2&1.0 & 4.9 & 2.5  & 48.82 & 13.37 \\
3&1.0 & 4.9 & 2.0 & 47.92 & 13.27 \\
4&1.0 & 4.9 & 1.0 & 46.8 & 13.21 \\
   5&1.5 & 5.7 & 1.8 & 46.4 & 13.35 \\
   6& 2.5 & 6.7 & 1.5 & 45.0 & 13.40\\
  7& 7.0 & 9.0 & 3.2 & 44.05 & 13.24 \\
   8& 7.0 & 8.0 & 2.1 & 43.88 & 13.22 \\
   9& 1.0 & 9.0 & 2.5 & 46.07 & 13.33 \\
   
    \bottomrule
    \end{tabular}
    \tablefoot{The leptonic parameters in all models are $R'_{\mathrm{b}}$ = $10^{15.57}$ cm, $B'$ = 2.2 G, $\Gamma_{\mathrm{b}}$ = 15.1, $\gamma'^{\textrm{min}}_{\mathrm{e}}$ = $10^{3.46}$, $\gamma'^{\textrm{max}}_{\mathrm{e}}$ = $10^{4.34}$, $\alpha'_{\mathrm{e}}$ =2.42, $L'_{\mathrm{e}}$ = $10^{42.65}$ erg~s$^{-1}$.}
    \label{tab:nu_degeneracy}
\end{table}

The collected data for the neutrino arrival SED are located in the typical frequency ranges of available instruments (NIR, optical, UV, X-rays, and gamma rays). They also constrain well the trough in the X-rays. The fact that the observed data are not sufficient to estimate the expected neutrino emission from the source is critical for modeling neutrino source candidates. The same photon energy ranges were used for obtaining the data for other sources associated with neutrino emission such as TXS~0506+056 or PKS~1502+106. The results of our modeling show that the SED alone is not sufficient to unambiguously predict the neutrino spectrum.  

Among the solutions with various hadronic contributions, we search for the one that predicts the highest number of detected neutrinos in IceCube. We do this by calculating the number of detected neutrino events as

\begin{equation}
    N_\nu = \frac{1}{3} \, T \, \int \Phi_\nu(E) A_{\mathrm{eff}} (E, \theta) dE,
\end{equation}
\noindent
where $A_{\mathrm{eff}} (E, \theta)$ is the effective area of the neutrino detector which depends on the neutrino energy and source declination, $\Phi_\nu(E)$ is the all-flavor neutrino flux, and $T$ is the exposure time. A coefficient $1/3$ is introduced to account for neutrino mixing during the propagation, while we only consider IceCube's muon track channel. We calculate the muon neutrino event rates in IceCube using the experiment's effective area \citep{2017ApJ...835..151A}. For the exposure time, we use the different flare periods as follows: beginning of the flare (neutrino arrival SED) -- 15 days, gamma flare peak -- 8 days, and post flare -- 27 days.

\begin{figure*}[htpb!]
\centering
 \includegraphics[width=0.9\textwidth]{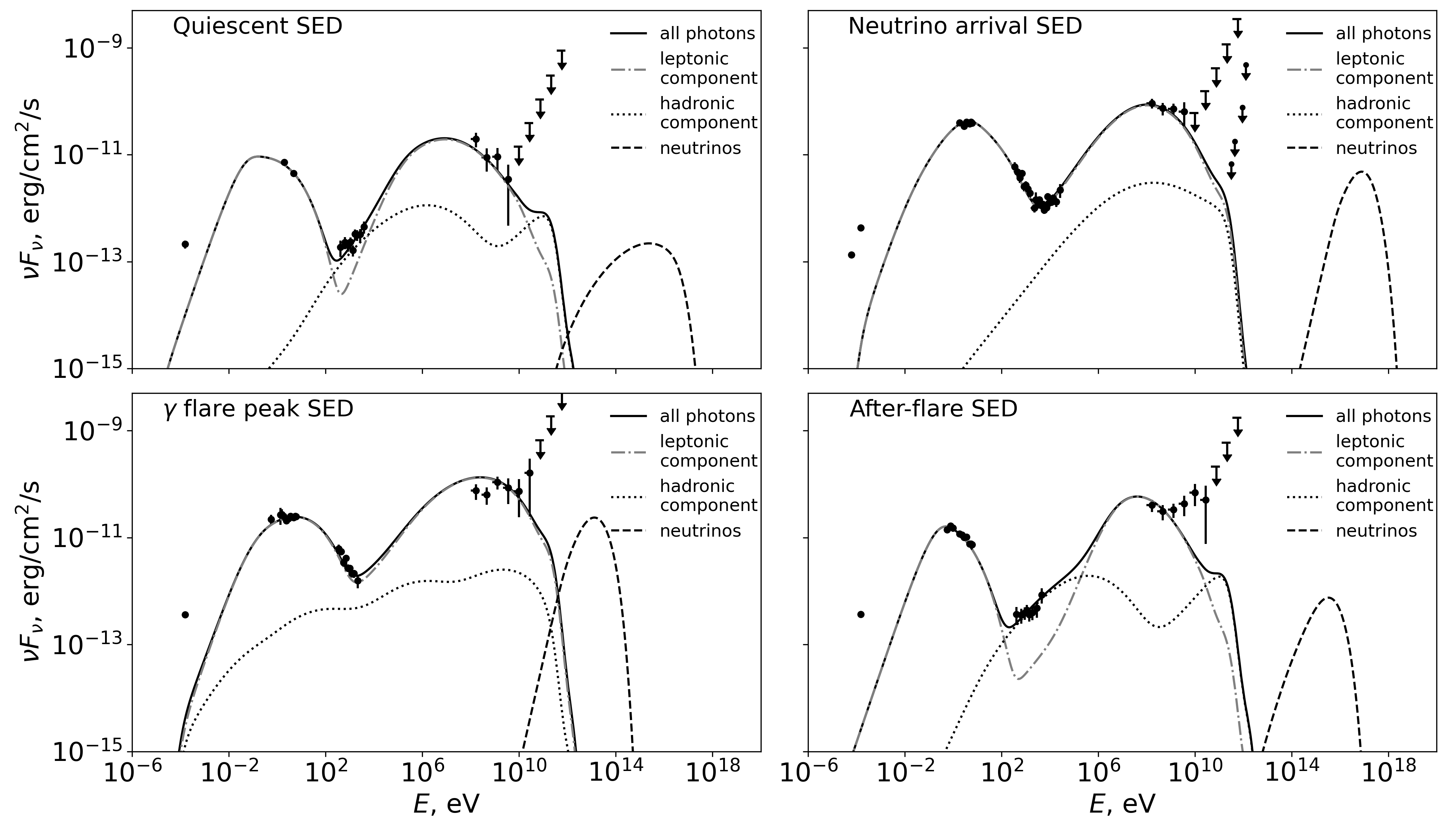}
  \caption{Best-fit models for the selected periods of PKS 0735+178 activity under the condition of maximization of neutrino events in IceCube.}
     \label{fig:lephad_models}
\end{figure*}

The expected number of neutrino events is calculated for all models with $\delta$($\chi^2$/n.d.f.) = $\chi_{\textrm{model}}^2$/n.d.f. $-$ $\chi_{\textrm{best}}^2$/n.d.f. < 1 corresponding to the various hadronic contributions. One model resulting in the maximum number of neutrino events is selected. Those maximal neutrino models for all analyzed periods of activity of PKS 0735+178 are shown in Figure \ref{fig:lephad_models}. The corresponding parameters of the leptohadronic models are shown in Table \ref{table:lephadr_pars}. 

Under the condition of maximization of the neutrino fluxes in IceCube, we predict at most 0.04 muon neutrino event per year during the quiescent state and 0.1 muon neutrinos during the 50 days 2020--2021 flare.

\begin{table}
\centering
\caption{Best-fit leptohadronic model parameters with neutrino event rate maximization condition}
\begin{adjustbox}{width=0.5\textwidth}
\begin{tabular}{lcccc}
\toprule
Parameters     & quiescent   & neutrino arrival  & $\gamma$ flare peak & post flare   \\
\midrule
$\log_{10} $R'$_{\mathrm{b}}$[cm]       & 16.17 & 15.57 & 15.54   & 16.3 \\
$B'$ [G] & 0.1 & 2.28 & 0.965   & 0.08\\

$\Gamma_{\mathrm{b}}$   & 24.0 & 15.0 & 17.0   & 21.0  \\
$\log_{10}\gamma'^{\textrm{min}}_{\mathrm{e}}$   & 3.0 & 3.5 & 3.00   & 3.44 \\
$\log_{10}\gamma'^{\textrm{max}}_{\mathrm{e}}$   & 4.17 & 4.3 & 4.11   & 4.21 \\

$\alpha_{\textrm{e}}$   & 2.8 & 2.42 & 1.26   & 2.9 \\

$\log_{10}L'_{\textrm{e}}$[erg/s]& 42.51 & 42.65 &  42.67&  42.63\\

$\log_{10}\gamma'^{\textrm{min}}_{\mathrm{p}}$   & 2.0 & 5.0 & 3.0   & 3.6 \\
$\log_{10}\gamma'^{\textrm{max}}_{\mathrm{p}}$   & 7.0 & 6.0 & 4.2   & 6.4 \\

$\alpha_{\textrm{p}}$   & 2.0 & 1.5 & 1.25   & 1.0 \\

$\log_{10}L'_{\textrm{p}}$[erg/s]& 46.9 & 46.0 &  48.33&  46.98\\

$\chi^2$/n.d.f.& 4.4 & 13.6 &  4.3&  5.3\\

$N_\nu$/ time interval & 0.04/yr & 0.03/15d & 0.07/8d & 0.004/27d \\

\bottomrule
\end{tabular}
\end{adjustbox}
\tablefoot{Parameter description: R$'_{\mathrm{b}}$ is the radius of the blob, $B'$ is the magnetic field strength in the emission region, $\Gamma_{\mathrm{b}}$ is blob Lorentz factor; $\gamma'^{\textrm{min}}_{\mathrm{e(p)}}$ and $\gamma'^{\textrm{max}}_{\mathrm{e(p)}}$ are the minimum and maximum Lorentz factor of the electrons (protons) respectively, $\alpha_{\textrm{e(p)}}$ is the spectral index of electrons (protons), $L'_{\textrm{e(p)}}$ is electron (proton) luminosity, $\chi^2$/n.d.f. is a value of the reduced $\chi^2$ function, and $N_\nu$/time interval is a number of expected neutrino events during the selected exposure time (for the flaring state corresponds to the duration of observed state).}
\label{table:lephadr_pars}
\end{table}

\section{Discussion} \label{sec:discuss}

Leptonic models successfully reproduce the SEDs not only for the quiescent state but also during periods of high activity. These models are based on the premise that an initial electron population upscatters their own synchrotron photons, generating gamma-ray emission through the synchrotron self-Compton mechanism (SSC). Initially, we applied this model to fit all the observed SEDs and achieved good agreement for three out of four blazar activity states with multiple equally good solutions found for each of them. However, purely leptonic models cannot account for neutrino emission. Neutrino production in a different region remains possible, yet requires additional assumptions on the geometry, physical conditions, and target photon fields thus increasing the number of free parameters. To avoid this complexity, we tested a simpler scenario of adding protons into the same region as the electrons. The addition of the protons to the blob does not impact the leading gamma-ray production mechanism which is SSC but suggests a significant hadronic contribution to  X-ray fluxes and explains neutrino emission. This approach yielded a good fit for most of the epochs and eliminated the need for more complicated models. 

Within a one-zone framework, we could not explain the observed gamma-ray fluxes in the post-flare SED. The fact that our algorithm failed to find any leptonic fit (and subsequently leptohadronic fit as well) that could simultaneously explain optical and gamma-ray data may be an indication of a need for a different kind of model. The hard shape of the gamma-ray spectrum, also observed in other sources, may be explained by synchrotron self-Compton and external Compton interactions with photons from the broad-line region \citep{2023ApJ...958L...2F}. Such a model requires the presence of external photon fields emitted by the broad-line region (BLR). It adds the least number of free parameters and is therefore preferred over multi-zone models in cases where leptohadronic models without external fields cannot explain the observed data. We do not consider an external photon field model in this work. In our models, introducing external photon fields created by the BLR adds an additional parameter (the disk temperature) and increases the complexity of numerical modeling with more interactions. This creates a computational challenge and is left for future work. However, if the explanation of after-flare SED requires an external photon field, this may be an indication of the emission region evolution. The emission zone could be located between the supermassive black hole~(SMBH) and the broad-line region~(BLR) at the moment of the neutrino detection and during the peak of the gamma-ray flare. With the blob moving along the jet in the post-flare period, the emission zone can enter the BLR. 

The model presented for the neutrino arrival SED in Fig. \ref{fig:lephad_models} predicts a slightly higher number of neutrino events than the hybrid model of \cite{10.1093/mnras/stac3607} which has a similar physical setup. Meanwhile, the proton luminosity in our model is four orders of magnitude lower. This is explained by the fact that the maximum proton energy in our models is not fixed. Protons of higher energy produce more neutrinos through photo-pion production, and the resulting cascades contribute mostly to the gamma-ray emission and thus are not limited by the X-ray emission. This compensates for the fact that the neutrino peak is shifted toward high energies in comparison with the IceCube maximum sensitivity which is around 100 TeV \citep{2019EPJC...79..234A}. 

Our best fit for the gamma-ray flare peak SED, on the contrary, produces almost three times more neutrino events in IceCube than the external field model of \cite{10.1093/mnras/stac3607}. In this case, lower proton energies and higher proton luminosities were preferable,  leading to a scenario in which more neutrinos are produced (proportional to the proton luminosity) in the energy range of IceCube's peak sensitivity. 

We note that the flaring states (neutrino arrival, gamma flare peak, and the post-flare SEDs) have higher minimal proton Lorentz factors and harder spectral indices. The hard values of spectral indices can be achieved if the particles are accelerated in the magnetic reconnection region with high magnetization. Since we do not model particle acceleration, particles could be initially pre-accelerated in the magnetic reconnection region before they enter the emission region. In addition, the highest magnetization in the jet is expected at the base of the jet, near SMBH, which dissipates to lower values further down the jet as particles cool down in the expanding blob. This also agrees with the previous hint about the emission zone location between the SMBH and BLR. In turn, a softer proton spectral index in the quiescent state may indicate that either the dominating emission region is located beyond the BLR or the dominating acceleration mechanism is diffuse shock acceleration during this period.

For the quiescent-state SED, even under the condition of maximization of the neutrino events in IceCube, the best fit can produce at most 0.04 neutrino events per year, consistent within Poisson fluctuations with 0.0 neutrino events found in the IceCube point-source analysis using 10~yr of IceCube data between 2008 and 2018 \citep{PhysRevLett.124.051103}. 

\cite{2023ApJ...954...70A} argue that the SSC model cannot explain the SED of PKS 0735+178, and instead propose a model with external photon fields which induces higher neutrino rates owing to more targets for p$\gamma$ interactions. As shown in the results, our SSC model for the neutrino arrival period explains the observed photon fluxes. It produces a factor of two fewer neutrino events than the external field model of \cite{2023ApJ...954...70A} but lowers the proton power requirement by two orders of magnitude. The neutrino emission during the gamma flare peak, on the contrary, produces a factor of two more neutrino events than the external field model by \cite{2023ApJ...954...70A}.

The predicted neutrino rates should be interpreted as a mean value of Poisson statistics. The statistical error of expected neutrino rates, however, is much lower than the uncertainties caused by the demonstrated degeneracy of hadronic parameters and the Eddington bias \citep{2019A&A...622L...9S}.

The leptohadronic one-zone models have a long-standing problem of too high proton power requirement to explain the observed neutrino emission \citep[see, e.g.,][]{gasparyan2022time, 2020ApJ...899..113P}. The fundamental jet luminosity is constrained by the Eddington limit $L_{\textrm{Edd}} = 1.3 \times 10^{38} (M/M_{\odot})$ erg~s$^{-1}$ \citep{10.1111/j.1365-2966.2009.15898.x}. The mass of the SMBH in PKS 0735+178 was estimated based on the optical intraday variability timescales \citep{GUPTA20128} as $1.89 \times 10^8 M_{\odot}$ (corresponding to $L_{\textrm{Edd}} = 2.5 \times 10^{46}$ erg~s$^{-1}$) assuming a jet Doppler factor of $\delta \approx 3.5$. If the Doppler factor of the emission zone is higher, which is the case for our leptohadronic models, the corresponding value of the SMBH mass increases by a factor of four, leading to $L_{\textrm{Edd}} \approx 10^{47}$ erg~s$^{-1}$. The quiescent and neutrino arrival SEDs are explained well with sub-Eddington proton luminosities, which shortly become super-Eddington ($L'_{\textrm{p}} \approx 10\,L_{\textrm{Edd}}$) in the gamma flare peak periods and return to the sub-Eddington value again in the post-flare period. The extreme value of $L'_{\textrm{p}} \approx 10\,L_{\textrm{Edd}}$ corresponds to the peak value during the flare. We note that the high value of proton luminosity during the quiescent state is obtained under the condition of neutrino rate maximization. It should rather be interpreted as an upper limit since no neutrinos from PKS 0735+178 were observed during the last ten years when the source activity was much lower compared to the recent extreme flare. The blob is characterized by close radius values during neutrino arrival and gamma flare peak but expands later during the post-flare phase. During the flare evolution, the value of the magnetic field strength constantly decreases with the highest value of $\sim$2 G during the neutrino arrival. The proton spectral index is also harder during periods of high activity than it is during the quiescent state. This can be interpreted as the same emission region dominating during the flaring states which agrees with our previous suggestions. However, the values of blob Lorentz factors differ for all four SEDs with the highest best-fit value for the quiescent state. This effect can be caused by independent fitting procedures for all states and when taking into account degeneracies a similarly good solution with the same Lorentz factor could be found. Alternatively, the difference could be caused by small changes in viewing angle ($\sim$0.5$^\circ$ between consequent states). In our modeling, we approximate the Doppler factor as $\delta_{\mathrm{D}} \approx \Gamma_{\mathrm{b}}$, therefore changes in the viewing angle could explain the variation in both the Doppler factor and the Lorentz factor. \cite{2010A&A...515A.105B} suggest that the effects of long-term radio variability for PKS~0735+178 can be explained by jet precession due to a binary supermassive black hole (best-fit precession speed $\Dot{\Omega}=15$ deg/year or 1.25 deg/month) which supports the viewing angle change hypothesis. During the one-month flare, the change of Lorentz factor from 15 to 21 would require a change in the viewing angle of $1.1^{\circ}$ which roughly agrees with the estimate from \cite{2010A&A...515A.105B}. 
Alternatively, if the dominating emission region during the flare is located closer to the jet base, the blob might accelerate while moving further down the jet. At the same time, the dominating emission region during the quiescent state might be a fully accelerated blob which could explain the evolution of Lorentz factors. Additionally, the jet with structures like banding and wobbling would also cause the regions where the blob movements would appear as accelerating or decelerating. The different Lorentz factors could also indicate that the emission comes from different blobs.

We also note that $\gamma'^{\textrm{min}}_{\mathrm{e}}\gg1$ and $\gamma'^{\textrm{min}}_{\mathrm{p}}\gg1$ in all our models. The high values of Lorentz factors are required to explain SEDs of extreme blazars \citep{2021A&A...654A..96Z}. Still, electrons and protons of lower energies can exist in the jet. For electrons, the lowering of the minimal Lorentz factor to one would lead to an incorrect shape of the synchrotron peak and failure of the SED fit even for hard spectral indices. Protons, in turn, can not produce any radiation since their energies are below the threshold for all hadronic processes. Such particles do not produce any observable signature in the SED but impact the energy budget. If the minimum proton Lorentz factor is set to one, the photons of hadronic origin peak at lower energies and are limited by the observed X-ray fluxes producing an order of magnitude lower neutrino rates than ones derived from the models shown in Fig. \ref{fig:lephad_models}. We set high values $\gamma'^{\textrm{min}}_{\mathrm{e}}$ and $\gamma'^{\textrm{min}}_{\mathrm{p}}$ to ensure the efficient interaction and radiation processes in the emission region and to estimate maximum neutrino rates. In addition to previously shown poor constraints on the $\gamma'^{\textrm{min}}_{\mathrm{e,p}}$, $\gamma'^{\textrm{max}}_{\mathrm{e,p}}$ and $\alpha_{\mathrm{e,p}}$ from the parameter space search, the possible presence of the particles with lower energies makes the uncertainty of $\gamma'^{\textrm{min}}_{\mathrm{e,p}}$ and $\alpha_{\mathrm{e,p}}$ even worse. Lower values of $\gamma'^{\textrm{min}}_{\mathrm{e,p}}$ and $\alpha_{\mathrm{e,p}}$ also imply that the required corresponding luminosities may be higher and may exceed the Eddington luminosity. Sub-Eddington luminosities can be achieved either by increasing the minimal proton energy or by hardening the spectral index, thus introducing a scenario in which the number of protons in the emission zone reduces as protons carry more energy leaving the energy density conserved.

We analyze the multi-messenger behavior of the source during the different activity states. For the purely leptonic models, the existence of multiple solutions does not allow us to trace the exact evolution of the radiation zone since the causal relation between different solutions at any epoch is not clear. However, we still see that for any solution the flare is characterized by higher values of the magnetic field strength and electron luminosity, simultaneous increase of which suggests energy equipartition. 

Under the condition of neutrino rates maximization, the different periods during the flare have enhanced magnetic field strength and more compact emission-zone regions. The neutrino flare could be linked to the increase of minimum proton energy and hardening of the proton spectrum leading to increased $p\gamma$ interaction rates. The production of $\sim$100 TeV neutrinos requires a pre-acceleration of protons to high energies. This can be interpreted as the transfer of a significant part of jet energy to protons followed by the neutrino arrival and flare which can have important implications regarding jet composition and dynamics.

\section{Summary and conclusions}
In this work, we modeled the multi-epoch emission of PKS~0735+178 including different stages of flare evolution. The grid-scan-based approach showed the degeneracy of leptonic and leptohadronic model parameters. Under the condition of maximization of the neutrino detection rates, we find leptohadronic solutions for different stages of the flare that produce 0.1 neutrino events during the 50~day flare, a higher number of neutrino events than any of the previously proposed models. We also show that in the quiescent state the maximum neutrino rate, which in our model is limited by the X-ray emission, is two orders of magnitude lower than during the flare. Still, the post-flare SED cannot be reproduced with a simple one-zone leptohadronic model and likely requires external field photons for the explanation of the gamma rays, which may be a sign of spatial evolution of the emission zone.

The demonstrated degeneracy of the hadronic parameters in the one-zone leptohadronic models creates a significant challenge for understanding potential neutrino sources. The properties of the proton spectrum cannot be constrained by the observed photon fluxes, leading to great uncertainty in the subsequent neutrino emission. To break this degeneracy, next-generation neutrino telescopes such as IceCube-Gen2 \citep{2021JPhG...48f0501A}, KM3NeT \citep{2016JPhG...43h4001A}, or P-ONE \citep{universe10020053} are needed. Additionally, multi-wavelength polarization can potentially constrain the hadronic component \citep{2024ApJ...967...93Z}, thus highlighting the importance of future X-ray and MeV polarimeters such as eXTP \citep{2016SPIE.9905E..1QZ}, COSI \citep{2019BAAS...51g..98T}, and AMEGO-X \citep{2022JATIS...8d4003C}.

%

\begin{acknowledgements}
    The authors express their appreciation to Xavier Rodrigues for useful discussions on the models and support with computing issues. We would like to thank the anonymous referee for insightful and constructive suggestions and comments that greatly improved the manuscript. This research has made use of data from the MOJAVE database that is maintained by the MOJAVE team (Lister et al. 2018). This publication makes use of data obtained at the Metsähovi Radio Observatory, operated by the Aalto University. A.O. was supported by DAAD funding program 57552340. A.F. acknowledges support from the DFG via the Collaborative Research Center SFB1491 \textit{Cosmic Interacting Matters -- From Source to Signal}. 
A.V.F.’s research group at UC Berkeley acknowledges financial assistance from the Christopher R. Redlich Fund, Gary and Cynthia Bengier, Clark and Sharon Winslow, Alan Eustace (W.Z. is a Bengier-Winslow-Eustace Specialist in Astronomy), and numerous other donors.
KAIT and its ongoing operation were made possible by donations from Sun Microsystems, Inc., the Hewlett-Packard Company, AutoScope Corporation, Lick Observatory, the U.S. National Science Foundation, the University of California, the Sylvia \& Jim Katzman Foundation, and the TABASGO Foundation. Research at Lick Observatory is partially supported by a generous gift from Google. 
    
\end{acknowledgements}


\bibliography{bibl}

\begin{thebibliography}{78}
\expandafter\ifx\csname natexlab\endcsname\relax\def\natexlab#1{#1}\fi

\bibitem[{Aartsen {et~al.}(2017)Aartsen, Ackermann, Adams, Aguilar, Ahlers,
  Ahrens, Altmann, Andeen, Anderson, Ansseau, Anton, Archinger, Argüelles,
  Auffenberg, Axani, Bai, Barwick, Baum, Bay, Beatty, Tjus, Becker, BenZvi,
  Berley, Bernardini, Bernhard, Besson, Binder, Bindig, Bissok, Blaufuss, Blot,
  Bohm, Börner, Bos, Bose, Böser, Botner, Braun, Brayeur, Bretz, Bron,
  Burgman, Carver, Casier, Cheung, Chirkin, Christov, Clark, Classen, Coenders,
  Collin, Conrad, Cowen, Cross, Day, de~Andr{\'{e}}, Clercq, del Pino~Rosendo,
  Dembinski, Ridder, Desiati, de~Vries, de~Wasseige, de~With, DeYoung,
  D{\'{\i}}az-V{\'{e}}lez, di~Lorenzo, Dujmovic, Dumm, Dunkman, Eberhardt,
  Ehrhardt, Eichmann, Eller, Euler, Evenson, Fahey, Fazely, Feintzeig, Felde,
  Filimonov, Finley, Flis, Fösig, Franckowiak, Friedman, Fuchs, Gaisser,
  Gallagher, Gerhardt, Ghorbani, Giang, Gladstone, Glauch, Glüsenkamp,
  Goldschmidt, Gonzalez, Grant, Griffith, Haack, Hallgren, Halzen, Hansen,
  Hansmann, Hanson, Hebecker, Heereman, Helbing, Hellauer, Hickford, Hignight,
  Hill, Hoffman, Hoffmann, Hoshina, Huang, Huber, Hultqvist, In, Ishihara,
  Jacobi, Japaridze, Jeong, Jero, Jones, Kang, Kappes, Karg, Karle, Katz,
  Kauer, Keivani, Kelley, Kheirandish, Kim, Kim, Kintscher, Kiryluk, Kittler,
  Klein, Kohnen, Koirala, Kolanoski, Konietz, Köpke, Kopper, Kopper, Koskinen,
  Kowalski, Krings, Kroll, Krückl, Krüger, Kunnen, Kunwar, Kurahashi,
  Kuwabara, Labare, Lanfranchi, Larson, Lauber, Lennarz, Lesiak-Bzdak,
  Leuermann, Lu, Lünemann, Madsen, Maggi, Mahn, Mancina, Mandelartz, Maruyama,
  Mase, Maunu, McNally, Meagher, Medici, Meier, Meli, Menne, Merino, Meures,
  Miarecki, Montaruli, Moulai, Nahnhauer, Naumann, Neer, Niederhausen, Nowicki,
  Nygren, Pollmann, Olivas, O'Murchadha, Palczewski, Pandya, Pankova, Peiffer,
  Penek, Pepper, de~los Heros, Pieloth, Pinat, Price, Przybylski, Quinnan,
  Raab, Rädel, Rameez, Rawlins, Reimann, Relethford, Relich, Resconi, Rhode,
  Richman, Riedel, Robertson, Rongen, Rott, Ruhe, Ryckbosch, Rysewyk,
  Sabbatini, Herrera, Sandrock, Sandroos, Sarkar, Satalecka, Schlunder,
  Schmidt, Schoenen, Schöneberg, Schumacher, Seckel, Seunarine, Soldin, Song,
  Spiczak, Spiering, Stanev, Stasik, Stettner, Steuer, Stezelberger, Stokstad,
  Stö{\ss}l, Ström, Strotjohann, Sullivan, Sutherland, Taavola, Taboada,
  Tatar, Tenholt, Ter-Antonyan, Terliuk, Te{\v{s}}i{\'{c}}, Tilav, Toale,
  Tobin, Toscano, Tosi, Tselengidou, Turcati, Unger, Usner, Vandenbroucke, van
  Eijndhoven, Vanheule, van Rossem, van Santen, Vehring, Voge, Vogel, Vraeghe,
  Walck, Wallace, Wallraff, Wandkowsky, Weaver, Weiss, Wendt, Westerhoff,
  Whelan, Wickmann, Wiebe, Wiebusch, Wille, Williams, Wills, Wolf, Wood,
  Woolsey, Woschnagg, Xu, Xu, Xu, Yanez, Yodh, Yoshida, \& Zoll}]{Aartsen_2017}
Aartsen, M., Ackermann, M., Adams, J., {et~al.} 2017, Astroparticle Physics,
  92, 30

\bibitem[{{Aartsen} {et~al.}(2021){Aartsen}, {Abbasi}, {Ackermann}, {Adams},
  {Aguilar}, {Ahlers}, {Ahrens}, {Alispach}, {Allison}, {Amin}, {Andeen},
  {Anderson}, {Ansseau}, {Anton}, {Arg{\"u}elles}, {Arlen}, {Auffenberg},
  {Axani}, {Bagherpour}, {Bai}, {Balagopal V}, {Barbano}, {Bartos}, {Bastian},
  {Basu}, {Baum}, {Baur}, {Bay}, {Beatty}, {Becker}, {Tjus}, {BenZvi},
  {Berley}, {Bernardini}, {Besson}, {Binder}, {Bindig}, {Blaufuss}, {Blot},
  {Bohm}, {Bohmer}, {B{\"o}ser}, {Botner}, {B{\"o}ttcher}, {Bourbeau},
  {Bourbeau}, {Bradascio}, {Braun}, {Bron}, {Brostean-Kaiser}, {Burgman},
  {Burley}, {Buscher}, {Busse}, {Bustamante}, {Campana}, {Carnie-Bronca},
  {Carver}, {Chen}, {Chen}, {Cheung}, {Chirkin}, {Choi}, {Clark}, {Clark},
  {Classen}, {Coleman}, {Collin}, {Connolly}, {Conrad}, {Coppin}, {Correa},
  {Cowen}, {Cross}, {Dave}, {Deaconu}, {De Clercq}, {DeLaunay}, {De Kockere},
  {Dembinski}, {Deoskar}, {De Ridder}, {Desai}, {Desiati}, {de Vries}, {de
  Wasseige}, {de With}, {DeYoung}, {Dharani}, {Diaz}, {D{\'\i}az-V{\'e}lez},
  {Dujmovic}, {Dunkman}, {DuVernois}, {Dvorak}, {Ehrhardt}, {Eller}, {Engel},
  {Evans}, {Evenson}, {Fahey}, {Farrag}, {Fazely}, {Felde}, {Fienberg},
  {Filimonov}, {Finley}, {Fischer}, {Fox}, {Franckowiak}, {Friedman}, {Fritz},
  {Gaisser}, {Gallagher}, {Ganster}, {Garcia-Fernandez}, {Garrappa}, {Gartner},
  {Gerhard}, {Gernhaeuser}, {Ghadimi}, {Glaser}, {Glauch}, {Gl{\"u}senkamp},
  {Goldschmidt}, {Gonzalez}, {Goswami}, {Grant}, {Gr{\'e}goire}, {Griffith},
  {Griswold}, {G{\"u}nd{\"u}z}, {Haack}, {Hallgren}, {Halliday}, {Halve},
  {Halzen}, {Hanson}, {Hanson}, {Hardin}, {Haugen}, {Haungs}, {Hauser},
  {Hebecker}, {Heinen}, {Heix}, {Helbing}, {Hellauer}, {Henningsen},
  {Hickford}, {Hignight}, {Hill}, {Hill}, {Hoffman}, {Hoffmann}, {Hoffmann},
  {Hoinka}, {Hokanson-Fasig}, {Holzapfel}, {Hoshina}, {Huang}, {Huber},
  {Huber}, {Huege}, {Hughes}, {Hultqvist}, {H{\"u}nnefeld}, {Hussain}, {In},
  {Iovine}, {Ishihara}, {Jansson}, {Japaridze}, {Jeong}, {Jones}, {Jonske},
  {Joppe}, {Kalekin}, {Kang}, {Kang}, {Kang}, {Kappes}, {Kappesser}, {Karg},
  {Karl}, {Karle}, {Katori}, {Katz}, {Kauer}, {Keivani}, {Kellermann},
  {Kelley}, {Kheirandish}, {Kim}, {Kin}, {Kintscher}, {Kiryluk}, {Kittler},
  {Kleifges}, {Klein}, {Koirala}, {Kolanoski}, {K{\"o}pke}, {Kopper}, {Kopper},
  {Koskinen}, {Koundal}, {Kovacevich}, {Kowalski}, {Krauss}, {Krings},
  {Kr{\"u}ckl}, {Kulacz}, {Kurahashi}, {Gualda}, {Lahmann}, {Lanfranchi},
  {Larson}, {Latif}, {Lauber}, {Lazar}, {Leonard}, {Leszczy{\'n}ska}, {Li},
  {Liu}, {Lohfink}, {LoSecco}, {Mariscal}, {Lu}, {Lucarelli}, {Ludwig},
  {L{\"u}nemann}, {Luszczak}, {Lyu}, {Ma}, {Madsen}, {Maggi}, {Mahn}, {Makino},
  {Mallik}, {Mancina}, {Mandalia}, {Mari{\c{s}}}, {Marka}, {Marka}, {Maruyama},
  {Mase}, {Maunu}, {McNally}, {Meagher}, {Medina}, {Meier}, {Meighen-Berger},
  {Merz}, {Meyers}, {Micallef}, {Mockler}, {Moment{\'e}}, {Montaruli}, {Moore},
  {Morse}, {Moulai}, {Muth}, {Naab}, {Nagai}, {Nam}, {Nauman}, {Necker},
  {Neer}, {Nelles}, {Nguyễn}, {Niederhausen}, {Nisa}, {Nowicki}, {Nygren},
  {Oberla}, {Pollmann}, {Oehler}, {Olivas}, {O'Sullivan}, {Pan}, {Pandya},
  {Pankova}, {Papp}, {Park}, {Parker}, {Paudel}, {Peiffer}, {P{\'e}rez de los
  Heros}, {Petersen}, {Philippen}, {Pieloth}, {Pieper}, {Pinfold}, {Pizzuto},
  {Plaisier}, {Plum}, {Popovych}, {Porcelli}, {Rodriguez}, {Price},
  {Przybylski}, {Raab}, {Raissi}, {Rameez}, {Rauch}, {Rawlins}, {Rea},
  {Rehman}, {Reimann}, {Renschler}, {Renzi}, {Resconi}, {Reusch}, {Rhode},
  {Richman}, {Riedel}, {Riegel}, {Roberts}, {Robertson}, {Roellinghoff},
  {Rongen}, {Rott}, {Ruhe}, {Ryckbosch}, {Cantu}, {Safa}, {Herrera},
  {Sandrock}, {Sandroos}, {Sandstrom}, {Santander}, {Sarkar}, {Sarkar},
  {Satalecka}, {Scharf}, {Schaufel}, {Schieler}, {Schlunder}, {Schmidt},
  {Schneider}, {Schneider}, {Schr{\"o}der}, {Schumacher}, {Sclafani}, {Seckel},
  {Seunarine}, {Shaevitz}, {Sharma}, {Shefali}, {Silva}, {Smith}, {Smithers},
  {Snihur}, {Soedingrekso}, {Soldin}, {S{\"o}ldner-Rembold}, {Song},
  {Southall}, {Spiczak}, {Spiering}, {Stachurska}, {Stamatikos}, {Stanev},
  {Stein}, {Stettner}, {Steuer}, {Stezelberger}, {Stokstad}, {Strotjohann},
  {St{\"u}rwald}, {Stuttard}, {Sullivan}, {Taboada}, {Taketa}, {Tanaka},
  {Tenholt}, {Ter-Antonyan}, {Terliuk}, {Tilav}, {Tollefson}, {Tomankova},
  {T{\"o}nnis}, {Torres}, {Toscano}, {Tosi}, {Trettin}, {Tselengidou}, {Tung},
  {Turcati}, {Turcotte}, {Turley}, {Twagirayezu}, {Ty}, {Unger}, {Elorrieta},
  {Vandenbroucke}, {van Eijk}, {van Eijndhoven}, {Vannerom}, {van Santen},
  {Veberic}, {Verpoest}, {Vieregg}, {Vraeghe}, {Walck}, {Watson}, {Weaver},
  {Weindl}, {Weinstock}, {Weiss}, {Weldert}, {Welling}, {Wendt}, {Werthebach},
  {Whitehorn}, {Wiebe}, {Wiebusch}, {Williams}, {Wissel}, {Wolf}, {Wood},
  {Woschnagg}, {Wrede}, {Wren}, {Wulff}, {Xu}, {Xu}, {Yanez}, {Yoshida},
  {Yuan}, {Zhang}, {Zierke}, \& {Z{\"o}cklein}}]{2021JPhG...48f0501A}
{Aartsen}, M.~G., {Abbasi}, R., {Ackermann}, M., {et~al.} 2021, Journal of
  Physics G Nuclear Physics, 48, 060501

\bibitem[{{Aartsen} {et~al.}(2017){Aartsen}, {Abraham}, {Ackermann}, {Adams},
  {Aguilar}, {Ahlers}, {Ahrens}, {Altmann}, {Andeen}, {Anderson}, {Ansseau},
  {Anton}, {Archinger}, {Arguelles}, {Arlen}, {Auffenberg}, {Axani}, {Bai},
  {Barwick}, {Baum}, {Bay}, {Beatty}, {Becker Tjus}, {Becker}, {BenZvi},
  {Berghaus}, {Berley}, {Bernardini}, {Bernhard}, {Besson}, {Binder}, {Bindig},
  {Bissok}, {Blaufuss}, {Blot}, {Boersma}, {Bohm}, {B{\"o}rner}, {Bos}, {Bose},
  {B{\"o}ser}, {Botner}, {Braun}, {Brayeur}, {Bretz}, {Burgman}, {Casey},
  {Casier}, {Cheung}, {Chirkin}, {Christov}, {Clark}, {Classen}, {Coenders},
  {Collin}, {Conrad}, {Cowen}, {Cruz Silva}, {Daughhetee}, {Davis}, {Day}, {de
  Andr{\'e}}, {De Clercq}, {del Pino Rosendo}, {Dembinski}, {De Ridder},
  {Desiati}, {de Vries}, {de Wasseige}, {de With}, {DeYoung},
  {D{\'\i}az-V{\'e}lez}, {di Lorenzo}, {Dujmovic}, {Dumm}, {Dunkman},
  {Eberhardt}, {Ehrhardt}, {Eichmann}, {Euler}, {Evenson}, {Fahey}, {Fazely},
  {Feintzeig}, {Felde}, {Filimonov}, {Finley}, {Flis}, {F{\"o}sig},
  {Franckowiak}, {Fuchs}, {Gaisser}, {Gaior}, {Gallagher}, {Gerhardt},
  {Ghorbani}, {Giang}, {Gladstone}, {Glagla}, {Gl{\"u}senkamp}, {Goldschmidt},
  {Golup}, {Gonzalez}, {G{\'o}ra}, {Grant}, {Griffith}, {Haack}, {Haj Ismail},
  {Hallgren}, {Halzen}, {Hansen}, {Hansmann}, {Hansmann}, {Hanson}, {Hebecker},
  {Heereman}, {Helbing}, {Hellauer}, {Hickford}, {Hignight}, {Hill}, {Hoffman},
  {Hoffmann}, {Holzapfel}, {Homeier}, {Hoshina}, {Huang}, {Huber}, {Huelsnitz},
  {Hultqvist}, {In}, {Ishihara}, {Jacobi}, {Japaridze}, {Jeong}, {Jero},
  {Jones}, {Jurkovic}, {Kappes}, {Karg}, {Karle}, {Katz}, {Kauer}, {Keivani},
  {Kelley}, {Kemp}, {Kheirandish}, {Kim}, {Kintscher}, {Kiryluk}, {Kittler},
  {Klein}, {Kohnen}, {Koirala}, {Kolanoski}, {Konietz}, {K{\"o}pke}, {Kopper},
  {Kopper}, {Koskinen}, {Kowalski}, {Krings}, {Kroll}, {Kr{\"u}ckl},
  {Kr{\"u}ger}, {Kunnen}, {Kunwar}, {Kurahashi}, {Kuwabara}, {Labare},
  {Lanfranchi}, {Larson}, {Lennarz}, {Lesiak-Bzdak}, {Leuermann}, {Leuner},
  {Lu}, {L{\"u}nemann}, {Madsen}, {Maggi}, {Mahn}, {Mancina}, {Mandelartz},
  {Maruyama}, {Mase}, {Maunu}, {McNally}, {Meagher}, {Medici}, {Meier}, {Meli},
  {Menne}, {Merino}, {Meures}, {Miarecki}, {Middell}, {Mohrmann}, {Montaruli},
  {Moulai}, {Nahnhauer}, {Naumann}, {Neer}, {Niederhausen}, {Nowicki},
  {Nygren}, {Obertacke Pollmann}, {Olivas}, {Omairat}, {O'Murchadha},
  {Palczewski}, {Pandya}, {Pankova}, {Penek}, {Pepper}, {P{\'e}rez de los
  Heros}, {Pfendner}, {Pieloth}, {Pinat}, {Posselt}, {Price}, {Przybylski},
  {Quinnan}, {Raab}, {R{\"a}del}, {Rameez}, {Rawlins}, {Reimann}, {Relich},
  {Resconi}, {Rhode}, {Richman}, {Riedel}, {Robertson}, {Rongen}, {Rott},
  {Ruhe}, {Ryckbosch}, {Rysewyk}, {Sabbatini}, {Sanchez Herrera}, {Sandrock},
  {Sandroos}, {Sarkar}, {Satalecka}, {Schimp}, {Schlunder}, {Schmidt},
  {Schoenen}, {Sch{\"o}neberg}, {Sch{\"o}nwald}, {Schumacher}, {Seckel},
  {Seunarine}, {Soldin}, {Song}, {Spiczak}, {Spiering}, {Stahlberg},
  {Stamatikos}, {Stanev}, {Stasik}, {Steuer}, {Stezelberger}, {Stokstad},
  {St{\"o}{\ss}l}, {Str{\"o}m}, {Strotjohann}, {Sullivan}, {Sutherland},
  {Taavola}, {Taboada}, {Tatar}, {Ter-Antonyan}, {Terliuk}, {Te{\v{s}}i{\'c}},
  {Tilav}, {Toale}, {Tobin}, {Toscano}, {Tosi}, {Tselengidou}, {Turcati},
  {Unger}, {Usner}, {Vallecorsa}, {Vandenbroucke}, {van Eijndhoven},
  {Vanheule}, {van Rossem}, {van Santen}, {Veenkamp}, {Vehring}, {Voge},
  {Vraeghe}, {Walck}, {Wallace}, {Wallraff}, {Wandkowsky}, {Weaver}, {Wendt},
  {Westerhoff}, {Whelan}, {Wickmann}, {Wiebe}, {Wiebusch}, {Wille}, {Williams},
  {Wills}, {Wissing}, {Wolf}, {Wood}, {Woolsey}, {Woschnagg}, {Xu}, {Xu}, {Xu},
  {Yanez}, {Yodh}, {Yoshida}, {Zoll}, \& {IceCube
  Collaboration}}]{2017ApJ...835...45A}
{Aartsen}, M.~G., {Abraham}, K., {Ackermann}, M., {et~al.} 2017, \apj, 835, 45

\bibitem[{Aartsen {et~al.}(2020)Aartsen, Ackermann, Adams, Aguilar, Ahlers,
  Ahrens, Alispach, Andeen, Anderson, Ansseau, Anton, Arg\"uelles, Auffenberg,
  Axani, Backes, Bagherpour, Bai, Balagopal, Barbano, Barwick, Bastian, Baum,
  Baur, Bay, Beatty, Becker, Becker~Tjus, BenZvi, Berley, Bernardini, Besson,
  Binder, Bindig, Blaufuss, Blot, Bohm, B\"orner, B\"oser, Botner, B\"ottcher,
  Bourbeau, Bourbeau, Bradascio, Braun, Bron, Brostean-Kaiser, Burgman,
  Buscher, Busse, Carver, Chen, Cheung, Chirkin, Choi, Clark, Classen, Coleman,
  Collin, Conrad, Coppin, Correa, Cowen, Cross, Dave, De~Clercq, DeLaunay,
  Dembinski, Deoskar, De~Ridder, Desiati, de~Vries, de~Wasseige, de~With,
  DeYoung, Diaz, D\'{\i}az-V\'elez, Dujmovic, Dunkman, Dvorak, Eberhardt,
  Ehrhardt, Eller, Engel, Evenson, Fahey, Fazely, Felde, Filimonov, Finley,
  Fox, Franckowiak, Friedman, Fritz, Gaisser, Gallagher, Ganster, Garrappa,
  Gerhardt, Ghorbani, Glauch, Gl\"usenkamp, Goldschmidt, Gonzalez, Grant,
  Griffith, Griswold, G\"under, G\"und\"uz, Haack, Hallgren, Halliday, Halve,
  Halzen, Hanson, Haungs, Hebecker, Heereman, Heix, Helbing, Hellauer,
  Henningsen, Hickford, Hignight, Hill, Hoffman, Hoffmann, Hoinka,
  Hokanson-Fasig, Hoshina, Huang, Huber, Huber, Hultqvist, H\"unnefeld,
  Hussain, In, Iovine, Ishihara, Japaridze, Jeong, Jero, Jones, Jonske, Joppe,
  Kang, Kang, Kappes, Kappesser, Karg, Karl, Karle, Katz, Kauer, Kelley,
  Kheirandish, Kim, Kintscher, Kiryluk, Kittler, Klein, Koirala, Kolanoski,
  K\"opke, Kopper, Kopper, Koskinen, Kowalski, Krings, Kr\"uckl, Kulacz,
  Kurahashi, Kyriacou, Labare, Lanfranchi, Larson, Lauber, Lazar, Leonard,
  Leszczy\ifmmode~\acute{n}\else \'{n}\fi{}ska, Leuermann, Liu, Lohfink,
  Lozano~Mariscal, Lu, Lucarelli, L\"unemann, Luszczak, Lyu, Ma, Madsen, Maggi,
  Mahn, Makino, Mallik, Mallot, Mancina, Mari\ifmmode~\mbox{\c{s}}\else
  \c{s}\fi{}, Maruyama, Mase, Matis, Maunu, McNally, Meagher, Medici, Medina,
  Meier, Meighen-Berger, Menne, Merino, Meures, Micallef, Mockler, Moment\'e,
  Montaruli, Moore, Morse, Moulai, Muth, Nagai, Naumann, Neer, Niederhausen,
  Nisa, Nowicki, Nygren, Obertacke~Pollmann, Oehler, Olivas, O'Murchadha,
  O'Sullivan, Palczewski, Pandya, Pankova, Park, Peiffer, P\'erez de~los Heros,
  Philippen, Pieloth, Pinat, Pizzuto, Plum, Porcelli, Price, Przybylski, Raab,
  Raissi, Rameez, Rauch, Rawlins, Rea, Reimann, Relethford, Renschler, Renzi,
  Resconi, Rhode, Richman, Robertson, Rongen, Rott, Ruhe, Ryckbosch, Rysewyk,
  Safa, Sanchez~Herrera, Sandrock, Sandroos, Santander, Sarkar, Sarkar,
  Satalecka, Schaufel, Schieler, Schlunder, Schmidt, Schneider, Schneider,
  Schr\"oder, Schumacher, Sclafani, Seckel, Seunarine, Shefali, Silva, Snihur,
  Soedingrekso, Soldin, Song, Spiczak, Spiering, Stachurska, Stamatikos,
  Stanev, Stein, Steinm\"uller, Stettner, Steuer, Stezelberger, Stokstad,
  St\"o\ss{}l, Strotjohann, St\"urwald, Stuttard, Sullivan, Taboada, Tenholt,
  Ter-Antonyan, Terliuk, Tilav, Tollefson, Tomankova, T\"onnis, Toscano, Tosi,
  Trettin, Tselengidou, Tung, Turcati, Turcotte, Turley, Ty, Unger,
  Unland~Elorrieta, Usner, Vandenbroucke, Van~Driessche, van Eijk, van
  Eijndhoven, Vanheule, van Santen, Vraeghe, Walck, Wallace, Wallraff,
  Wandkowsky, Watson, Weaver, Weindl, Weiss, Weldert, Wendt, Werthebach,
  Whelan, Whitehorn, Wiebe, Wiebusch, Wille, Williams, Wills, Wolf, Wood, Wood,
  Woschnagg, Wrede, Xu, Xu, Xu, Yanez, Yodh, Yoshida, Yuan, \&
  Z\"ocklein}]{PhysRevLett.124.051103}
Aartsen, M.~G., Ackermann, M., Adams, J., {et~al.} 2020, Phys. Rev. Lett., 124,
  051103

\bibitem[{{Aartsen} {et~al.}(2019){Aartsen}, {Ackermann}, {Adams}, {Aguilar},
  {Ahlers}, {Ahrens}, {Altmann}, {Andeen}, {Anderson}, {Ansseau}, {Anton},
  {Arg{\"u}elles}, {Auffenberg}, {Axani}, {Backes}, {Bagherpour}, {Bai},
  {Barbano}, {Barron}, {Barwick}, {Baum}, {Bay}, {Beatty}, {Becker Tjus},
  {Becker}, {BenZvi}, {Berley}, {Bernardini}, {Besson}, {Binder}, {Bindig},
  {Blaufuss}, {Blot}, {Bohm}, {B{\"o}rner}, {Bos}, {B{\"o}ser}, {Botner},
  {Bourbeau}, {Bourbeau}, {Bradascio}, {Braun}, {Bretz}, {Bron},
  {Brostean-Kaiser}, {Burgman}, {Busse}, {Carver}, {Chen}, {Cheung}, {Chirkin},
  {Clark}, {Classen}, {Collin}, {Conrad}, {Coppin}, {Correa}, {Cowen}, {Cross},
  {Dave}, {Day}, {de Andr{\'e}}, {De Clercq}, {DeLaunay}, {Dembinski},
  {Deoskar}, {De Ridder}, {Desiati}, {de Vries}, {de Wasseige}, {de With},
  {DeYoung}, {D{\'\i}az-V{\'e}lez}, {Dujmovic}, {Dunkman}, {Dvorak},
  {Eberhardt}, {Ehrhardt}, {Eichmann}, {Eller}, {Evenson}, {Fahey}, {Fazely},
  {Felde}, {Filimonov}, {Finley}, {Franckowiak}, {Friedman}, {Fritz},
  {Gaisser}, {Gallagher}, {Ganster}, {Garrappa}, {Gerhardt}, {Ghorbani},
  {Giang}, {Glauch}, {Gl{\"u}senkamp}, {Goldschmidt}, {Gonzalez}, {Grant},
  {Griffith}, {Haack}, {Hallgren}, {Halve}, {Halzen}, {Hanson}, {Hebecker},
  {Heereman}, {Helbing}, {Hellauer}, {Hickford}, {Hignight}, {Hill}, {Hoffman},
  {Hoffmann}, {Hoinka}, {Hokanson-Fasig}, {Hoshina}, {Huang}, {Huber},
  {Hultqvist}, {H{\"u}nnefeld}, {Hussain}, {In}, {Iovine}, {Ishihara},
  {Jacobi}, {Japaridze}, {Jeong}, {Jero}, {Jones}, {Kalaczynski}, {Kang},
  {Kappes}, {Kappesser}, {Karg}, {Karle}, {Katz}, {Kauer}, {Keivani}, {Kelley},
  {Kheirandish}, {Kim}, {Kintscher}, {Kiryluk}, {Kittler}, {Klein}, {Koirala},
  {Kolanoski}, {K{\"o}pke}, {Kopper}, {Kopper}, {Koskinen}, {Kowalski},
  {Krings}, {Kroll}, {Kr{\"u}ckl}, {Kunwar}, {Kurahashi}, {Kyriacou}, {Labare},
  {Lanfranchi}, {Larson}, {Lauber}, {Leonard}, {Leuermann}, {Liu}, {Lohfink},
  {Mariscal}, {Lu}, {L{\"u}nemann}, {Luszczak}, {Madsen}, {Maggi}, {Mahn},
  {Makino}, {Mancina}, {Mari{\c{s}}}, {Maruyama}, {Mase}, {Maunu}, {Meagher},
  {Medici}, {Meier}, {Menne}, {Merino}, {Meures}, {Miarecki}, {Micallef},
  {Moment{\'e}}, {Montaruli}, {Moore}, {Moulai}, {Nagai}, {Nahnhauer},
  {Nakarmi}, {Naumann}, {Neer}, {Niederhausen}, {Nowicki}, {Nygren}, {Obertacke
  Pollmann}, {Olivas}, {O'Murchadha}, {O'Sullivan}, {Palczewski}, {Pandya},
  {Pankova}, {Peiffer}, {P{\'e}rez de los Heros}, {Pieloth}, {Pinat},
  {Pizzuto}, {Plum}, {Price}, {Przybylski}, {Raab}, {Rameez}, {Rauch},
  {Rawlins}, {Rea}, {Reimann}, {Relethford}, {Renzi}, {Resconi}, {Rhode},
  {Richman}, {Robertson}, {Rongen}, {Rott}, {Ruhe}, {Ryckbosch}, {Rysewyk},
  {Safa}, {Sanchez Herrera}, {Sandrock}, {Sandroos}, {Santander}, {Sarkar},
  {Sarkar}, {Satalecka}, {Schaufel}, {Schlunder}, {Schmidt}, {Schneider},
  {Schneider}, {Sch{\"o}neberg}, {Schumacher}, {Sclafani}, {Seckel},
  {Seunarine}, {Soedingrekso}, {Soldin}, {Song}, {Spiczak}, {Spiering},
  {Stachurska}, {Stamatikos}, {Stanev}, {Stasik}, {Stein}, {Stettner},
  {Steuer}, {Stezelberger}, {Stokstad}, {St{\"o}{\ss}l}, {Strotjohann},
  {Stuttard}, {Sullivan}, {Sutherland}, {Taboada}, {Tenholt}, {Ter-Antonyan},
  {Terliuk}, {Tilav}, {Tobin}, {T{\"o}nnis}, {Toscano}, {Tosi}, {Tselengidou},
  {Tung}, {Turcati}, {Turcotte}, {Turley}, {Ty}, {Unger}, {Unland Elorrieta},
  {Usner}, {Vandenbroucke}, {Van Driessche}, {van Eijk}, {van Eijndhoven},
  {Vanheule}, {van Santen}, {Vraeghe}, {Walck}, {Wallace}, {Wallraff},
  {Wandler}, {Wandkowsky}, {Watson}, {Weaver}, {Weiss}, {Wendt}, {Werthebach},
  {Westerhoff}, {Whelan}, {Whitehorn}, {Wiebe}, {Wiebusch}, {Wille},
  {Williams}, {Wills}, {Wolf}, {Wood}, {Wood}, {Woolsey}, {Woschnagg}, {Wrede},
  {Xu}, {Xu}, {Xu}, {Yanez}, {Yodh}, {Yoshida}, \&
  {Yuan}}]{2019EPJC...79..234A}
{Aartsen}, M.~G., {Ackermann}, M., {Adams}, J., {et~al.} 2019, European
  Physical Journal C, 79, 234

\bibitem[{{Abbasi} {et~al.}(2023){Abbasi}, {Ackermann}, {Adams}, {Agarwalla},
  {Aguilar}, {Ahlers}, {Alameddine}, {Amin}, {Andeen}, {Anton},
  {Arg{\"u}elles}, {Ashida}, {Athanasiadou}, {Axani}, {Bai}, {Balagopal},
  {Baricevic}, {Barwick}, {Basu}, {Bay}, {Beatty}, {Becker}, {Becker Tjus},
  {Beise}, {Bellenghi}, {Benning}, {BenZvi}, {Berley}, {Bernardini}, {Besson},
  {Binder}, {Blaufuss}, {Blot}, {Bontempo}, {Book}, {Meneguolo}, {B{\"o}ser},
  {Botner}, {B{\"o}ttcher}, {Bourbeau}, {Braun}, {Brinson}, {Brostean-Kaiser},
  {Burley}, {Busse}, {Butterfield}, {Campana}, {Carloni}, {Carnie-Bronca},
  {Chattopadhyay}, {Chau}, {Chen}, {Chen}, {Chirkin}, {Choi}, {Clark},
  {Classen}, {Coleman}, {Collin}, {Connolly}, {Conrad}, {Coppin}, {Correa},
  {Countryman}, {Cowen}, {Dave}, {De Clercq}, {DeLaunay}, {Delgado},
  {Dembinski}, {Deng}, {Deoskar}, {Desai}, {Desiati}, {de Vries}, {de
  Wasseige}, {DeYoung}, {Diaz}, {D{\'\i}az-V{\'e}lez}, {Dittmer}, {Domi},
  {Dujmovic}, {DuVernois}, {Ehrhardt}, {Eller}, {El Mentawi}, {Engel},
  {Erpenbeck}, {Evans}, {Evenson}, {Fan}, {Fang}, {Farrag}, {Fazely},
  {Fedynitch}, {Feigl}, {Fiedlschuster}, {Finley}, {Fischer}, {Fox},
  {Franckowiak}, {Friedman}, {Fritz}, {F{\"u}rst}, {Gaisser}, {Gallagher},
  {Ganster}, {Garcia}, {Gerhardt}, {Ghadimi}, {Glaser}, {Glauch},
  {Gl{\"u}senkamp}, {Goehlke}, {Gonzalez}, {Goswami}, {Grant}, {Gray}, {Gries},
  {Griffin}, {Griswold}, {G{\"u}nther}, {Gutjahr}, {Haack}, {Hallgren},
  {Halliday}, {Halve}, {Halzen}, {Hamdaoui}, {Minh}, {Hanson}, {Hardin},
  {Harnisch}, {Hatch}, {Haungs}, {Helbing}, {Hellrung}, {Henningsen},
  {Heuermann}, {Heyer}, {Hickford}, {Hidvegi}, {Hill}, {Hill}, {Hoffman},
  {Hori}, {Hoshina}, {Hou}, {Huber}, {Hultqvist}, {H{\"u}nnefeld}, {Hussain},
  {Hymon}, {In}, {Ishihara}, {Jacquart}, {Janik}, {Jansson}, {Japaridze},
  {Jayakumar}, {Jeong}, {Jin}, {Jones}, {Kang}, {Kang}, {Kang}, {Kappes},
  {Kappesser}, {Kardum}, {Karg}, {Karl}, {Karle}, {Katz}, {Kauer}, {Kelley},
  {Zathul}, {Kheirandish}, {Kiryluk}, {Klein}, {Kochocki}, {Koirala},
  {Kolanoski}, {Kontrimas}, {K{\"o}pke}, {Kopper}, {Koskinen}, {Koundal},
  {Kovacevich}, {Kowalski}, {Kozynets}, {Kruiswijk}, {Krupczak}, {Kumar},
  {Kun}, {Kurahashi}, {Lad}, {Lagunas Gualda}, {Lamoureux}, {Larson},
  {Latseva}, {Lauber}, {Lazar}, {Lee}, {Leonard DeHolton}, {Leszczy{\'n}ska},
  {Lincetto}, {Liu}, {Liubarska}, {Lohfink}, {Love}, {Mariscal}, {Lu},
  {Lucarelli}, {Ludwig}, {Luszczak}, {Lyu}, {Madsen}, {Mahn}, {Makino},
  {Manao}, {Mancina}, {Sainte}, {Mari{\c{s}}}, {Marka}, {Marka}, {Marsee},
  {Martinez-Soler}, {Maruyama}, {Mayhew}, {McElroy}, {McNally}, {Mead},
  {Meagher}, {Mechbal}, {Medina}, {Meier}, {Merckx}, {Merten}, {Micallef},
  {Montaruli}, {Moore}, {Morii}, {Morse}, {Moulai}, {Mukherjee}, {Naab},
  {Nagai}, {Nakos}, {Naumann}, {Necker}, {Neumann}, {Niederhausen}, {Nisa},
  {Noell}, {Nowicki}, {Obertacke Pollmann}, {O'Dell}, {Oehler}, {Oeyen},
  {Olivas}, {Orsoe}, {Osborn}, {O'Sullivan}, {Pandya}, {Park}, {Parker},
  {Paudel}, {Paul}, {P{\'e}rez de los Heros}, {Peterson}, {Philippen},
  {Pieper}, {Pizzuto}, {Plum}, {Pont{\'e}n}, {Popovych}, {Prado Rodriguez},
  {Pries}, {Procter-Murphy}, {Przybylski}, {Rack-Helleis}, {Rawlins}, {Rechav},
  {Rehman}, {Reichherzer}, {Renzi}, {Resconi}, {Reusch}, {Rhode}, {Richman},
  {Riedel}, {Rifaie}, {Roberts}, {Robertson}, {Rodan}, {Roellinghoff},
  {Rongen}, {Rott}, {Ruhe}, {Ruohan}, {Ryckbosch}, {Safa}, {Saffer},
  {Salazar-Gallegos}, {Sampathkumar}, {Sanchez Herrera}, {Sandrock},
  {Santander}, {Sarkar}, {Sarkar}, {Savelberg}, {Savina}, {Schaufel},
  {Schieler}, {Schindler}, {Schlickmann}, {Schl{\"u}ter}, {Schl{\"u}ter},
  {Schmidt}, {Schneider}, {Schr{\"o}der}, {Schumacher}, {Schwefer}, {Sclafani},
  {Seckel}, {Seikh}, {Seunarine}, {Shah}, {Sharma}, {Shefali}, {Shimizu},
  {Silva}, {Skrzypek}, {Smithers}, {Snihur}, {Soedingrekso}, {S{\o}gaard},
  {Soldin}, {Soldin}, {Sommani}, {Spannfellner}, {Spiczak}, {Spiering},
  {Stamatikos}, {Stanev}, {Stezelberger}, {St{\"u}rwald}, {Stuttard},
  {Sullivan}, {Taboada}, {Ter-Antonyan}, {Thiesmeyer}, {Thompson}, {Thwaites},
  {Tilav}, {Tollefson}, {T{\"o}nnis}, {Toscano}, {Tosi}, {Trettin}, {Tung},
  {Turcotte}, {Twagirayezu}, {Ty}, {Unland Elorrieta}, {Upadhyay}, {Upshaw},
  {Valtonen-Mattila}, {Vandenbroucke}, {van Eijndhoven}, {Vannerom}, {van
  Santen}, {Vara}, {Veitch-Michaelis}, {Venugopal}, {Vereecken}, {Verpoest},
  {Veske}, {Walck}, {Watson}, {Weaver}, {Weigel}, {Weindl}, {Weldert}, {Wendt},
  {Werthebach}, {Weyrauch}, {Whitehorn}, {Wiebusch}, {Willey}, {Williams},
  {Wolf}, {Wolf}, {Wrede}, {Xu}, {Yanez}, {Yildizci}, {Yoshida}, {Young}, {Yu},
  {Yu}, {Yuan}, {Zhang}, {Zhelnin}, \& {IceCube
  Collaboration}}]{2023ApJ...954...75A}
{Abbasi}, R., {Ackermann}, M., {Adams}, J., {et~al.} 2023, \apj, 954, 75

\bibitem[{{Acharyya} {et~al.}(2023){Acharyya}, {Adams}, {Archer}, {Bangale},
  {Bartkoske}, {Batista}, {Benbow}, {Brill}, {Buckley}, {Christiansen},
  {Chromey}, {Errando}, {Falcone}, {Feng}, {Foote}, {Fortson}, {Furniss},
  {Gallagher}, {Hanlon}, {Hanna}, {Hervet}, {Hinrichs}, {Hoang}, {Holder},
  {Humensky}, {Jin}, {Kaaret}, {Kertzman}, {Kherlakian}, {Kieda}, {Kleiner},
  {Korzoun}, {Kumar}, {Lang}, {Lundy}, {Maier}, {McGrath}, {Millard}, {Millis},
  {Mooney}, {Moriarty}, {Mukherjee}, {O'Brien}, {Ong}, {Pohl}, {Pueschel},
  {Quinn}, {Ragan}, {Reynolds}, {Ribeiro}, {Roache}, {Sadeh}, {Sadun}, {Saha},
  {Santander}, {Sembroski}, {Shang}, {Splettstoesser}, {Talluri}, {Tucci},
  {Vassiliev}, {Weinstein}, {Williams}, {Wong}, {Woo}, {Aharonian},
  {Aschersleben}, {Backes}, {Martins}, {Batzofin}, {Becherini}, {Berge},
  {Bernl{\"o}hr}, {Bi}, {B{\"o}ttcher}, {Boisson}, {Bolmont}, {de Bony de
  Lavergne}, {Borowska}, {Bouyahiaoui}, {Bradascio}, {Breuhaus}, {Brose},
  {Brun}, {Bruno}, {Bulik}, {Burger-Scheidlin}, {Caroff}, {Casanova}, {Cecil},
  {Celic}, {Cerruti}, {Chand}, {Chandra}, {Chen}, {Chibueze}, {Chibueze},
  {Cotter}, {Dai}, {Mbarubucyeye}, {Djannati-Ata{\"\i}}, {Dmytriiev},
  {Doroshenko}, {Einecke}, {Ernenwein}, {de Clairfontaine}, {Filipovic},
  {Fontaine}, {F{\"u}{\ss}ling}, {Funk}, {Gabici}, {Ghafourizadeh}, {Giavitto},
  {Glawion}, {Glicenstein}, {Goswami}, {Grolleron}, {Haerer}, {Hinton},
  {Holch}, {Holler}, {Horns}, {Jamrozy}, {Jankowsky}, {Joshi}, {Jung-Richardt},
  {Kasai}, {Katarzy{\'n}ski}, {Khatoon}, {Kh{\'e}lifi}, {Klepser},
  {Klu{\'z}niak}, {Kosack}, {Kostunin}, {Lang}, {Le Stum}, {Lemi{\`e}re},
  {Lenain}, {Leuschner}, {Lohse}, {Luashvili}, {Lypova}, {Mackey}, {Malyshev},
  {Marandon}, {Marchegiani}, {Marcowith}, {Mart{\'\i}-Devesa}, {Marx},
  {Mitchell}, {Moderski}, {Mohrmann}, {Montanari}, {Moulin}, {Murach},
  {Nakashima}, {Niemiec}, {Noel}, {O'Brien}, {Olivera-Nieto}, {de Ona
  Wilhelmi}, {Ostrowski}, {Panny}, {Panter}, {Peron}, {Prokhorov},
  {P{\"u}hlhofer}, {Punch}, {Quirrenbach}, {Reichherzer}, {Reimer}, {Reimer},
  {Ren}, {Renaud}, {Rieger}, {Rudak}, {Ruiz-Velasco}, {Sahakian}, {Santangelo},
  {Sasaki}, {Sch{\"a}fer}, {Sch{\"u}ssler}, {Schutte}, {Schwanke}, {Shapopi},
  {Specovius}, {Spencer}, {Stawarz}, {Steenkamp}, {Steinmassl}, {Sushch},
  {Suzuki}, {Takahashi}, {Tanaka}, {Terrier}, {van Eldik}, {Vecchi}, {Veh},
  {Venter}, {Vink}, {White}, {Wierzcholska}, {Wong}, {Zacharias}, {Zargaryan},
  {Zdziarski}, {Zech}, {Zouari}, {{\.Z}ywucka}, {Mori}, \& {H.~E.~S.~S.
  Collaboration}}]{2023ApJ...954...70A}
{Acharyya}, A., {Adams}, C.~B., {Archer}, A., {et~al.} 2023, \apj, 954, 70

\bibitem[{{Adri{\'a}n-Mart{\'\i}nez} {et~al.}(2016){Adri{\'a}n-Mart{\'\i}nez},
  {Ageron}, {Aharonian}, {Aiello}, {Albert}, {Ameli}, {Anassontzis}, {Andre},
  {Androulakis}, {Anghinolfi}, {Anton}, {Ardid}, {Avgitas}, {Barbarino},
  {Barbarito}, {Baret}, {Barrios-Mart{\'\i}}, {Belhorma}, {Belias}, {Berbee},
  {van den Berg}, {Bertin}, {Beurthey}, {van Beveren}, {Beverini}, {Biagi},
  {Biagioni}, {Billault}, {Bond{\`\i}}, {Bormuth}, {Bouhadef}, {Bourlis},
  {Bourret}, {Boutonnet}, {Bouwhuis}, {Bozza}, {Bruijn}, {Brunner}, {Buis},
  {Busto}, {Cacopardo}, {Caillat}, {Calamai}, {Calvo}, {Capone}, {Caramete},
  {Cecchini}, {Celli}, {Champion}, {Cherkaoui El Moursli}, {Cherubini},
  {Chiarusi}, {Circella}, {Classen}, {Cocimano}, {Coelho}, {Coleiro},
  {Colonges}, {Coniglione}, {Cordelli}, {Cosquer}, {Coyle}, {Creusot},
  {Cuttone}, {D'Amico}, {De Bonis}, {De Rosa}, {De Sio}, {Di Capua}, {Di
  Palma}, {D{\'\i}az Garc{\'\i}a}, {Distefano}, {Donzaud}, {Dornic},
  {Dorosti-Hasankiadeh}, {Drakopoulou}, {Drouhin}, {Drury}, {Durocher},
  {Eberl}, {Eichie}, {van Eijk}, {El Bojaddaini}, {El Khayati}, {Elsaesser},
  {Enzenh{\"o}fer}, {Fassi}, {Favali}, {Fermani}, {Ferrara}, {Filippidis},
  {Frascadore}, {Fusco}, {Gal}, {Galat{\`a}}, {Garufi}, {Gay}, {Gebyehu},
  {Giordano}, {Gizani}, {Gracia}, {Graf}, {Gr{\'e}goire}, {Grella}, {Habel},
  {Hallmann}, {van Haren}, {Harissopulos}, {Heid}, {Heijboer}, {Heine},
  {Henry}, {Hern{\'a}ndez-Rey}, {Hevinga}, {Hofest{\"a}dt}, {Hugon},
  {Illuminati}, {James}, {Jansweijer}, {Jongen}, {de Jong}, {Kadler},
  {Kalekin}, {Kappes}, {Katz}, {Keller}, {Kieft}, {Kie{\ss}ling}, {Koffeman},
  {Kooijman}, {Kouchner}, {Kulikovskiy}, {Lahmann}, {Lamare}, {Leisos},
  {Leonora}, {Clark}, {Liolios}, {Llorens Alvarez}, {Lo Presti}, {L{\"o}hner},
  {Lonardo}, {Lotze}, {Loucatos}, {Maccioni}, {Mannheim}, {Margiotta},
  {Marinelli}, {Mari{\c{s}}}, {Markou}, {Mart{\'\i}nez-Mora}, {Martini},
  {Mele}, {Melis}, {Michael}, {Migliozzi}, {Migneco}, {Mijakowski}, {Miraglia},
  {Mollo}, {Mongelli}, {Morganti}, {Moussa}, {Musico}, {Musumeci}, {Navas},
  {Nicolau}, {Olcina}, {Olivetto}, {Orlando}, {Papaikonomou}, {Papaleo},
  {P{\u{a}}v{\u{a}}la{\c{s}}}, {Peek}, {Pellegrino}, {Perrina}, {Pfutzner},
  {Piattelli}, {Pikounis}, {Poma}, {Popa}, {Pradier}, {Pratolongo},
  {P{\"u}hlhofer}, {Pulvirenti}, {Quinn}, {Racca}, {Raffaelli}, {Randazzo},
  {Rapidis}, {Razis}, {Real}, {Resvanis}, {Reubelt}, {Riccobene}, {Rossi},
  {Rovelli}, {Salda{\~n}a}, {Salvadori}, {Samtleben}, {S{\'a}nchez
  Garc{\'\i}a}, {S{\'a}nchez Losa}, {Sanguineti}, {Santangelo}, {Santonocito},
  {Sapienza}, {Schimmel}, {Schmelling}, {Sciacca}, {Sedita}, {Seitz}, {Sgura},
  {Simeone}, {Siotis}, {Sipala}, {Spisso}, {Spurio}, {Stavropoulos},
  {Steijger}, {Stellacci}, {Stransky}, {Taiuti}, {Tayalati}, {T{\'e}zier},
  {Theraube}, {Thompson}, {Timmer}, {T{\"o}nnis}, {Trasatti}, {Trovato},
  {Tsirigotis}, {Tzamarias}, {Tzamariudaki}, {Vallage}, {Van Elewyck},
  {Vermeulen}, {Vicini}, {Viola}, {Vivolo}, {Volkert}, {Voulgaris}, {Wiggers},
  {Wilms}, {de Wolf}, {Zachariadou}, {Zornoza}, \&
  {Z{\'u}{\~n}iga}}]{2016JPhG...43h4001A}
{Adri{\'a}n-Mart{\'\i}nez}, S., {Ageron}, M., {Aharonian}, F., {et~al.} 2016,
  Journal of Physics G Nuclear Physics, 43, 084001

\bibitem[{{Bellenghi} {et~al.}(2023){Bellenghi}, {Padovani}, {Resconi}, \&
  {Giommi}}]{2023ApJ...955L..32B}
{Bellenghi}, C., {Padovani}, P., {Resconi}, E., \& {Giommi}, P. 2023, \apjl,
  955, L32

\bibitem[{{Bharathan} {et~al.}(2024){Bharathan}, {Stalin}, {Sahayanathan},
  {Bhattacharyya}, \& {Mathew}}]{2024MNRAS.529.3503B}
{Bharathan}, A.~M., {Stalin}, C.~S., {Sahayanathan}, S., {Bhattacharyya}, S.,
  \& {Mathew}, B. 2024, \mnras, 529, 3503

\bibitem[{{B{\"o}ttcher} {et~al.}(2013){B{\"o}ttcher}, {Reimer}, {Sweeney}, \&
  {Prakash}}]{2013ApJ...768...54B}
{B{\"o}ttcher}, M., {Reimer}, A., {Sweeney}, K., \& {Prakash}, A. 2013, \apj,
  768, 54

\bibitem[{{Britzen} {et~al.}(2010){Britzen}, {Witzel}, {Gong}, {Zhang},
  {Gopal-Krishna}, {Goyal}, {Aller}, {Aller}, \&
  {Zensus}}]{2010A&A...515A.105B}
{Britzen}, S., {Witzel}, A., {Gong}, B.~P., {et~al.} 2010, \aap, 515, A105

\bibitem[{{Burrows} {et~al.}(2004){Burrows}, {Hill}, {Nousek}, {Wells},
  {Chincarini}, {Abbey}, {Beardmore}, {Bosworth}, {Br{\"a}uninger}, {Burkert},
  {Campana}, {Capalbi}, {Chang}, {Citterio}, {Freyberg}, {Giommi}, {Hartner},
  {Killough}, {Kittle}, {Klar}, {Mangels}, {McMeekin}, {Miles}, {Moretti},
  {Mori}, {Morris}, {Mukerjee}, {Osborne}, {Short}, {Tagliaferri},
  {Tamburelli}, {Watson}, {Willingale}, \& {Zugger}}]{2004SPIE.5165..201B}
{Burrows}, D.~N., {Hill}, J.~E., {Nousek}, J.~A., {et~al.} 2004, in \procspie,
  Vol. 5165, X-Ray and Gamma-Ray Instrumentation for Astronomy XIII, ed. K.~A.
  {Flanagan} \& O.~H.~W. {Siegmund}, 201--216

\bibitem[{{Buson} {et~al.}(2023){Buson}, {Tramacere}, {Oswald}, {Barbano},
  {Fichet de Clairfontaine}, {Pfeiffer}, {Azzollini}, {Baghmanyan}, \&
  {Ajello}}]{2023arXiv230511263B}
{Buson}, S., {Tramacere}, A., {Oswald}, L., {et~al.} 2023, arXiv e-prints,
  arXiv:2305.11263

\bibitem[{{Buson} {et~al.}(2022){Buson}, {Tramacere}, {Pfeiffer}, {Oswald}, {de
  Menezes}, {Azzollini}, \& {Ajello}}]{2022ApJ...933L..43B}
{Buson}, S., {Tramacere}, A., {Pfeiffer}, L., {et~al.} 2022, \apjl, 933, L43

\bibitem[{{Caputo} {et~al.}(2022){Caputo}, {Ajello}, {Kierans}, {Perkins},
  {Racusin}, {Baldini}, {Baring}, {Bissaldi}, {Burns}, {Cannady}, {Charles},
  {da Silva}, {Fang}, {Fleischhack}, {Fryer}, {Fukazawa}, {Grove}, {Hartmann},
  {Howell}, {Jadhav}, {Karwin}, {Kocevski}, {Kurahashi}, {Latronico}, {Lewis},
  {Leys}, {Lien}, {Marcotulli}, {Martinez-Castellanos}, {Mazziotta}, {McEnery},
  {Metcalfe}, {Murase}, {Negro}, {Parker}, {Phlips}, {Prescod-Weinstein},
  {Razzaque}, {Shawhan}, {Sheng}, {Shutt}, {Shy}, {Sleator}, {Steinhebel},
  {Striebig}, {Suda}, {Tak}, {Tajima}, {Valverde}, {Venters}, {Wadiasingh},
  {Woolf}, {Wulf}, {Zhang}, \& {Zoglauer}}]{2022JATIS...8d4003C}
{Caputo}, R., {Ajello}, M., {Kierans}, C.~A., {et~al.} 2022, Journal of
  Astronomical Telescopes, Instruments, and Systems, 8, 044003

\bibitem[{{Carrasco} {et~al.}(2021){Carrasco}, {Recillas}, {Escobedo},
  {Porras}, {Chavushyan}, \& {Mayya}}]{2021ATel15148....1C}
{Carrasco}, L., {Recillas}, E., {Escobedo}, G., {et~al.} 2021, The Astronomer's
  Telegram, 15148, 1

\bibitem[{{Carswell} {et~al.}(1974){Carswell}, {Strittmatter}, {Williams},
  {Kinman}, \& {Serkowski}}]{Carswell1974}
{Carswell}, R.~F., {Strittmatter}, P.~A., {Williams}, R.~E., {Kinman}, T.~D.,
  \& {Serkowski}, K. 1974, \apjl, 190, L101

\bibitem[{{Cash}(1979)}]{1979ApJ...228..939C}
{Cash}, W. 1979, \apj, 228, 939

\bibitem[{{Cerruti} {et~al.}(2019){Cerruti}, {Zech}, {Boisson}, {Emery},
  {Inoue}, \& {Lenain}}]{2019MNRAS.483L..12C}
{Cerruti}, M., {Zech}, A., {Boisson}, C., {et~al.} 2019, \mnras, 483, L12

\bibitem[{Cerruti {et~al.}(2015)Cerruti, Zech, Boisson, \&
  Inoue}]{10.1093/mnras/stu2691}
Cerruti, M., Zech, A., Boisson, C., \& Inoue, S. 2015, Monthly Notices of the
  Royal Astronomical Society, 448, 910

\bibitem[{{Dembinski} {et~al.}(2020){Dembinski}, {Ongmongkolkul}, {Deil},
  {Schreiner}, {Feickert}, {Burr}, {Watson}, {Rost}, {Pearce}, \&
  {Geiger}}]{iminuit}
{Dembinski}, H., {Ongmongkolkul}, P., {Deil}, C., {et~al.} 2020

\bibitem[{{Dom{\'\i}nguez} {et~al.}(2011){Dom{\'\i}nguez}, {Primack},
  {Rosario}, {Prada}, {Gilmore}, {Faber}, {Koo}, {Somerville},
  {P{\'e}rez-Torres}, {P{\'e}rez-Gonz{\'a}lez}, {Huang}, {Davis},
  {Guhathakurta}, {Barmby}, {Conselice}, {Lozano}, {Newman}, \&
  {Cooper}}]{2011MNRAS.410.2556D}
{Dom{\'\i}nguez}, A., {Primack}, J.~R., {Rosario}, D.~J., {et~al.} 2011,
  \mnras, 410, 2556

\bibitem[{{Dzhilkibaev} {et~al.}(2021){Dzhilkibaev}, {Suvorova}, \& {Baikal-GVD
  Collaboration}}]{2021ATel15112....1D}
{Dzhilkibaev}, Z.~A., {Suvorova}, O., \& {Baikal-GVD Collaboration}. 2021, The
  Astronomer's Telegram, 15112, 1

\bibitem[{{Evans} {et~al.}(2009){Evans}, {Beardmore}, {Page}, {Osborne},
  {O'Brien}, {Willingale}, {Starling}, {Burrows}, {Godet}, {Vetere}, {Racusin},
  {Goad}, {Wiersema}, {Angelini}, {Capalbi}, {Chincarini}, {Gehrels}, {Kennea},
  {Margutti}, {Morris}, {Mountford}, {Pagani}, {Perri}, {Romano}, \&
  {Tanvir}}]{2009MNRAS.397.1177E}
{Evans}, P.~A., {Beardmore}, A.~P., {Page}, K.~L., {et~al.} 2009, \mnras, 397,
  1177

\bibitem[{{Fallah Ramazani} {et~al.}(2017){Fallah Ramazani}, {Lindfors}, \&
  {Nilsson}}]{2017A&A...608A..68F}
{Fallah Ramazani}, V., {Lindfors}, E., \& {Nilsson}, K. 2017, \aap, 608, A68

\bibitem[{{Falomo} {et~al.}(2021){Falomo}, {Treves}, \& {Paiano}}]{Falomo2021}
{Falomo}, R., {Treves}, A., \& {Paiano}, S. 2021, The Astronomer's Telegram,
  15132, 1

\bibitem[{{Fichet de Clairfontaine} {et~al.}(2023){Fichet de Clairfontaine},
  {Buson}, {Pfeiffer}, {Marchesi}, {Azzollini}, {Baghmanyan}, {Tramacere},
  {Barbano}, \& {Oswald}}]{2023ApJ...958L...2F}
{Fichet de Clairfontaine}, G., {Buson}, S., {Pfeiffer}, L., {et~al.} 2023,
  \apjl, 958, L2

\bibitem[{{Filippenko} {et~al.}(2001){Filippenko}, {Li}, {Treffers}, \&
  {Modjaz}}]{2001ASPC..246..121F}
{Filippenko}, A.~V., {Li}, W.~D., {Treffers}, R.~R., \& {Modjaz}, M. 2001, in
  Astronomical Society of the Pacific Conference Series, Vol. 246, IAU Colloq.
  183: Small Telescope Astronomy on Global Scales, ed. B.~{Paczynski}, W.-P.
  {Chen}, \& C.~{Lemme}, 121

\bibitem[{{Filippini} {et~al.}(2022){Filippini}, {Illuminati}, {Heijboer},
  {Gatius}, {Muller}, {Dornic}, {Huang}, {Le Stum}, {Palacios Gonz{\'a}lez},
  {Celli}, {Zegarelli}, {Coniglione}, {Samtleben}, {Kovalev}, \&
  {Plavin}}]{2022ATel15290....1F}
{Filippini}, F., {Illuminati}, G., {Heijboer}, A., {et~al.} 2022, The
  Astronomer's Telegram, 15290, 1

\bibitem[{Franckowiak {et~al.}(2020)Franckowiak, Garrappa, Paliya, Shappee,
  Stein, Strotjohann, Kowalski, Buson, Kiehlmann, Max-Moerbeck, \&
  Angioni}]{Franckowiak_2020}
Franckowiak, A., Garrappa, S., Paliya, V., {et~al.} 2020, The Astrophysical
  Journal, 893, 162

\bibitem[{{Gao} {et~al.}(2017){Gao}, {Pohl}, \& {Winter}}]{2017ApJ...843..109G}
{Gao}, S., {Pohl}, M., \& {Winter}, W. 2017, \apj, 843, 109

\bibitem[{{Garrappa} {et~al.}(2019){Garrappa}, {Buson}, {Franckowiak},
  {Fermi-LAT Collaboration}, {Shappee}, {Beacom}, {Dong}, {Holoien},
  {Kochanek}, {Prieto}, {Stanek}, {Thompson}, {ASAS-SN Collaboration},
  {Aartsen}, {Ackermann}, {Adams}, {Aguilar}, {Ahlers}, {Ahrens}, {Alispach},
  {Andeen}, {Anderson}, {Ansseau}, {Anton}, {Arg{\"u}elles}, {Auffenberg},
  {Axani}, {Backes}, {Bagherpour}, {Bai}, {Barbano}, {Barwick}, {Baum}, {Bay},
  {Beatty}, {Becker}, {Becker Tjus}, {BenZvi}, {Berley}, {Bernardini},
  {Besson}, {Binder}, {Bindig}, {Blaufuss}, {Blot}, {Bohm}, {B{\"o}rner},
  {B{\"o}ser}, {Botner}, {Bourbeau}, {Bourbeau}, {Bradascio}, {Braun}, {Bretz},
  {Bron}, {Brostean-Kaiser}, {Burgman}, {Busse}, {Carver}, {Chen}, {Cheung},
  {Chirkin}, {Clark}, {Classen}, {Collin}, {Conrad}, {Coppin}, {Correa},
  {Cowen}, {Cross}, {Dave}, {de Andr{\'e}}, {De Clercq}, {DeLaunay},
  {Dembinski}, {Deoskar}, {De Ridder}, {Desiati}, {de Vries}, {de Wasseige},
  {de With}, {DeYoung}, {Diaz}, {D{\'\i}az-V{\'e}lez}, {Dujmovic}, {Dunkman},
  {Dvorak}, {Eberhardt}, {Ehrhardt}, {Eller}, {Evenson}, {Fahey}, {Fazely},
  {Felde}, {Filimonov}, {Finley}, {Franckowiak}, {Friedman}, {Fritz},
  {Gaisser}, {Gallagher}, {Ganster}, {Garrappa}, {Gerhardt}, {Ghorbani},
  {Glauch}, {Gl{\"u}senkamp}, {Goldschmidt}, {Gonzalez}, {Grant}, {Griffith},
  {G{\"u}nder}, {G{\"u}nd{\"u}z}, {Haack}, {Hallgren}, {Halve}, {Halzen},
  {Hanson}, {Hebecker}, {Heereman}, {Helbing}, {Hellauer}, {Henningsen},
  {Hickford}, {Hignight}, {Hill}, {Hoffman}, {Hoffmann}, {Hoinka},
  {Hokanson-Fasig}, {Hoshina}, {Huang}, {Huber}, {Hultqvist}, {H{\"u}nnefeld},
  {Hussain}, {In}, {Iovine}, {Ishihara}, {Jacobi}, {Japaridze}, {Jeong},
  {Jero}, {Jones}, {Kang}, {Kappes}, {Kappesser}, {Karg}, {Karl}, {Karle},
  {Katz}, {Kauer}, {Keivani}, {Kelley}, {Kheirandish}, {Kim}, {Kintscher},
  {Kiryluk}, {Kittler}, {Klein}, {Koirala}, {Kolanoski}, {K{\"o}pke}, {Kopper},
  {Kopper}, {Koskinen}, {Kowalski}, {Krings}, {Kr{\"u}ckl}, {Kulacz}, {Kunwar},
  {Kurahashi}, {Kyriacou}, {Labare}, {Lanfranchi}, {Larson}, {Lauber}, {Lazar},
  {Leonard}, {Leuermann}, {Liu}, {Lohfink}, {Lozano Mariscal}, {Lu},
  {Lucarelli}, {L{\"u}nemann}, {Luszczak}, {Madsen}, {Maggi}, {Mahn}, {Makino},
  {Mallot}, {Mancina}, {Mari{\c{s}}}, {Maruyama}, {Mase}, {Maunu}, {Meagher},
  {Medici}, {Medina}, {Meier}, {Meighen-Berger}, {Menne}, {Merino}, {Meures},
  {Miarecki}, {Micallef}, {Moment{\'e}}, {Montaruli}, {Moore}, {Moulai},
  {Nagai}, {Nahnhauer}, {Nakarmi}, {Naumann}, {Neer}, {Niederhausen},
  {Nowicki}, {Nygren}, {Obertacke Pollmann}, {Olivas}, {O'Murchadha},
  {O'Sullivan}, {Palczewski}, {Pandya}, {Pankova}, {Park}, {Peiffer},
  {P{\'e}rez de los Heros}, {Pieloth}, {Pinat}, {Pizzuto}, {Plum}, {Price},
  {Przybylski}, {Raab}, {Raissi}, {Rameez}, {Rauch}, {Rawlins}, {Rea},
  {Reimann}, {Relethford}, {Renzi}, {Resconi}, {Rhode}, {Richman}, {Robertson},
  {Rongen}, {Rott}, {Ruhe}, {Ryckbosch}, {Rysewyk}, {Safa}, {Sanchez Herrera},
  {Sandrock}, {Sandroos}, {Santander}, {Sarkar}, {Sarkar}, {Satalecka},
  {Schaufel}, {Schlunder}, {Schmidt}, {Schneider}, {Schneider}, {Schumacher},
  {Sclafani}, {Seckel}, {Seunarine}, {Silva}, {Snihur}, {Soedingrekso},
  {Soldin}, {Song}, {Spiczak}, {Spiering}, {Stachurska}, {Stamatikos},
  {Stanev}, {Stasik}, {Stein}, {Stettner}, {Steuer}, {Stezelberger},
  {Stokstad}, {St{\"o}{\ss}l}, {Strotjohann}, {Stuttard}, {Sullivan},
  {Sutherland}, {Taboada}, {Tenholt}, {Ter-Antonyan}, {Terliuk}, {Tilav},
  {Tomankova}, {T{\"o}nnis}, {Toscano}, {Tosi}, {Tselengidou}, {Tung},
  {Turcati}, {Turcotte}, {Turley}, {Ty}, {Unger}, {Unland Elorrieta}, {Usner},
  {Vandenbroucke}, {Van Driessche}, {van Eijk}, {van Eijndhoven}, {Vanheule},
  {van Santen}, {Vraeghe}, {Walck}, {Wallace}, {Wallraff}, {Wandkowsky},
  {Watson}, {Weaver}, {Weiss}, {Weldert}, {Wendt}, {Werthebach}, {Westerhoff},
  {Whelan}, {Whitehorn}, {Wiebe}, {Wiebusch}, {Wille}, {Williams}, {Wills},
  {Wolf}, {Wood}, {Wood}, {Woschnagg}, {Wrede}, {Xu}, {Xu}, {Xu}, {Yanez},
  {Yodh}, {Yoshida}, {Yuan}, \& {IceCube Collaboration}}]{2019ApJ...880..103G}
{Garrappa}, S., {Buson}, S., {Franckowiak}, A., {et~al.} 2019, \apj, 880, 103

\bibitem[{{Garrappa} {et~al.}(2024){Garrappa}, {Buson}, {Sinapius},
  {Franckowiak}, {Liodakis}, {Bartolini}, {Giroletti}, {Nanci}, {Principe}, \&
  {Venters}}]{Garrappa2024}
{Garrappa}, S., {Buson}, S., {Sinapius}, J., {et~al.} 2024, arXiv e-prints,
  arXiv:2401.06666

\bibitem[{Gasparyan {et~al.}(2022)Gasparyan, B{\'e}gu{\'e}, \&
  Sahakyan}]{gasparyan2022time}
Gasparyan, S., B{\'e}gu{\'e}, D., \& Sahakyan, N. 2022, Monthly Notices of the
  Royal Astronomical Society, 509, 2102

\bibitem[{Ghisellini {et~al.}(2010)Ghisellini, Tavecchio, Foschini, Ghirlanda,
  Maraschi, \& Celotti}]{10.1111/j.1365-2966.2009.15898.x}
Ghisellini, G., Tavecchio, F., Foschini, L., {et~al.} 2010, Monthly Notices of
  the Royal Astronomical Society, 402, 497

\bibitem[{{Giommi} \& {Padovani}(2021)}]{2021Univ....7..492G}
{Giommi}, P. \& {Padovani}, P. 2021, Universe, 7, 492

\bibitem[{Gupta {et~al.}(2012)Gupta, Pandey, Singh, Rani, Pan, Fan, \&
  Gupta}]{GUPTA20128}
Gupta, S., Pandey, U., Singh, K., {et~al.} 2012, New Astronomy, 17, 8

\bibitem[{Han \& {Neumann}(2006)}]{doi:10.1080/10556780512331318290}
Han, L. \& {Neumann}, M. 2006, Optimization Methods and Software, 21, 1

\bibitem[{{IceCube Collaboration}(2013)}]{doi:10.1126/science.1242856}
{IceCube Collaboration}. 2013, Science, 342, 1242856

\bibitem[{{IceCube Collaboration}(2017)}]{2017ApJ...835..151A}
{IceCube Collaboration}. 2017, The Astrophysical Journal, 835, 151

\bibitem[{{IceCube Collaboration}(2021)}]{2021GCN.31191....1I}
{IceCube Collaboration}. 2021, GRB Coordinates Network, 31191, 1

\bibitem[{{IceCube Collaboration} {et~al.}(2018{\natexlab{a}}){IceCube
  Collaboration}, {Aartsen}, {Ackermann}, {Adams}, {Aguilar}, {Ahlers},
  {Ahrens}, {Al Samarai}, {Altmann}, {Andeen}, {Anderson}, {Ansseau}, {Anton},
  {Arg{\"u}elles}, {Auffenberg}, {Axani}, {Bagherpour}, {Bai}, {Barron},
  {Barwick}, {Baum}, {Bay}, {Beatty}, {Becker Tjus}, {Becker}, {BenZvi},
  {Berley}, {Bernardini}, {Besson}, {Binder}, {Bindig}, {Blaufuss}, {Blot},
  {Bohm}, {B{\"o}rner}, {Bos}, {B{\"o}ser}, {Botner}, {Bourbeau}, {Bourbeau},
  {Bradascio}, {Braun}, {Brenzke}, {Bretz}, {Bron}, {Brostean-Kaiser},
  {Burgman}, {Busse}, {Carver}, {Cheung}, {Chirkin}, {Christov}, {Clark},
  {Classen}, {Coenders}, {Collin}, {Conrad}, {Coppin}, {Correa}, {Cowen},
  {Cross}, {Dave}, {Day}, {de Andr{\'e}}, {De Clercq}, {DeLaunay}, {Dembinski},
  {De Ridder}, {Desiati}, {de Vries}, {de Wasseige}, {de With}, {DeYoung},
  {D{\'\i}az-V{\'e}lez}, {di Lorenzo}, {Dujmovic}, {Dumm}, {Dunkman}, {Dvorak},
  {Eberhardt}, {Ehrhardt}, {Eichmann}, {Eller}, {Evenson}, {Fahey}, {Fazely},
  {Felde}, {Filimonov}, {Finley}, {Flis}, {Franckowiak}, {Friedman}, {Fritz},
  {Gaisser}, {Gallagher}, {Gerhardt}, {Ghorbani}, {Glauch}, {Gl{\"u}senkamp},
  {Goldschmidt}, {Gonzalez}, {Grant}, {Griffith}, {Haack}, {Hallgren},
  {Halzen}, {Hanson}, {Hebecker}, {Heereman}, {Helbing}, {Hellauer},
  {Hickford}, {Hignight}, {Hill}, {Hoffman}, {Hoffmann}, {Hoinka},
  {Hokanson-Fasig}, {Hoshina}, {Huang}, {Huber}, {Hultqvist}, {H{\"u}nnefeld},
  {Hussain}, {In}, {Iovine}, {Ishihara}, {Jacobi}, {Japaridze}, {Jeong},
  {Jero}, {Jones}, {Kalaczynski}, {Kang}, {Kappes}, {Kappesser}, {Karg},
  {Karle}, {Katz}, {Kauer}, {Keivani}, {Kelley}, {Kheirandish}, {Kim}, {Kim},
  {Kintscher}, {Kiryluk}, {Kittler}, {Klein}, {Koirala}, {Kolanoski},
  {K{\"o}pke}, {Kopper}, {Kopper}, {Koschinsky}, {Koskinen}, {Kowalski},
  {Krings}, {Kroll}, {Kr{\"u}ckl}, {Kunwar}, {Kurahashi}, {Kuwabara},
  {Kyriacou}, {Labare}, {Lanfranchi}, {Larson}, {Lauber}, {Leonard},
  {Lesiak-Bzdak}, {Leuermann}, {Liu}, {Lozano Mariscal}, {Lu}, {L{\"u}nemann},
  {Luszczak}, {Madsen}, {Maggi}, {Mahn}, {Mancina}, {Maruyama}, {Mase},
  {Maunu}, {Meagher}, {Medici}, {Meier}, {Menne}, {Merino}, {Meures},
  {Miarecki}, {Micallef}, {Moment{\'e}}, {Montaruli}, {Moore}, {Morse},
  {Moulai}, {Nahnhauer}, {Nakarmi}, {Naumann}, {Neer}, {Niederhausen},
  {Nowicki}, {Nygren}, {Obertacke Pollmann}, {Olivas}, {O'Murchadha},
  {O'Sullivan}, {Palczewski}, {Pandya}, {Pankova}, {Peiffer}, {Pepper},
  {P{\'e}rez de los Heros}, {Pieloth}, {Pinat}, {Plum}, {Price}, {Przybylski},
  {Raab}, {R{\"a}del}, {Rameez}, {Rauch}, {Rawlins}, {Rea}, {Reimann},
  {Relethford}, {Relich}, {Resconi}, {Rhode}, {Richman}, {Robertson}, {Rongen},
  {Rott}, {Ruhe}, {Ryckbosch}, {Rysewyk}, {Safa}, {S{\"a}lzer}, {Sanchez
  Herrera}, {Sandrock}, {Sandroos}, {Santander}, {Sarkar}, {Sarkar},
  {Satalecka}, {Schlunder}, {Schmidt}, {Schneider}, {Schoenen},
  {Sch{\"o}neberg}, {Schumacher}, {Sclafani}, {Seckel}, {Seunarine},
  {Soedingrekso}, {Soldin}, {Song}, {Spiczak}, {Spiering}, {Stachurska},
  {Stamatikos}, {Stanev}, {Stasik}, {Stein}, {Stettner}, {Steuer},
  {Stezelberger}, {Stokstad}, {St{\"o}{\ss}l}, {Strotjohann}, {Stuttard},
  {Sullivan}, {Sutherland}, {Taboada}, {Tatar}, {Tenholt}, {Ter-Antonyan},
  {Terliuk}, {Tilav}, {Toale}, {Tobin}, {Toennis}, {Toscano}, {Tosi},
  {Tselengidou}, {Tung}, {Turcati}, {Turley}, {Ty}, {Unger}, {Usner},
  {Vandenbroucke}, {Van Driessche}, {van Eijk}, {van Eijndhoven}, {Vanheule},
  {van Santen}, {Vogel}, {Vraeghe}, {Walck}, {Wallace}, {Wallraff}, {Wandler},
  {Wandkowsky}, {Waza}, {Weaver}, {Weiss}, {Wendt}, {Werthebach}, {Westerhoff},
  {Whelan}, {Whitehorn}, {Wiebe}, {Wiebusch}, {Wille}, {Williams}, {Wills},
  {Wolf}, {Wood}, {Wood}, {Woschnagg}, {Xu}, {Xu}, {Xu}, {Yanez}, {Yodh},
  {Yoshida}, {Yuan}, {Fermi-LAT Collaboration}, {Abdollahi}, {Ajello},
  {Angioni}, {Baldini}, {Ballet}, {Barbiellini}, {Bastieri}, {Bechtol},
  {Bellazzini}, {Berenji}, {Bissaldi}, {Blandford}, {Bonino}, {Bottacini},
  {Bregeon}, {Bruel}, {Buehler}, {Burnett}, {Burns}, {Buson}, {Cameron},
  {Caputo}, {Caraveo}, {Cavazzuti}, {Charles}, {Chen}, {Cheung}, {Chiang},
  {Chiaro}, {Ciprini}, {Cohen-Tanugi}, {Conrad}, {Costantin}, {Cutini},
  {D'Ammando}, {de Palma}, {Digel}, {Di Lalla}, {Di Mauro}, {Di Venere},
  {Dom{\'\i}nguez}, {Favuzzi}, {Franckowiak}, {Fukazawa}, {Funk}, {Fusco},
  {Gargano}, {Gasparrini}, {Giglietto}, {Giomi}, {Giommi}, {Giordano},
  {Giroletti}, {Glanzman}, {Green}, {Grenier}, {Grondin}, {Guiriec}, {Harding},
  {Hayashida}, {Hays}, {Hewitt}, {Horan}, {J{\'o}hannesson}, {Kadler},
  {Kensei}, {Kocevski}, {Krauss}, {Kreter}, {Kuss}, {La Mura}, {Larsson},
  {Latronico}, {Lemoine-Goumard}, {Li}, {Longo}, {Loparco}, {Lovellette},
  {Lubrano}, {Magill}, {Maldera}, {Malyshev}, {Manfreda}, {Mazziotta},
  {McEnery}, {Meyer}, {Michelson}, {Mizuno}, {Monzani}, {Morselli},
  {Moskalenko}, {Negro}, {Nuss}, {Ojha}, {Omodei}, {Orienti}, {Orlando},
  {Palatiello}, {Paliya}, {Perkins}, {Persic}, {Pesce-Rollins}, {Piron},
  {Porter}, {Principe}, {Rain{\`o}}, {Rando}, {Rani}, {Razzano}, {Razzaque},
  {Reimer}, {Reimer}, {Renault-Tinacci}, {Ritz}, {Rochester}, {Saz Parkinson},
  {Sgr{\`o}}, {Siskind}, {Spandre}, {Spinelli}, {Suson}, {Tajima}, {Takahashi},
  {Tanaka}, {Thayer}, {Thompson}, {Tibaldo}, {Torres}, {Torresi}, {Tosti},
  {Troja}, {Valverde}, {Vianello}, {Vogel}, {Wood}, {Wood}, {Zaharijas}, {MAGIC
  Collaboration}, {Ahnen}, {Ansoldi}, {Antonelli}, {Arcaro}, {Baack},
  {Babi{\'c}}, {Banerjee}, {Bangale}, {Barres de Almeida}, {Barrio}, {Becerra
  Gonz{\'a}lez}, {Bednarek}, {Bernardini}, {Berti}, {Bhattacharyya}, {Biland},
  {Blanch}, {Bonnoli}, {Carosi}, {Carosi}, {Ceribella}, {Chatterjee}, {Colak},
  {Colin}, {Colombo}, {Contreras}, {Cortina}, {Covino}, {Cumani}, {Da Vela},
  {Dazzi}, {De Angelis}, {De Lotto}, {Delfino}, {Delgado}, {Di Pierro},
  {Dom{\'\i}nguez}, {Dominis Prester}, {Dorner}, {Doro}, {Einecke},
  {Elsaesser}, {Fallah Ramazani}, {Fern{\'a}ndez-Barral}, {Fidalgo}, {Foffano},
  {Pfrang}, {Fonseca}, {Font}, {Franceschini}, {Fruck}, {Galindo}, {Gallozzi},
  {Garc{\'\i}a L{\'o}pez}, {Garczarczyk}, {Gaug}, {Giammaria}, {Godinovi{\'c}},
  {Gora}, {Guberman}, {Hadasch}, {Hahn}, {Hassan}, {Hayashida}, {Herrera},
  {Hose}, {Hrupec}, {Inoue}, {Ishio}, {Konno}, {Kubo}, {Kushida}, {Lelas},
  {Lindfors}, {Lombardi}, {Longo}, {L{\'o}pez}, {Maggio}, {Majumdar},
  {Makariev}, {Maneva}, {Manganaro}, {Mannheim}, {Maraschi}, {Mariotti},
  {Mart{\'\i}nez}, {Masuda}, {Mazin}, {Minev}, {M}, {Mirzoyan}, {Moralejo},
  {Moreno}, {Moretti}, {Nagayoshi}, {Neustroev}, {Niedzwiecki}, {Nievas
  Rosillo}, {Nigro}, {Nilsson}, {Ninci}, {Nishijima}, {Noda}, {Nogu{\'e}s},
  {Paiano}, {Palacio}, {Paneque}, {Paoletti}, {Paredes}, {Pedaletti},
  {Peresano}, {Persic}, {Prada Moroni}, {Prandini}, {Puljak}, {Rodriguez
  Garcia}, {Reichardt}, {Rhode}, {Rib{\'o}}, {Rico}, {Righi}, {Rugliancich},
  {Saito}, {Satalecka}, {Schweizer}, {Sitarek}, {{\v{S}}nidaric
  {\textasciiacute}}, {Sobczynska}, {Stamerra}, {Strzys}, {Suri{\'c}},
  {Takahashi}, {Tavecchio}, {Temnikov}, {Terzi{\'c}}, {Teshima},
  {Torres-Alb{\`a}}, {Treves}, {Tsujimoto}, {Vanzo}, {Vazquez Acosta}, {Vovk},
  {Ward}, {Will}, {S}, {Zaric {\textasciiacute}}, {AGILE Team}, {Lucarelli},
  {Tavani}, {Piano}, {Donnarumma}, {Pittori}, {Verrecchia}, {Barbiellini},
  {Bulgarelli}, {Caraveo}, {Cattaneo}, {Colafrancesco}, {Costa}, {Di Cocco},
  {Ferrari}, {Gianotti}, {Giuliani}, {Lipari}, {Mereghetti}, {Morselli},
  {Pacciani}, {Paoletti}, {Parmiggiani}, {Pellizzoni}, {Picozza}, {Pilia},
  {Rappoldi}, {Trois}, {Vercellone}, {Vittorini}, {ASAS-SN Team}, {Stanek},
  {Franckowiak}, {Kochanek}, {Beacom}, {Thompson}, {Holoien}, {Dong}, {Prieto},
  {Shappee}, {Holmbo}, {HAWC Collaboration}, {Abeysekara}, {Albert}, {Alfaro},
  {Alvarez}, {Arceo}, {Arteaga-Vel{\'a}zquez}, {Avila Rojas}, {Ayala Solares},
  {Becerril}, {Belmont-Moreno}, {Bernal}, {Caballero-Mora}, {Capistr{\'a}n},
  {Carrami{\~n}ana}, {Casanova}, {Castillo}, {Cotti}, {Cotzomi}, {Couti{\~n}o
  de Le{\'o}n}, {De Le{\'o}n}, {De la Fuente}, {Diaz Hernandez}, {Dichiara},
  {Dingus}, {DuVernois}, {D{\'\i}az-V{\'e}lez}, {Ellsworth}, {Engel},
  {Fiorino}, {Fleischhack}, {Fraija}, {Garc{\'\i}a-Gonz{\'a}lez}, {Garfias},
  {Gonz{\'a}lez Mu{\~n}oz}, {Gonz{\'a}lez}, {Goodman}, {Hampel-Arias},
  {Harding}, {Hernandez}, {Hona}, {Hueyotl-Zahuantitla}, {Hui},
  {H{\"u}ntemeyer}, {Iriarte}, {Jardin-Blicq}, {Joshi}, {Kaufmann}, {Kunde},
  {Lara}, {Lauer}, {Lee}, {Lennarz}, {Le{\'o}n Vargas}, {Linnemann},
  {Longinotti}, {Luis-Raya}, {Luna-Garc{\'\i}a}, {Malone}, {Marinelli},
  {Martinez}, {Martinez-Castellanos}, {Mart{\'\i}nez-Castro},
  {Mart{\'\i}nez-Huerta}, {Matthews}, {Miranda-Romagnoli}, {Moreno},
  {Mostaf{\'a}}, {Nayerhoda}, {Nellen}, {Newbold}, {Nisa}, {Noriega-Papaqui},
  {Pelayo}, {Pretz}, {P{\'e}rez-P{\'e}rez}, {Ren}, {Rho}, {Rivi{\`e}re},
  {Rosa-Gonz{\'a}lez}, {Rosenberg}, {Ruiz-Velasco}, {Ruiz-Velasco}, {Salesa
  Greus}, {Sandoval}, {Schneider}, {Schoorlemmer}, {Sinnis}, {Smith},
  {Springer}, {Surajbali}, {Tibolla}, {Tollefson}, {Torres}, {Villase{\~n}or},
  {Weisgarber}, {Werner}, {Yapici}, {Gaurang}, {Zepeda}, {Zhou}, {{\'A}lvarez},
  {H.~E.~S.~S. Collaboration}, {Abdalla}, {Ang{\"u}ner}, {Armand}, {Backes},
  {Becherini}, {Berge}, {B{\"o}ttcher}, {Boisson}, {Bolmont}, {Bonnefoy},
  {Bordas}, {Brun}, {B{\"u}chele}, {Bulik}, {Caroff}, {Carosi}, {Casanova},
  {Cerruti}, {Chakraborty}, {Chandra}, {Chen}, {Colafrancesco}, {Davids},
  {Deil}, {Devin}, {Djannati-Ata{\"\i}}, {Egberts}, {Emery}, {Eschbach},
  {Fiasson}, {Fontaine}, {Funk}, {F{\"u}{\ss}ling}, {Gallant}, {Gat{\'e}},
  {Giavitto}, {Glawion}, {Glicenstein}, {Gottschall}, {Grondin}, {Haupt},
  {Henri}, {Hinton}, {Hoischen}, {Holch}, {Huber}, {Jamrozy}, {Jankowsky},
  {Jankowsky}, {Jouvin}, {Jung-Richardt}, {Kerszberg}, {Kh{\'e}lifi}, {King},
  {Klepser}, {Kluz {\textasciiacute}niak}, {Komin}, {Kraus}, {Lefaucheur},
  {Lemi{\`e}re}, {Lemoine-Goumard}, {Lenain}, {Leser}, {Lohse},
  {L{\'o}pez-Coto}, {Lorentz}, {Lypova}, {Marandon}, {Guillem
  Mart{\'\i}-Devesa}, {Maurin}, {Mitchell}, {Moderski}, {Mohamed}, {Mohrmann},
  {Moulin}, {Murach}, {de Naurois}, {Niederwanger}, {Niemiec}, {Oakes},
  {O'Brien}, {Ohm}, {Ostrowski}, {Oya}, {Panter}, {Parsons}, {Perennes},
  {Piel}, {Pita}, {Poireau}, {Priyana Noel}, {Prokoph}, {P{\"u}hlhofer},
  {Quirrenbach}, {Raab}, {Rauth}, {Renaud}, {Rieger}, {Rinchiuso}, {Romoli},
  {Rowell}, {Rudak}, {Sasaki}, {Sanchez}, {Schlickeiser}, {Sch{\"u}ssler},
  {Schulz}, {Schwanke}, {Seglar-Arroyo}, {Shafi}, {Simoni}, {Sol}, {Stegmann},
  {Steppa}, {Tavernier}, {Taylor}, {Tiziani}, {Trichard}, {Tsirou}, {van
  Eldik}, {van Rensburg}, {van Soelen}, {Veh}, {Vincent}, {Voisin}, {Wagner},
  {Wagner}, {Wierzcholska}, {Zanin}, {Zdziarski}, {Zech}, {Ziegler}, {Zorn},
  {{\.Z}ywucka}, {INTEGRAL Team}, {Savchenko}, {Ferrigno}, {Bazzano}, {Diehl},
  {Kuulkers}, {Laurent}, {Mereghetti}, {Natalucci}, {Panessa}, {Rodi},
  {Ubertini}, {Kanata}, Teams, {Morokuma}, {Ohta}, {Tanaka}, {Mori},
  {Yamanaka}, {Kawabata}, {Utsumi}, {Nakaoka}, {Kawabata}, {Nagashima},
  {Yoshida}, {Matsuoka}, {Itoh}, {Kapteyn Team}, {Keel}, {Liverpool Telescope
  Team}, {Copperwheat}, {Steele}, {Swift/NuSTAR Team}, {Cenko}, {Cowen},
  {DeLaunay}, {Evans}, {Fox}, {Keivani}, {Kennea}, {Marshall}, {Osborne},
  {Santander}, {Tohuvavohu}, {Turley}, {VERITAS Collaboration}, {Abeysekara},
  {Archer}, {Benbow}, {Bird}, {Brill}, {Brose}, {Buchovecky}, {Buckley},
  {Bugaev}, {Christiansen}, {Connolly}, {Cui}, {Daniel}, {Errando}, {Falcone},
  {Feng}, {Finley}, {Fortson}, {Furniss}, {Gueta}, {H{\"u}tten}, {Hervet},
  {Hughes}, {Humensky}, {Johnson}, {Kaaret}, {Kar}, {Kelley-Hoskins},
  {Kertzman}, {Kieda}, {Krause}, {Krennrich}, {Kumar}, {Lang}, {Lin}, {Maier},
  {McArthur}, {Moriarty}, {Mukherjee}, {Nieto}, {O'Brien}, {Ong}, {Otte},
  {Park}, {Petrashyk}, {Pohl}, {Popkow}, {Pueschel}, {Quinn}, {Ragan},
  {Reynolds}, {Richards}, {Roache}, {Rulten}, {Sadeh}, {Santander}, {Scott},
  {Sembroski}, {Shahinyan}, {Sushch}, {Tr{\'e}panier}, {Tyler}, {Vassiliev},
  {Wakely}, {Weinstein}, {Wells}, {Wilcox}, {Wilhelm}, {Williams}, {Zitzer},
  {VLA/B Team}, {Tetarenko}, {Kimball}, {Miller-Jones}, \&
  {Sivakoff}}]{2018Sci...361.1378I}
{IceCube Collaboration}, {Aartsen}, M.~G., {Ackermann}, M., {et~al.}
  2018{\natexlab{a}}, Science, 361, eaat1378

\bibitem[{{IceCube Collaboration} {et~al.}(2018{\natexlab{b}}){IceCube
  Collaboration}, {Aartsen}, {Ackermann}, {Adams}, {Aguilar}, {Ahlers},
  {Ahrens}, {Samarai}, {Altmann}, {Andeen}, {Anderson}, {Ansseau}, {Anton},
  {Arg{\"u}elles}, {Arsioli}, {Auffenberg}, {Axani}, {Bagherpour}, {Bai},
  {Barron}, {Barwick}, {Baum}, {Bay}, {Beatty}, {Becker Tjus}, {Becker},
  {BenZvi}, {Berley}, {Bernardini}, {Besson}, {Binder}, {Bindig}, {Blaufuss},
  {Blot}, {Bohm}, {B{\"o}rner}, {Bos}, {B{\"o}ser}, {Botner}, {Bourbeau},
  {Bourbeau}, {Bradascio}, {Braun}, {Brenzke}, {Bretz}, {Bron},
  {Brostean-Kaiser}, {Burgman}, {Busse}, {Carver}, {Cheung}, {Chirkin},
  {Christov}, {Clark}, {Classen}, {Coenders}, {Collin}, {Conrad}, {Coppin},
  {Correa}, {Cowen}, {Cross}, {Dave}, {Day}, {de Andr{\'e}}, {De Clercq},
  {DeLaunay}, {Dembinski}, {DeRidder}, {Desiati}, {de Vries}, {de Wasseige},
  {de With}, {DeYoung}, {D{\'\i}az-V{\'e}lez}, {di Lorenzo}, {Dujmovic},
  {Dumm}, {Dunkman}, {Dvorak}, {Eberhardt}, {Ehrhardt}, {Eichmann}, {Eller},
  {Evenson}, {Fahey}, {Fazely}, {Felde}, {Filimonov}, {Finley}, {Flis},
  {Franckowiak}, {Friedman}, {Fritz}, {Gaisser}, {Gallagher}, {Gerhardt},
  {Ghorbani}, {Giommi}, {Glauch}, {Gl{\"u}senkamp}, {Goldschmidt}, {Gonzalez},
  {Grant}, {Griffith}, {Haack}, {Hallgren}, {Halzen}, {Hanson}, {Hebecker},
  {Heereman}, {Helbing}, {Hellauer}, {Hickford}, {Hignight}, {Hill}, {Hoffman},
  {Hoffmann}, {Hoinka}, {Hokanson-Fasig}, {Hoshina}, {Huang}, {Huber},
  {Hultqvist}, {H{\"u}nnefeld}, {Hussain}, {In}, {Iovine}, {Ishihara},
  {Jacobi}, {Japaridze}, {Jeong}, {Jero}, {Jones}, {Kalaczynski}, {Kang},
  {Kappes}, {Kappesser}, {Karg}, {Karle}, {Katz}, {Kauer}, {Keivani}, {Kelley},
  {Kheirandish}, {Kim}, {Kim}, {Kintscher}, {Kiryluk}, {Kittler}, {Klein},
  {Koirala}, {Kolanoski}, {K{\"o}pke}, {Kopper}, {Kopper}, {Koschinsky},
  {Koskinen}, {Kowalski}, {Krammer}, {Krings}, {Kroll}, {Kr{\"u}ckl}, {Kunwar},
  {Kurahashi}, {Kuwabara}, {Kyriacou}, {Labare}, {Lanfranchi}, {Larson},
  {Lauber}, {Leonard}, {Lesiak-Bzdak}, {Leuermann}, {Liu}, {Lozano Mariscal},
  {Lu}, {L{\"u}nemann}, {Luszczak}, {Madsen}, {Maggi}, {Mahn}, {Mancina},
  {Maruyama}, {Mase}, {Maunu}, {Meagher}, {Medici}, {Meier}, {Menne}, {Merino},
  {Meures}, {Miarecki}, {Micallef}, {Moment{\'e}}, {Montaruli}, {Moore},
  {Morse}, {Moulai}, {Nahnhauer}, {Nakarmi}, {Naumann}, {Neer}, {Niederhausen},
  {Nowicki}, {Nygren}, {Obertacke Pollmann}, {Olivas}, {O'Murchadha},
  {O'Sullivan}, {Padovani}, {Palczewski}, {Pandya}, {Pankova}, {Peiffer},
  {Pepper}, {P{\'e}rez de los Heros}, {Pieloth}, {Pinat}, {Plum}, {Price},
  {Przybylski}, {Raab}, {R{\"a}del}, {Rameez}, {Rawlins}, {Rea}, {Reimann},
  {Relethford}, {Relich}, {Resconi}, {Rhode}, {Richman}, {Robertson}, {Rongen},
  {Rott}, {Ruhe}, {Ryckbosch}, {Rysewyk}, {Safa}, {Sahakyan}, {S{\"a}lzer},
  {Sanchez Herrera}, {Sandrock}, {Sandroos}, {Santander}, {Sarkar}, {Sarkar},
  {Satalecka}, {Schlunder}, {Schmidt}, {Schneider}, {Schoenen},
  {Sch{\"o}neberg}, {Schumacher}, {Sclafani}, {Seckel}, {Seunarine},
  {Soedingrekso}, {Soldin}, {Song}, {Spiczak}, {Spiering}, {Stachurska},
  {Stamatikos}, {Stanev}, {Stasik}, {Stettner}, {Steuer}, {Stezelberger},
  {Stokstad}, {St{\"o}{\ss}l}, {Strotjohann}, {Stuttard}, {Sullivan},
  {Sutherland}, {Taboada}, {Tatar}, {Tenholt}, {Ter-Antonyan}, {Terliuk},
  {Tilav}, {Toale}, {Tobin}, {Toennis}, {Toscano}, {Tosi}, {Tselengidou},
  {Tung}, {Turcati}, {Turley}, {Ty}, {Unger}, {Usner}, {Vandenbroucke}, {Van
  Driessche}, {van Eijk}, {van Eijndhoven}, {Vanheule}, {van Santen}, {Vogel},
  {Vraeghe}, {Walck}, {Wallace}, {Wallraff}, {Wandler}, {Wandkowsky}, {Waza},
  {Weaver}, {Weiss}, {Wendt}, {Werthebach}, {Westerhoff}, {Whelan},
  {Whitehorn}, {Wiebe}, {Wiebusch}, {Wille}, {Williams}, {Wills}, {Wolf},
  {Wood}, {Wood}, {Woschnagg}, {Xu}, {Xu}, {Xu}, {Yanez}, {Yodh}, {Yoshida}, \&
  {Yuan}}]{2018Sci...361..147I}
{IceCube Collaboration}, {Aartsen}, M.~G., {Ackermann}, M., {et~al.}
  2018{\natexlab{b}}, Science, 361, 147

\bibitem[{{James} \& {Roos}(1975)}]{1975CoPhC..10..343J}
{James}, F. \& {Roos}, M. 1975, Computer Physics Communications, 10, 343

\bibitem[{{Kadler} {et~al.}(2016){Kadler}, {Krau{\ss}}, {Mannheim}, {Ojha},
  {M{\"u}ller}, {Schulz}, {Anton}, {Baumgartner}, {Beuchert}, {Buson},
  {Carpenter}, {Eberl}, {Edwards}, {Eisenacher Glawion}, {Els{\"a}sser},
  {Gehrels}, {Gr{\"a}fe}, {Gulyaev}, {Hase}, {Horiuchi}, {James}, {Kappes},
  {Kappes}, {Katz}, {Kreikenbohm}, {Kreter}, {Kreykenbohm}, {Langejahn},
  {Leiter}, {Litzinger}, {Longo}, {Lovell}, {McEnery}, {Natusch}, {Phillips},
  {Pl{\"o}tz}, {Quick}, {Ros}, {Stecker}, {Steinbring}, {Stevens}, {Thompson},
  {Tr{\"u}stedt}, {Tzioumis}, {Weston}, {Wilms}, \&
  {Zensus}}]{2016NatPh..12..807K}
{Kadler}, M., {Krau{\ss}}, F., {Mannheim}, K., {et~al.} 2016, Nature Physics,
  12, 807

\bibitem[{Klinger {et~al.}(2023)Klinger, Rudolph, Rodrigues, Yuan,
  de~Clairfontaine, Fedynitch, Winter, Pohl, \& Gao}]{klinger2023am3}
Klinger, M., Rudolph, A., Rodrigues, X., {et~al.} 2023, AM$^3$: An Open-Source
  Tool for Time-Dependent Lepto-Hadronic Modeling of Astrophysical Sources

\bibitem[{Li {et~al.}(2003)Li, Filippenko, Chornock, \& Jha}]{Li_2003}
Li, W., Filippenko, A.~V., Chornock, R., \& Jha, S. 2003, Publications of the
  Astronomical Society of the Pacific, 115, 844

\bibitem[{{Lindfors} {et~al.}(2021){Lindfors}, {Hovatta}, {Pursimo}, {Pinter},
  {Warren}, {Andradi-Brown}, {Xie}, {Banados}, {Fugazza}, \&
  {Molinari}}]{2021ATel15136....1L}
{Lindfors}, E., {Hovatta}, T., {Pursimo}, T., {et~al.} 2021, The Astronomer's
  Telegram, 15136, 1

\bibitem[{{Lister} {et~al.}(2018){Lister}, {Aller}, {Aller}, {Hodge}, {Homan},
  {Kovalev}, {Pushkarev}, \& {Savolainen}}]{2018ApJS..234...12L}
{Lister}, M.~L., {Aller}, M.~F., {Aller}, H.~D., {et~al.} 2018, \apjs, 234, 12

\bibitem[{Malecki(2024)}]{universe10020053}
Malecki, P. 2024, Universe, 10

\bibitem[{{Meyer} {et~al.}(2019){Meyer}, {Scargle}, \&
  {Blandford}}]{2019ApJ...877...39M}
{Meyer}, M., {Scargle}, J.~D., \& {Blandford}, R.~D. 2019, \apj, 877, 39

\bibitem[{{Nilsson} {et~al.}(2018){Nilsson}, {Lindfors}, {Takalo}, {Reinthal},
  {Berdyugin}, {Sillanp{\"a}{\"a}}, {Ciprini}, {Halkola}, {Hein{\"a}m{\"a}ki},
  {Hovatta}, {Kadenius}, {Nurmi}, {Ostorero}, {Pasanen}, {Rekola}, {Saarinen},
  {Sainio}, {Tuominen}, {Villforth}, {Vornanen}, \& {Zaprudin}}]{nilsson_kva}
{Nilsson}, K., {Lindfors}, E., {Takalo}, L.~O., {et~al.} 2018, \aap, 620, A185

\bibitem[{{Nilsson} {et~al.}(2012){Nilsson}, {Pursimo}, {Villforth},
  {Lindfors}, {Takalo}, \& {Sillanp{\"a}{\"a}}}]{2012A&A...547A...1N}
{Nilsson}, K., {Pursimo}, T., {Villforth}, C., {et~al.} 2012, \aap, 547, A1

\bibitem[{Nilsson {et~al.}(2012)Nilsson, Pursimo, Villforth, Lindfors, Takalo,
  \& Sillanpää}]{Nilsson2012}
Nilsson, K., Pursimo, T., Villforth, C., {et~al.} 2012, Astronomy {\&}
  Astrophysics, 547, A1

\bibitem[{{Oikonomou} {et~al.}(2019){Oikonomou}, {Murase}, {Padovani},
  {Resconi}, \& {M{\'e}sz{\'a}ros}}]{2019MNRAS.489.4347O}
{Oikonomou}, F., {Murase}, K., {Padovani}, P., {Resconi}, E., \&
  {M{\'e}sz{\'a}ros}, P. 2019, \mnras, 489, 4347

\bibitem[{{Petkov} {et~al.}(2021){Petkov}, {Novoseltsev}, {Novoseltseva}, \&
  {Baksan Underground Scintillation Telescope Group}}]{2021ATel15143....1P}
{Petkov}, V.~B., {Novoseltsev}, Y.~F., {Novoseltseva}, R.~V., \& {Baksan
  Underground Scintillation Telescope Group}. 2021, The Astronomer's Telegram,
  15143, 1

\bibitem[{{Petropoulou} {et~al.}(2020{\natexlab{a}}){Petropoulou}, {Murase},
  {Santander}, {Buson}, {Tohuvavohu}, {Kawamuro}, {Vasilopoulos}, {Negoro},
  {Ueda}, {Siegel}, {Keivani}, {Kawai}, {Mastichiadis}, \&
  {Dimitrakoudis}}]{2020ApJ...891..115P}
{Petropoulou}, M., {Murase}, K., {Santander}, M., {et~al.} 2020{\natexlab{a}},
  \apj, 891, 115

\bibitem[{{Petropoulou} {et~al.}(2020{\natexlab{b}}){Petropoulou}, {Oikonomou},
  {Mastichiadis}, {Murase}, {Padovani}, {Vasilopoulos}, \&
  {Giommi}}]{2020ApJ...899..113P}
{Petropoulou}, M., {Oikonomou}, F., {Mastichiadis}, A., {et~al.}
  2020{\natexlab{b}}, \apj, 899, 113

\bibitem[{{Plavin} {et~al.}(2021){Plavin}, {Kovalev}, {Kovalev}, \&
  {Troitsky}}]{2021ApJ...908..157P}
{Plavin}, A.~V., {Kovalev}, Y.~Y., {Kovalev}, Y.~A., \& {Troitsky}, S.~V. 2021,
  \apj, 908, 157

\bibitem[{{Plavin} {et~al.}(2023){Plavin}, {Kovalev}, {Kovalev}, \&
  {Troitsky}}]{2023MNRAS.523.1799P}
{Plavin}, A.~V., {Kovalev}, Y.~Y., {Kovalev}, Y.~A., \& {Troitsky}, S.~V. 2023,
  \mnras, 523, 1799

\bibitem[{Prince {et~al.}(2023)Prince, Das, Gupta, Majumdar, \&
  Czerny}]{10.1093/mnras/stad3804}
Prince, R., Das, S., Gupta, N., Majumdar, P., \& Czerny, B. 2023, Monthly
  Notices of the Royal Astronomical Society, 527, 8746

\bibitem[{Rodrigues {et~al.}(2021)Rodrigues, Garrappa, Gao, Paliya,
  Franckowiak, \& Winter}]{Rodrigues_2021}
Rodrigues, X., Garrappa, S., Gao, S., {et~al.} 2021, The Astrophysical Journal,
  912, 54

\bibitem[{{Rodrigues} {et~al.}(2024{\natexlab{a}}){Rodrigues}, {Karl},
  {Padovani}, {Giommi}, {Paiano}, {Falomo}, {Petropoulou}, \&
  {Oikonomou}}]{2024arXiv240606667R}
{Rodrigues}, X., {Karl}, M., {Padovani}, P., {et~al.} 2024{\natexlab{a}}, arXiv
  e-prints, arXiv:2406.06667

\bibitem[{{Rodrigues} {et~al.}(2024{\natexlab{b}}){Rodrigues}, {Paliya},
  {Garrappa}, {Omeliukh}, {Franckowiak}, \&
  {Winter}}]{rodrigues2023leptohadronic}
{Rodrigues}, X., {Paliya}, V.~S., {Garrappa}, S., {et~al.} 2024{\natexlab{b}},
  \aap, 681, A119

\bibitem[{{Roming} {et~al.}(2005){Roming}, {Kennedy}, {Mason}, {Nousek}, {Ahr},
  {Bingham}, {Broos}, {Carter}, {Hancock}, {Huckle}, {Hunsberger}, {Kawakami},
  {Killough}, {Koch}, {McLelland}, {Smith}, {Smith}, {Soto}, {Boyd},
  {Breeveld}, {Holland}, {Ivanushkina}, {Pryzby}, {Still}, \&
  {Stock}}]{2005SSRv..120...95R}
{Roming}, P. W.~A., {Kennedy}, T.~E., {Mason}, K.~O., {et~al.} 2005, \ssr, 120,
  95

\bibitem[{Sahakyan {et~al.}(2022)Sahakyan, Giommi, Padovani, Petropoulou,
  Bégué, Boccardi, \& Gasparyan}]{10.1093/mnras/stac3607}
Sahakyan, N., Giommi, P., Padovani, P., {et~al.} 2022, Monthly Notices of the
  Royal Astronomical Society, 519, 1396

\bibitem[{{Scargle} {et~al.}(2013){Scargle}, {Norris}, {Jackson}, \&
  {Chiang}}]{ScargleBB}
{Scargle}, J.~D., {Norris}, J.~P., {Jackson}, B., \& {Chiang}, J. 2013, \apj,
  764, 167

\bibitem[{{Schlafly} \& {Finkbeiner}(2011)}]{2011ApJSchlafly}
{Schlafly}, E.~F. \& {Finkbeiner}, D.~P. 2011, \apj, 737, 103

\bibitem[{{Stahl} {et~al.}(2019){Stahl}, {Zheng}, {de Jaeger}, {Filippenko},
  {Bigley}, {Blanchard}, {Blanchard}, {Brink}, {Cargill}, {Casper}, {Channa},
  {Choi}, {Choksi}, {Chu}, {Clubb}, {Cohen}, {Ellison}, {Falcon}, {Fazeli},
  {Fuller}, {Ganeshalingam}, {Gates}, {Gould}, {Halevi}, {Hayakawa},
  {Hestenes}, {Jeffers}, {Joubert}, {Kandrashoff}, {Kim}, {Kim}, {Kislak},
  {Kleiser}, {Kong}, {de Kouchkovsky}, {Krishnan}, {Kumar}, {Leja}, {Leonard},
  {Li}, {Li}, {Lu}, {Mason}, {Molloy}, {Pina}, {Rex}, {Ross}, {Stegman},
  {Tang}, {Thrasher}, {Wang}, {Wilkins}, {Yuk}, {Yunus}, \&
  {Zhang}}]{2019MNRAS.490.3882S}
{Stahl}, B.~E., {Zheng}, W., {de Jaeger}, T., {et~al.} 2019, \mnras, 490, 3882

\bibitem[{{Strotjohann} {et~al.}(2019){Strotjohann}, {Kowalski}, \&
  {Franckowiak}}]{2019A&A...622L...9S}
{Strotjohann}, N.~L., {Kowalski}, M., \& {Franckowiak}, A. 2019, \aap, 622, L9

\bibitem[{{Ter\"asranta} {et~al.}(1998){Ter\"asranta}, {Tornikoski}, {Mujunen},
  {Karlamaa}, {Valtonen}, {Henelius}, {Urpo}, {Lainela}, {Pursimo}, {Nilsson},
  {Wiren}, {Laehteenmaeki}, {Korpi}, {Rekola}, {Heinaemaeki}, {Hanski},
  {Nurmi}, {Kokkonen}, {Keinaenen}, {Joutsamo}, {Oksanen}, {Pietilae},
  {Valtaoja}, {Valtonen}, \& {Koenoenen}}]{1998A&AS..132..305T}
{Ter\"asranta}, H., {Tornikoski}, M., {Mujunen}, A., {et~al.} 1998, \aaps, 132,
  305

\bibitem[{{Tomsick} {et~al.}(2019){Tomsick}, {Zoglauer}, {Sleator}, {Lazar},
  {Beechert}, {Boggs}, {Roberts}, {Siegert}, {Lowell}, {Wulf}, {Grove},
  {Phlips}, {Brandt}, {Smale}, {Kierans}, {Burns}, {Hartmann}, {Leising},
  {Ajello}, {Fryer}, {Amman}, {Chang}, {Jean}, \& {von
  Ballmoos}}]{2019BAAS...51g..98T}
{Tomsick}, J., {Zoglauer}, A., {Sleator}, C., {et~al.} 2019, in Bulletin of the
  American Astronomical Society, Vol.~51, 98

\bibitem[{{Tonry} {et~al.}(2012){Tonry}, {Stubbs}, {Lykke}, {Doherty},
  {Shivvers}, {Burgett}, {Chambers}, {Hodapp}, {Kaiser}, {Kudritzki},
  {Magnier}, {Morgan}, {Price}, \& {Wainscoat}}]{2012ApJ...750...99T}
{Tonry}, J.~L., {Stubbs}, C.~W., {Lykke}, K.~R., {et~al.} 2012, \apj, 750, 99

\bibitem[{{Willingale} {et~al.}(2013){Willingale}, {Starling}, {Beardmore},
  {Tanvir}, \& {O'Brien}}]{2013MNRAS.431..394W}
{Willingale}, R., {Starling}, R.~L.~C., {Beardmore}, A.~P., {Tanvir}, N.~R., \&
  {O'Brien}, P.~T. 2013, \mnras, 431, 394

\bibitem[{{Zech} \& {Lemoine}(2021)}]{2021A&A...654A..96Z}
{Zech}, A. \& {Lemoine}, M. 2021, \aap, 654, A96

\bibitem[{{Zhang} {et~al.}(2024){Zhang}, {B{\"o}ttcher}, \&
  {Liodakis}}]{2024ApJ...967...93Z}
{Zhang}, H., {B{\"o}ttcher}, M., \& {Liodakis}, I. 2024, \apj, 967, 93

\bibitem[{{Zhang} {et~al.}(2016){Zhang}, {Feroci}, {Santangelo}, {Dong},
  {Feng}, {Lu}, {Nandra}, {Wang}, {Zhang}, {Bozzo}, {Brandt}, {De Rosa}, {Gou},
  {Hernanz}, {van der Klis}, {Li}, {Liu}, {Orleanski}, {Pareschi}, {Pohl},
  {Poutanen}, {Qu}, {Schanne}, {Stella}, {Uttley}, {Watts}, {Xu}, {Yu}, {in 't
  Zand}, {Zane}, {Alvarez}, {Amati}, {Baldini}, {Bambi}, {Basso},
  {Bhattacharyya S.}, {}, {Belloni}, {Bellutti}, {Bianchi}, {Brez}, {Bursa},
  {Burwitz}, {Budtz-J{\o}rgensen}, {Caiazzo}, {Campana}, {Cao}, {Casella},
  {Chen}, {Chen}, {Chen}, {Chen}, {Chen}, {Chen}, {Civitani}, {Coti Zelati},
  {Cui}, {Cui}, {Dai}, {Del Monte}, {de Martino}, {Di Cosimo}, {Diebold},
  {Dovciak}, {Donnarumma}, {Doroshenko}, {Esposito}, {Evangelista}, {Favre},
  {Friedrich}, {Fuschino}, {Galvez}, {Gao}, {Ge}, {Gevin}, {Goetz}, {Han},
  {Heyl}, {Horak}, {Hu}, {Huang}, {Huang}, {Hudec}, {Huppenkothen}, {Israel},
  {Ingram}, {Karas}, {Karelin}, {Jenke}, {Ji}, {Korpela}, {Kunneriath},
  {Labanti}, {Li}, {Li}, {Li}, {Liang}, {Limousin}, {Lin}, {Ling}, {Liu},
  {Liu}, {Liu}, {Lu}, {Lund}, {Lai}, {Luo}, {Luo}, {Ma}, {Mahmoodifar},
  {Marisaldi}, {Martindale}, {Meidinger}, {Men}, {Michalska}, {Mignani},
  {Minuti}, {Motta}, {Muleri}, {Neilsen}, {Orlandini}, {Pan}, {Patruno},
  {Perinati}, {Picciotto}, {Piemonte}, {Pinchera}, {Rachevski A.}, {Rapisarda},
  {Rea}, {Rossi}, {Rubini}, {Sala}, {Shu}, {Sgro}, {Shen}, {Soffitta}, {Song},
  {Spandre}, {Stratta}, {Strohmayer}, {Sun}, {Svoboda}, {Tagliaferri},
  {Tenzer}, {Hong}, {Taverna}, {Torok}, {Turolla}, {Vacchi}, {Wang}, {Walton},
  {Wang}, {Wang}, {Wang}, {Wang}, {Weng}, {Wilms}, {Winter}, {Wu}, {Wu},
  {Xiong}, {Xu}, {Xue}, {Yan}, {Yang}, {Yang}, {Yang}, {Yuan}, {Yuan}, {Yuan},
  {Zampa}, {Zampa}, {Zdziarski}, {Zhang}, {Zhang}, {Zhang}, {Zhang}, {Zhang},
  {Zhang}, {Zheng}, {Zhou}, \& {Zhou X.~L.}}]{2016SPIE.9905E..1QZ}
{Zhang}, S.~N., {Feroci}, M., {Santangelo}, A., {et~al.} 2016, in Society of
  Photo-Optical Instrumentation Engineers (SPIE) Conference Series, Vol. 9905,
  Space Telescopes and Instrumentation 2016: Ultraviolet to Gamma Ray, ed.
  J.-W.~A. {den Herder}, T.~{Takahashi}, \& M.~{Bautz}, 99051Q

\end{thebibliography}

\clearpage

\begin{appendix}

\section{Tables of parameter space boundaries and model parameters}

For the search of the best-fit parameters of leptonic one-zone models, a grid-scan method was used. The region of parameter space that was studied is defined by the boundaries presented in Table \ref{tab:lep_parspace}. The intervals where equally divided into 10 points (including the boundaries) in linear space for $B'$, $\Gamma_{\textrm{b}}$, and $\alpha_e$, and in decimal logarithmic space for all other parameters. 

\begin{table}[H]
    \centering
    \caption{List of leptonic model parameters and boundaries of the parameter space.}
    \begin{tabular}{ll}
    \toprule
    Parameter & Value range\\
    \midrule
    $R^\prime_\mathrm{blob}$, cm & [$10^{15}$, $10^{17.5}$]\\
    $B^\prime$, gauss & [0.1, 5]\\
    $\Gamma_\mathrm{b}$ & [3.0, 30.0]\\
    $\gamma_\mathrm{e}^{\prime\mathrm{min}}$ & [$10^{3.0}$, $10^{3.95}$]\\
    $\gamma_\mathrm{e}^{\prime\mathrm{max}}$ & [$10^{4}$, $10^{5}$]\\
    $\alpha_\mathrm{e}$ & [0.5, 3.5]\\
    $L^\prime_\mathrm{e}$ / erg s$^{-1}$ & [$10^{42}$,$10^{47}$]\\
    \bottomrule
    \end{tabular}
    
    \label{tab:lep_parspace}
\end{table}

During the search for best-fit parameters of one-zone leptohadronic models (see Section \ref{method_lephad}), the grid scan of four hadronic parameters was performed. The subregion of the parameter space that was studied is defined by the boundaries presented in Table \ref{tab:had_parspace}. The intervals of $\gamma_\mathrm{p}^{\prime\mathrm{min}}$ and $\gamma_\mathrm{p}^{\prime\mathrm{max}}$ where equally divided into 30 points (including the boundaries) in decimal logarithmic space. The $L^\prime_\mathrm{p}$ interval was divided into 29 points (30 together with zero) in decimal logarithmic space. Five equally distant in linear space points from the range of $\alpha_\mathrm{p}$ were selected including boundaries. The unphysical combinations where $\gamma_\mathrm{p}^{\prime\mathrm{min}} > \gamma_\mathrm{p}^{\prime\mathrm{max}}$ were excluded.

\begin{table}[h]
    \centering
    \caption{List of hadronic model parameters and boundaries of the parameter space.}
    \begin{tabular}{ll}
    \toprule
    Parameter & Value range\\
    \midrule
    $\gamma_\mathrm{p}^{\prime\mathrm{min}}$ & [$10^{1}$, $10^{6}$]\\
    $\gamma_\mathrm{p}^{\prime\mathrm{max}}$ & [$10^{1}$, $10^{9}$]\\
    $\alpha_\mathrm{p}$ & [1.0, 3.0]\\
    $L^\prime_\mathrm{p}$ / erg s$^{-1}$ & [$10^{40}$,$10^{48}$] $\cup \{0\}$\\
    \bottomrule
    \end{tabular}
    
    \label{tab:had_parspace}
\end{table}

Table \ref{tab:nu_degeneracy} shows values of hadronic parameters of the models presented in Fig.\ref{fig:nu_spectra}. The model number corresponds to the legend of  Fig. \ref{fig:nu_spectra}.

Table \ref{table:lephadr_pars} presents the best-fit parameter values of the leptohadronic models under the condition of neutrino rate maximization. They correspond to the models shown in Fig. \ref{fig:lephad_models}.

\section{Paramater space of one-zone leptonic models} \label{appB}

Similarly to Fig. \ref{fig:sed2_fast_cornerplots}, Figures \ref{fig:sed2_slow_cornerplots}--\ref{fig:sed4_slow_cornerplots} show the value of reduced $\chi^2$ for variations of every pair of leptonic model parameters when the remaining parameters are fixed to the best fit. The best-fit value is marked with a pink circle in all plots.

\begin{figure}[htpb!]
\centering
 \includegraphics[width=0.5\textwidth]{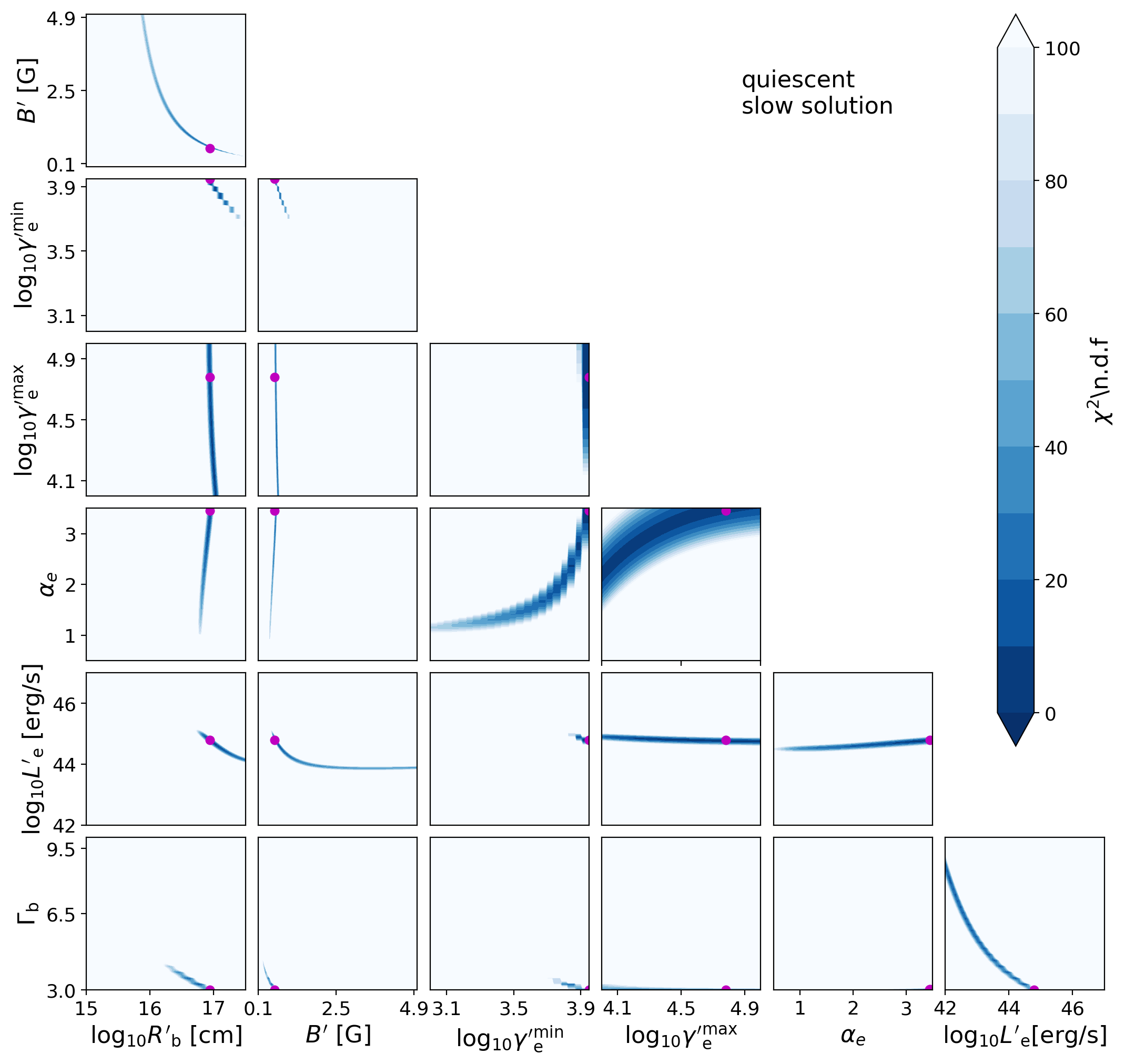}
  \caption{Values of the reduced $\chi^2$ around best-fit slow solution for the quiescent state SED.}
     \label{fig:sed1_slow_cornerplots}
\end{figure}

\begin{figure}[htpb!]
\centering
 \includegraphics[width=0.5\textwidth]{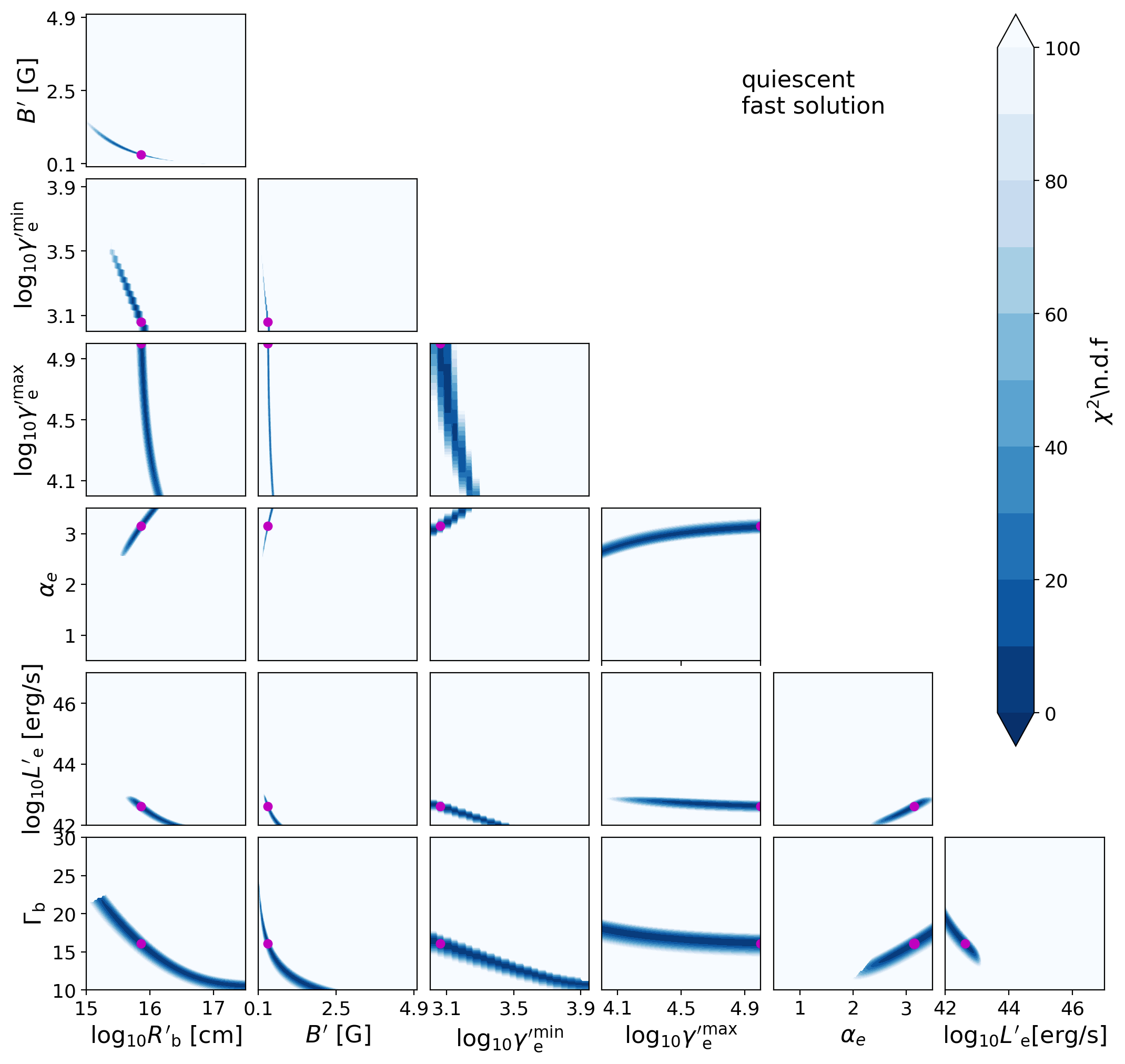}
  \caption{Values of the reduced $\chi^2$ around best-fit fast solution for the quiescent state SED.}
     \label{fig:sed1_fast_cornerplots}
\end{figure}

\begin{figure}[htpb!]
\centering
 \includegraphics[width=0.5\textwidth]{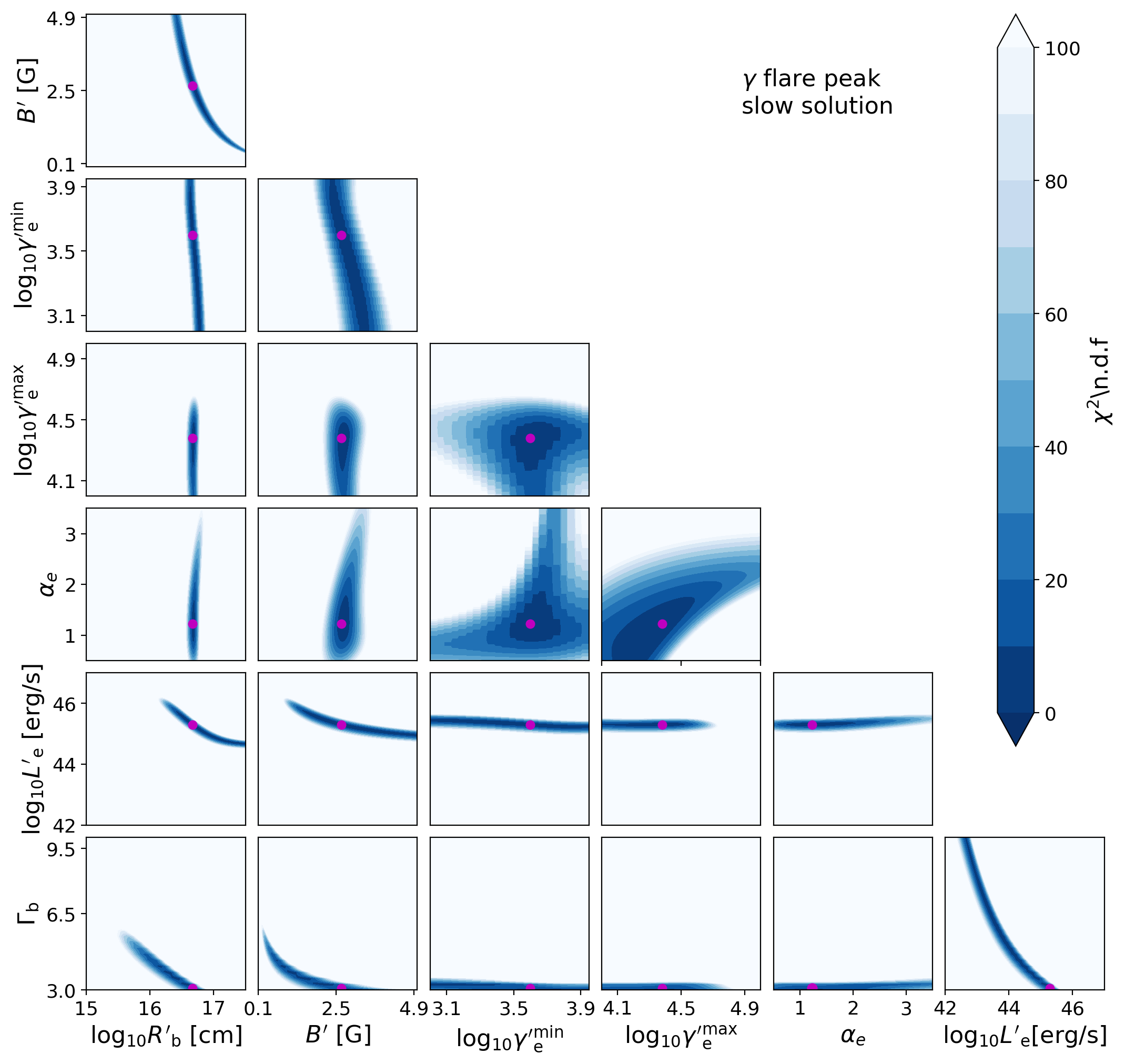}
  \caption{Values of the reduced $\chi^2$ around best-fit slow solution for the $\gamma$ ray flare peak SED.}
     \label{fig:sed3_slow_cornerplots}
\end{figure}

\begin{figure}[htpb!]
\centering
 \includegraphics[width=0.5\textwidth]{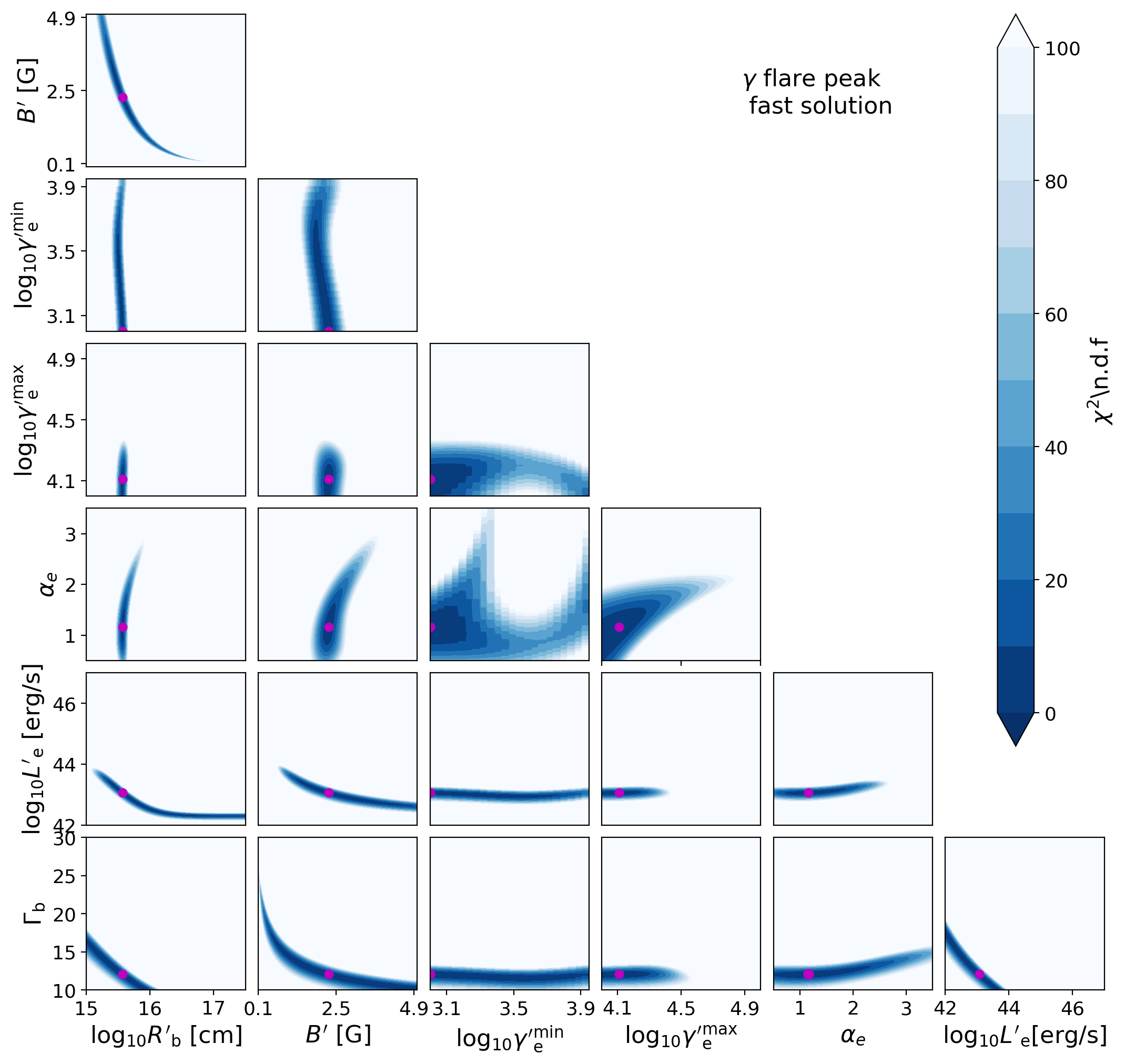}
  \caption{Values of the reduced $\chi^2$ around best-fit fast solution for the $\gamma$ ray flare peak SED.}
     \label{fig:sed3_fast_cornerplots}
\end{figure}

\begin{figure}[htpb!]
\centering
 \includegraphics[width=0.5\textwidth]{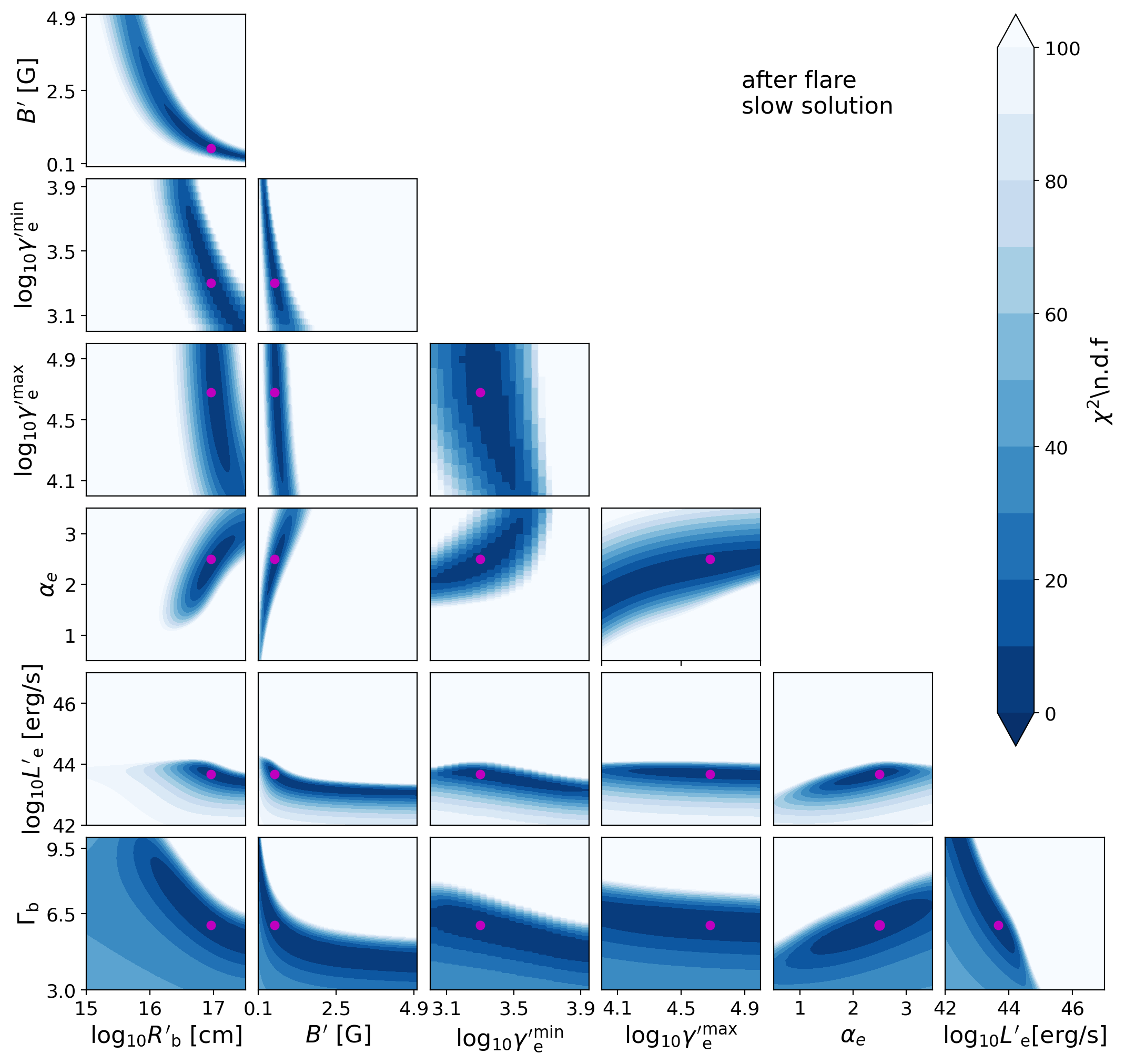}
  \caption{Values of the reduced $\chi^2$ around best-fit slow solution for the after-flare SED.}
     \label{fig:sed4_slow_cornerplots}
\end{figure}

\begin{figure}[htpb!]
\centering
 \includegraphics[width=0.5\textwidth]{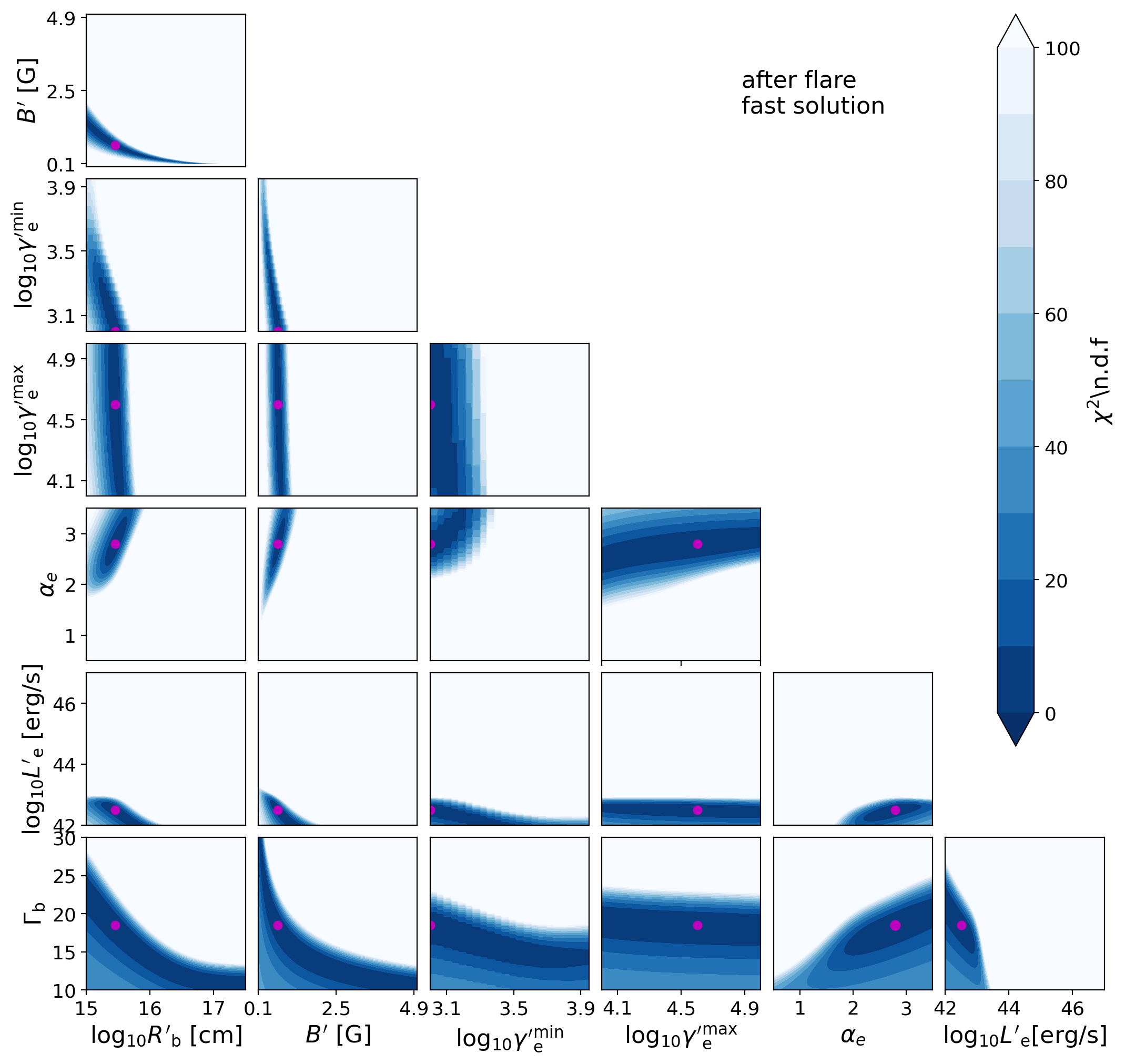}
  \caption{Values of the reduced $\chi^2$ around best-fit fast solution for the after-flare SED.}
     \label{fig:sed4_fast_cornerplots}
\end{figure}

\begin{figure}[htpb!]
\centering
 \includegraphics[width=0.5\textwidth]{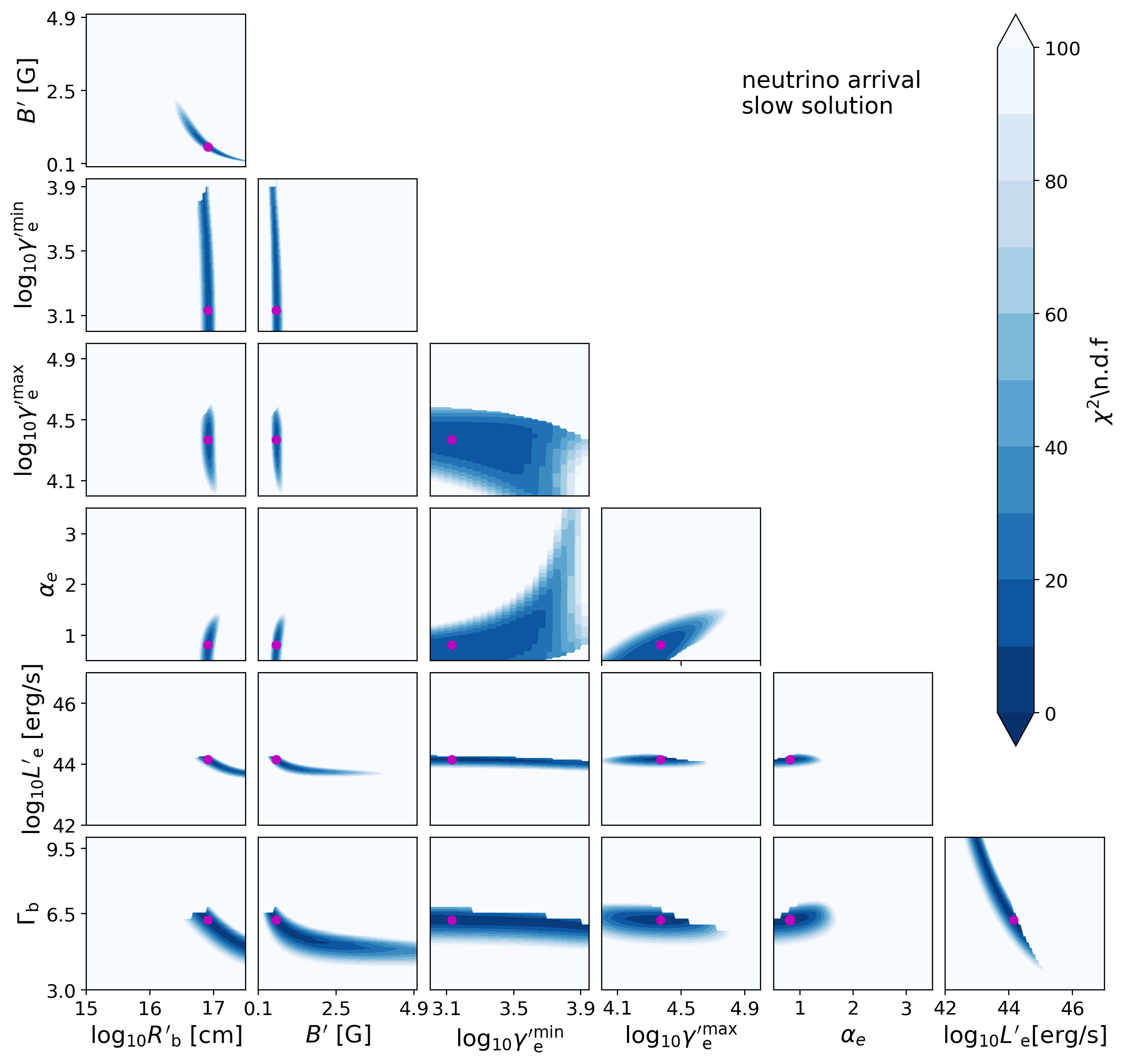}
  \caption{Values of the reduced $\chi^2$ around best-fit slow solution for the neutrino arrival SED.}
     \label{fig:sed2_slow_cornerplots}
\end{figure}

\end{appendix}

\end{document}